\documentclass{emulateapj}
\usepackage{epsfig,natbib}
\usepackage{graphicx,hyperref,amsmath}
\newcommand{\ms}{\mbox{m\,s$^{-1}~$}}

\newcommand{\mse}{\mbox{m\,s$^{-1}$}}

\newcommand{\rsun}{R$_{\odot}~$}

\newcommand{\lsune}{L$_{\odot}$}

\newcommand{\mearth}{$M_\earth$~}
\newcommand{\mearthe}{$M_\earth$}

\newcommand{\rearthe}{$R_\earth$}

\newcommand{\mstar}{\ensuremath{M_{\star}}}
\newcommand{\rstar}{\ensuremath{R_{\star}}}

\newcommand{\feh}{\ensuremath{[\mbox{Fe}/\mbox{H}]}}

\newcommand{\rphk}{\ensuremath{R'_{\mbox{\scriptsize HK}}}}
\newcommand{\shk}{\ensuremath{S_{\mbox{\scriptsize HK}}}}
\newcommand{\lrphk}{\ensuremath{\log{\rphk}}}
\newcommand{\caii}{\ion{Ca}{2} H \& K}

\newcommand{\msini}{\ensuremath{M \sin i}}

\newcommand{\teff}{\ensuremath{T_{\rm eff}}}

\newcommand{\logg}{\ensuremath{\log{g}}}
\newcommand{\vsini}{\ensuremath{v \sin{i}}}

\newcommand{\bjdtdb}{\ensuremath{\rm {BJD_{TDB}}}}

\newcommand{\specmatch}{\texttt{SpecMatch}\ } 
\newcommand{\mfourtwo}{$15.4 \pm 2.4$ \mearthe} 
\newcommand{\monesix}{$12.9 \pm 1.6$ \mearthe} 
\newcommand{\monefour}{$25 \pm 2$ \mearthe} 

\shortauthors{Fulton {et~al.}}
\shorttitle{Three Neptunes}
\submitted{In Preparation for ApJ}
\begin{document}
\pagenumbering{arabic}

\title{Three Temperate Neptunes Orbiting Nearby Stars\altaffilmark{1}}
\author{
Benjamin J.\ Fulton\altaffilmark{2,17,*},
Andrew W. Howard\altaffilmark{2},
Lauren M.\ Weiss\altaffilmark{3,18},
Evan Sinukoff\altaffilmark{2,19},
Erik A.\ Petigura\altaffilmark{5,20},
Howard Isaacson\altaffilmark{3},
Lea Hirsch\altaffilmark{3},
Geoffrey W. Marcy\altaffilmark{3},
Gregory W. Henry\altaffilmark{4},
Samuel K. Grunblatt\altaffilmark{2},
Daniel Huber\altaffilmark{7,9,10},
Kaspar von Braun\altaffilmark{6},
Tabetha S. Boyajian\altaffilmark{8},
Stephen R. Kane\altaffilmark{11},
Justin Wittrock\altaffilmark{11},
Elliott P. Horch\altaffilmark{12},
David R. Ciardi\altaffilmark{13},
Steve B. Howell\altaffilmark{14},
Jason T. Wright\altaffilmark{15,16},
Eric B. Ford\altaffilmark{15,16}
}

\altaffiltext{1}{Based on observations obtained at the W.\,M.\,Keck Observatory, 
                      which is operated jointly by the University of California and the 
                      California Institute of Technology.  
                      Keck time was granted for this project by the University of Hawai`i, the University of California, and NASA.
                      }  
\altaffiltext{2}{Institute for Astronomy, University of Hawai`i, 2680 Woodlawn Drive, Honolulu, HI 96822, USA} 
\altaffiltext{3}{Department of Astronomy, University of California, Berkeley, CA 94720, USA}
\altaffiltext{4}{Center of Excellence in Information Systems, Tennessee State University, 3500 John A. Merritt Blvd., Box 9501, Nashville, TN 37209}
\altaffiltext{5}{California Institute of Technology, Pasadena, California, U.S.A.}
\altaffiltext{6}{Lowell Observatory, Flagstaff, AZ 86001, USA}
\altaffiltext{7}{SETI Institute, 189 Bernardo Avenue, Mountain View, CA 94043, USA}
\altaffiltext{8}{Yale University, New Haven, CT 06520, USA}
\altaffiltext{9}{Sydney Institute for Astronomy (SIfA), School of Physics, University of Sydney, NSW 2006}
\altaffiltext{10}{Stellar Astrophysics Centre, Department of Physics and Astronomy, Aarhus University, Ny Munkegade 120, DK-8000 Aarhus C, Denmark}
\altaffiltext{11}{Department of Physics and Astronomy, San Francisco State University, 1600 Holloway Avenue, San Francisco, CA 94132, USA}
\altaffiltext{12}{Department of Physics, Southern Connecticut State University, New Haven, CT 06515, USA}
\altaffiltext{13}{NASA Exoplanet Science Institute, Caltech, MS 100-22, 770 South Wilson Avenue, Pasadena, CA 91125, USA}
\altaffiltext{14}{NASA Ames Research Center, Moffett Field, CA 94035, US}
\altaffiltext{15}{Center for Exoplanets and Habitable Worlds, The Pennsylvania State University, University Park, PA 16802}
\altaffiltext{16}{Department of Astronomy \& Astrophysics, The Pennsylvania State University, University Park, PA 16802}

\altaffiltext{17}{National Science Foundation Graduate Research Fellow}
\altaffiltext{18}{Ken and Gloria Levy Graduate Student Research Fellow}
\altaffiltext{19}{Natural Sciences and Engineering Research Council of Canada Graduate Student Fellow}
\altaffiltext{20}{Hubble Fellow}
\altaffiltext{*}{bfulton@hawaii.edu}

\begin{abstract}
We present the discovery of three modestly-irradiated, roughly Neptune-mass planets orbiting three nearby Solar-type stars. HD 42618 b has a minimum mass of \mfourtwo, a semi-major axis of 0.55 AU, an equilibrium temperature of 337 K, and is the first planet discovered to orbit the solar analogue host star, HD 42618. We also discover new planets orbiting the known exoplanet host stars HD 164922 and HD 143761 ($\rho$ CrB). The new planet orbiting HD 164922 has a minimum mass of \monesix\ and orbits interior to the previously known Jovian mass planet orbiting at 2.1 AU. HD 164922 c has a semi-major axis of 0.34 AU and an equilibrium temperature of 418 K. HD 143761 c orbits with a semi-major axis of  0.44 AU, has a minimum mass of \monefour, and is the warmest of the three new planets with an equilibrium temperature of 445 K. It orbits exterior to the previously known warm Jupiter in the system. A transit search using space-based CoRoT data and ground-based photometry from the Automated Photometric Telescopes (APTs) at Fairborn Observatory failed to detect any transits, but the precise, high-cadence APT photometry helped to disentangle planetary-reflex motion from stellar activity. These planets were discovered as part of an ongoing radial velocity survey of bright, nearby, chromospherically-inactive stars using the Automated Planet Finder (APF) telescope at Lick Observatory. The high-cadence APF data combined with nearly two decades of radial velocity data from Keck Observatory and gives unprecedented sensitivity to both short-period low-mass, and long-period intermediate-mass planets. 
\end{abstract}


\section{Introduction}
\label{sec:intro}
The mass function of extrasolar planets potentially offers rich clues to the processes that shape their growth and evolution.  This mass function is known to rise with decreasing planet mass based on the discovery and characterization of planets orbiting nearby stars \citep{Howard10,Mayor11}.  
Although the \emph{Kepler} mission discovered thousands of transiting exoplanets allowing for detailed characterization of the planet radius distribution \citep{Howard12,Fressin13,Petigura13b}, we have a much sparser sample of planets orbiting nearby stars.  We only know of 17 confirmed planets with measured minimum masses $\msini < 30 $ \mearth and orbital periods  $P > 75$ days\footnote{Based on a query of exoplanets.org on 28 June 2016 \citep{Han14}}.
%we still know of only 52 confirmed planets with minimum masses (\msini) less than 30 \mearth\ and orbital periods (P) greater than 75 days\footnote{Based on a query of exoplanets.org on 18 November 2015.}.
These low-mass, temperate planets reside in an region of parameter space that must be explored in order to understand the formation of the extremely abundant population of close-in super-Earths.

The discovery that the occurrence rate of Jovian planets increases at orbital distances of 1--3 AU \citep{Cumming08} has been suggested to be a sign that the ice line is important to the formation of Jovian planets. The increased abundance of solids in the protoplanetary disk beyond the ice-line is expected to speed up the coagulation of planetesimals and $\sim10$ \mearth cores that can undergo runaway gas accretion before the gas in the disk is dissipated \citep{Ida08}. 

The formation of close-in intermediate mass planets known as super-Earths or mini-Neptunes presents some challenges for planet formation theory. It was initially thought that such planets should either remain small terrestrial planets, or if they grow to large cores then they should quickly accrete substantial nebular gas and grow to be gas giants \citep{Ida04, Mordasini09}. Alternatively, planets that form in situ likely require either a protoplanetary disk that is much more massive than the minimum mass solar nebula \citep{Hansen12, Chiang13}, an extremely metal rich disk, or fine tuning of the formation timescales \citep{Lee14}. New models featuring gas-drag driven ``pebble accretion" which offers a mechanism to transport solids in the disc \citep{Chatterjee14}, or the delayed formation of super-Earth cores until the gas disk begins to dissipate \citep{Lee15} could facilitate \emph{in situ} formation of super-Earths and/or mini Neptunes.

Alternatively, migration from further out in the disk where there is plenty of material to form massive planet cores or super-Earths could explain the presence of these planets near their host stars. We would expect to see multi-planet systems in resonant chains if a slow, smooth migration were the dominant mechanism and this is not observed \citep{Veras12, Fabrycky14}. Another indication that migration is at play would be an increased occurrence rate of super-Earths at large orbital separations. The occurrence rates of super-Earths as a function of orbital period appears fairly flat to periods as long as $\sim$200 days \citep{Petigura13b}, however the sensitivity to long-period planets from transit surveys is very low. In addition, the planet radius distribution can be influenced by a variety of different factors including stellar irradiation, and thermal evolution \citep{Lopez14}. Only a very small percentage by mass of volatiles can significantly inflate a planet's radius and hide the fundamental properties of the planet that encode information about the formation mechanism \citep{Lopez13}. RV surveys are now starting to discover a statistically useful sample of super-Earth to Neptune mass planets at larger 
orbital separations and lower stellar irradiance that will help to map out the details of the mass function for long-period super-Earth to Neptune mass planets.

The Eta-Earth RV survey of nearby stars \citep{Howard10} was conducted using the HIRES spectrograph at Keck observatory \citep{Vogt94}. They searched for planets in a volume-limited sample of 166 nearby G and K dwarfs. With the catalogue of planets detected in this survey \citep{Howard09,Howard11a,Howard11b,Howard14}, and the completeness limits calculated for each star they were able to measure the occurrence rate of small, short-period planets and show that planets with masses of 3-30 \mearth are much more common than planets larger than 30 \mearthe. Although these low-mass planets are common, they are still very difficult to detect, requiring $>$100 measurements per star with $\lesssim2 \ms$ precision. The number of nearby stars for which large, high-precision radial velocity datasets exist is increasing thanks to the proliferation of dedicated and robotic radial velocity facilities such as the APF and MINERVA \citep{Swift15} and the ongoing long-term surveys from Keck, HARPS-N \citep{Motalebi15}, and HARPS \citep{Pepe04}.

We are currently using the APF telescope to conduct a RV survey of 51 of the brightest, and least chromospherically active, stars from the Eta-Earth survey. We capitalize on the robotic nature of the telescope to monitor the stars at high cadence for the entire four year duration of the survey.
This survey builds on the Eta-Earth Survey, but with improved Doppler precision due to the high observing cadence and larger number of measurements.
We will measure the occurrence rate and mass function of small planets in our local neighborhood using the new planets discovered by the APF-50 survey and the set of planets already known to orbit stars in our sample. With a larger sample of planets with measured masses in the 3-30 \mearth range we will measure the mass function for small planets with higher mass resolution. Combining the mass function from this survey with the size distribution from \emph{Kepler}, we will probe the density and core mass properties of super-Earths to inform formation theories of the galaxy's most abundant planets.

%Due to their brightness, nearby stars are the easiest stars in the sky to characterize in great detail. With the relatively large photon flux received from the star we can obtain extremely high precision radial velocities to fully characterize the planetary system down to very low mass planets, and extremely high precision photometry to identify transiting planets and stellar activity. The high precision photometry and/or velocities allow for the inference of the bulk stellar structure using asteroseismology. The improved knowledge of the stellar properties allows for a better understanding of the properties of the planets that orbit those stars. Planets that transit these stars will be the first, and ultimately most well characterized by transmission, emission, and direct spectroscopy.
Nearby G and K dwarf stars are observationally advantageous.  High SNR spectroscopy can often be obtained with relatively short exposure times, facilitating many precise time series RV measurements with sensitivity to planets with small Doppler amplitudes.  The stars can be characterized precisely using spectroscopy and asteroseismology (when available) that improve estimates of the star and planet properties.  The advantages of nearby, bright targets are critical for characterization of the planets atmospheres transmission, emission, and direct spectroscopy.

%It is critical to locate nearby planets in order to guide the next generation of direct imaging surveys. A Neptune size planet orbiting at 0.5 AU at a distance of 20 pc will be at a projected separation of 25 milliarcseconds during the periods of maximum elongation. This is right at the limit of advanced AO systems planned for TMT and E-ELT \citep{Macintosh06,Kasper10}. As radial velocity detections push out to lower mass planets are larger separations we will be building a target catalogue for direct atmospheric characterization using future 30 m class telescopes. 

Here we present the discovery of three roughly Neptune mass planets orbiting bright stars within 25 pc.
%using data from the Automated Planet Finder (APF) telescope at Lick observatory and Keck/HIRES \citep{Vogt94,Vogt14,Radovan14}. Two of these planets are sub-Neptune mass with \msini = \mfourtwo\ and \msini = \monesix, and the third is a super-Neptune with \msini = \monefour. These planets orbit G dwarf stars that are brighter than $V=7$. HD 42618 is considered a ``solar twin" due to its similarity in mass, radius, and composition to that of the Sun \citep{Ramirez12} and HD 42618 b is the first planet discovered orbiting this star. HD 42618 was also the target of the CoRoT mission to search for asteroseismic oscillations \citep{Barban13}. HD 143761 c  ($\rho$ CrB c) and HD 164922 c both join one previously known Jovian mass planet in each of those systems \citep{Noyes97, Butler06}. These planets contribute to the relatively limited knowledge of long-period, Neptune-mass systems in the local solar neighborhood. 
This paper is structured as follows. Our observational setup and RV measurements are described in Section \ref{sec:measurements}. In Section \ref{sec:stellar_props} we discuss our derived stellar properties for each of the three stars and compare with previous literature studies. We describe our methods used to discover these planets in Section \ref{sec:discovery}. We describe our modeling procedure used to obtain the final adopted parameters and their associated uncertainties and our various tests to ensure that the signals are planetary in nature in Section \ref{sec:characterization}. We analyze photometry of each of the three systems in Section \ref{sec:photometry}, and conclude with a summary and discussion in Section \ref{sec:discussion}.

\section{Radial Velocity Measurements}
\label{sec:measurements}
%\subsection{Keck/HIRES Spectroscopy}
\label{sec:keckdata}
We collected 571 RV measurements of HD 42618 using Keck/HIRES \citep{Vogt94} and 35 measurements using the Levy spectrograph on the Automated Planet Finder \citep[APF,][]{Vogt14,Radovan14} over the past 19 years starting in 1996. For each RV measurement the starlight is passed through a cell of gaseous iodine that imprints a dense forest of molecular absorption lines onto the stellar spectrum and serves as both a wavelength and point spread function (PSF) reference. We also collected a single set of iodine-free observations of this star that was deconvolved with the instrumental PSF and used as a model of the intrinsic stellar spectrum. Each observation was forward modeled as the intrinsic stellar spectrum doppler shifted by an arbitrary amount, then multiplied by the transmission of iodine, and convolved with the instrumental PSF modeled as a sum of 13 Gaussians with fixed widths and positions but heights free to vary \citep{Butler96b}. The Levy slit-fed spectrograph also relies on an iodine cell for precise RVs and our observational setup is described in detail in \citet{Fulton15a}.

Our setup was identical for the three stars. We collected 328 Keck/HIRES measurements and 73 APF measurements for HD 164922 over the past 19 years. All of the APF measurements and 244 of the Keck measurements are new since the publication of \citet{Butler06}. For HD 143761 we obtained 519 RV measurements using Keck/HIRES and 157 measurements using the Levy spectrograph on the APF. The Keck observations of HD 143761 started in 2006 and the total observational baseline is 8 years. We do not include the Lick 3.0 m data from \citet{Noyes97} and \citet{Butler06} in our analysis. We find that including the Lick data does not significantly improve the uncertainties on any of the orbital parameters and it adds an additional source of systematic uncertainty that is less well characterized and understood.

%After the forward modeling process is complete we have an RV time series along with the best fit amplitudes of the 13 satellite Gaussians used to model the instrumental PSF corresponding to each RV measurement.
Our Doppler pipeline has been tuned in small ways over the years to improve RV precision.  Here we describe a new pipeline improvement that decorrelates the measured RVs with nuisance parameters from the model and spectrometer state parameters.  This decorrelation offers modest improvements in Doppler precision ($\sim$1 \ms) and is only applied to time series RV of stars for which the number of spectra greatly exceeds the number of potential decorrelation parameters.  The nuisance parameters in Doppler analysis include descriptions of the point spread function (PSF) over the spectral format and the wavelength solution.  The PSF is parameterized as a sum Gaussians with fixed widths and centers, but variable amplitudes. We also have a wealth of information about the weather and environment inside and outside the spectrograph extracted from the FITS headers of the raw APF spectra. Some environmental information is available in the FITS headers for the Keck data, but we have not yet implemented a system to extract these values for the tens of thousands of Keck spectra taken over the last 20 years. For Keck data we only include the nuisance parameters that are part of the forward modeling process. We clean the RVs of systematic trends by removing any correlations that these parameters show with the final RVs. We search for significant correlations of the RVs with all of the PSF parameters by calculating the Spearman rank correlation coefficient \citep{Spearman1904} and flag any parameters that show correlation coefficients greater than 0.1. The flagged parameters are included in a multivariate ordinary least squares linear regression using the \texttt{STATSMODELS}\footnote{https://pypi.python.org/pypi/statsmodels} package in Python. This model for RV as a function of all parameters included in the fit is then subtracted from the raw RVs. This process is done blindly in the planet discovery/identification phase but once planet candidates are identified in a given dataset we first model the system to find the best fit N-planet Keplerian model then perform the detrending procedure on the residuals to this model. The detection of all newly discovered planets in this work does not depend on this detrending and they are easily identified at high significance in either the detrended or non-detrended datasets. 

We reject measurements with low signal-to-noise ratios (SNR$<~60$ per pixel) and/or uncertainties more then 9 $\sigma$ larger then the median uncertainty, which results in the omission of $<$1\% of the data for each of the three stars. Since these stars are exceptionally bright we almost always collect three consecutive measurements in order to average out RV shifts caused by p-mode oscillations \citep{Dumusque11}. The three measurements are then binned together before the stellar jitter is added in quadrature during the modeling process (see Section \ref{sec:characterization}). This effectively reduces the weight of the three measurements to that of a single measurement, but averages out some of the astrophysical noise in the process and prevents time-correlated instrumental systematic noise from biasing the results. We also extract the \caii\ activity index (\shk) using the technique of \citet{Isaacson10} for every RV measurement on both Keck and APF, however there may be an arbitrary zero point offset in the \shk\ values between the Keck and APF values. The uncertainties for the \shk\ measurements are systematically limited to 0.002 for Keck and 0.004 for APF. This was estimated by measuring the standard deviation of all measurements of the extremely chromospherically quiet star, HD 10700. All RV measurements and the associated \shk\ values can be found in Table \ref{tab:rv}. We include only the detrended velocities in Table \ref{tab:rv} but the full set of environmental and PSF parameters for each observation along with the non-detrended velocities can be downloaded from \url{https://github.com/bjfultn/three\_neptunes}.

\begin{deluxetable}{cccccc}
\tabletypesize{\footnotesize}
\tablecaption{Radial Velocities}
\tablewidth{245pt}
\tablehead{ 
    \colhead{HD}   &   \colhead{\bjdtdb}               & \colhead{RV\tablenotemark{1}}  & \colhead{Unc.} & \colhead{Inst.\tablenotemark{2}} &  \colhead{\ensuremath{S_{\mbox{\scriptsize HK}}}\tablenotemark{3}} \\
    \colhead{}        &    \colhead{(-- 2440000)}  & \colhead{(\mse)}            & \colhead{(\mse)}       & \colhead{} & \colhead{}
}
\startdata
42618  &  2450366.126333 & \phn$+$2.85 & 1.12 & k & \nodata  \\
42618  &  2453694.093412 & \phn$+$1.17 & 1.13 & j & 0.161  \\
164922 &  2454777.744397 & \phn$-$9.19 & 1.04 & j & 0.152  \\
164922 &  2457267.662041 & \phn$-$4.43 & 1.90 & a & 0.145  \\
143761 &  2455455.762546 &    $-$41.98 & 1.27 & j & 0.149  \\
143761 &  2457292.685768 & \phn$-$8.08 & 2.92 & a & 0.135 
\enddata

\tablenotetext{}{(This table is available in its entirety in a machine-readable form in the online journal. A portion is shown here for guidance regarding its form and content.)}
\tablenotetext{1}{Zero point offsets between instruments have not been removed and must be fit as free parameters when analyzing this dataset}
\tablenotetext{2}{k = pre-upgrade Keck/HIRES, j = post-upgrade Keck/HIRES, a = APF}
\tablenotetext{3}{Uncertainties on \shk\ are 0.002 for all Keck measurements and 0.004 for all APF measurements}
\vspace{10pt}

\label{tab:rv}
\end{deluxetable}

\section{Stellar Properties}
\label{sec:stellar_props}

\subsection{HD 42618}
HD 42618, also known as HIP 29432 and Gl 3387, is a well studied solar analogue located at a distance of 23.5 pc \citep{vanLeeuwen07}. The star was not previously known to host any exoplanets. We analyzed 5 high SNR Keck-HIRES spectra (described below) using  \texttt{SpecMatch} \citep{Petigura15} to obtain the mean spectroscopic parameters listed in Table \ref{tab:stellar_params}. \specmatch uses trilinear interpolation to synthesize high resolution model spectra from the \citet{Coelho14} grid of models for any set of arbitrary stellar parameters (\teff, \logg, \feh, and \vsini) that are contained within the limits of the model grid. The interpolated models are then compared to the observed spectrum. We maximize the likelihood ($\mathcal{L}=e^{-\chi^{2}/2}$) to determine the optimal stellar parameters, where $\chi^2$ is summed over the extracted spectral pixels and normalized by the flux uncertainties.

Our spectral analysis is consistent with the results of \citet{Valenti05} who extracted spectroscopic parameters using Spectroscopy Made Easy (SME) and found $\teff=5747\pm44$ K, $\logg=4.43\pm0.06$, and $\feh=-0.11\pm0.03$. HD 42618 is chromospherically quiet with $\lrphk=-5.01$\citep{Isaacson10}. It was deemed to be a good solar-analog based on a very similar chemical abundance pattern to the Sun \citep{Morel13}. Those authors also derive $\teff=5765\pm17$ K, and $\logg=4.48\pm0.04$, consistent with our \specmatch results. \citet{Ramirez14} measured the fundamental parameters of HD 42618 differentially relative to the Sun which allowed them to obtain highly precise values for $\teff=5758\pm5$ K, $\logg=4.44\pm0.01$, and $\feh=-0.096\pm0.005$ that show good agreement with our results. It has also been noted that HD 42618 shows a low lithium abundance of $A_{\rm Li}=1.22$ \citep{Ramirez12} similar to that of the Sun \citep[$A_{\rm Li}=1.05\pm0.1$][]{Asplund09}. Our adopted stellar mass and radius for HD 42618 are based on the relations of \citet{Torres10} using our spectroscopic constraints on \teff, \logg, and \feh. HD 42618 was also target of the CoRoT mission \citep{Baglin09}, with a preliminary detection of solar-like oscillations presented by \citet{Barban13}. We performed an independent asteroseismic analysis of the CoRoT photometry (see Section \ref{sec:asteroseismology}), which yielded a mass and radius in agreement with our adopted values.
%HD 42618 was a target of the CoRoT mission \citep{Baglin09} to search for solar-like oscillations with a preliminary detection \citep{Barban13}. We performed a similar asteroseismic analysis of the CoRoT photometry (see Section \ref{sec:corot_phot}) to constrain the stellar mass and radius. We calculate the stellar mass and radius for HD 42618 using the relations of \citet{Torres10} and our spectroscopic constraints on \teff, \logg, and \feh.

\begin{deluxetable*}{lrrr}
\tabletypesize{\footnotesize}
\tablecaption{Adopted Stellar Properties}
\tablewidth{0.5\textwidth}
\tablehead{
  \colhead{Parameter}   & 
  \colhead{HD 42618} &
  \colhead{HD 164922} &
  \colhead{HD 143761} 
}
\startdata
Spectral type ~~~~~~&         G4V\tablenotemark{1} &                          G9V\tablenotemark{6}  &                        G0V\tablenotemark{8} \\
$B-V$ (mag) &                       0.657\tablenotemark{2} &    	                   0.800\tablenotemark{7}  &                      0.600  \\
$V$ (mag)   &                         6.839\tablenotemark{2}  &                       6.99\tablenotemark{7}  &                        5.41\tablenotemark{8} \\
$J$  (mag)  &                          5.701 $\pm$ 0.023\tablenotemark{3}  &  5.553 $\pm$ 0.026\tablenotemark{3} &                      4.09\tablenotemark{3} \\
$H$  (mag)  &                         5.385 $\pm$ 0.024\tablenotemark{3}  &  5.203 $\pm$ 0.017\tablenotemark{3}  &                      3.99\tablenotemark{3} \\
$K$ (mag)   &                         5.301 $\pm$ 0.020\tablenotemark{3} &   5.113 $\pm$ 0.020\tablenotemark{3}  &                      3.89 $\pm$ 0.05\tablenotemark{8} \\
Distance (pc) &                      23.50$\pm$0.30\tablenotemark{4}  &     22.13$\pm$0.27\tablenotemark{4}  &  17.236$\pm$0.024\tablenotemark{4} \\
$T_\mathrm{eff}$ (K) &          5727 $\pm$ 60  &                                     5293 $\pm$ 32  &                                   5627 $\pm$ 54\tablenotemark{11} \\
log\,$g$ (cgs) &                      4.44 $\pm$ 0.07 &                                    4.387 $\pm$ 0.014  &                                 4.121 $\pm$ 0.018\\
\feh\ (dex) &                            $-$0.09 $\pm$ 0.04 &                        $+$0.16 $\pm$ 0.05  &                                 $-$0.31 $\pm$ 0.05 \\
$v$\,sin\,$i$ (km\,s$^{-1}$) &  $\leq2$  &                                                 $\leq2$  &                                               $\leq2$ \\
$L_{\star}$ ($L_{\sun}$) &       0.98 $\pm$ 0.17 &                                0.703 $\pm$ 0.017  &                             1.706 $\pm$ 0.042\tablenotemark{11} \\
\mstar\ ($M_{\sun}$) &            $1.015 \pm 0.061$\tablenotemark{12} &            0.874 $\pm$ 0.012   &                             0.889 $\pm$ 0.030 \\
\rstar\ ($R_{\sun}$) &              0.999 $\pm$ 0.087\tablenotemark{12}  &           0.999 $\pm$ 0.017 &   							$1.3617 \pm 0.0262$\tablenotemark{11}  \\
\lrphk  &                                  $-$5.01\tablenotemark{5} &                      $-$5.06\tablenotemark{5} &                    $-$5.05\tablenotemark{5} \\
$S_\mathrm{HK}$  &              0.157\tablenotemark{5} &                         0.154\tablenotemark{5}  &                       0.150\tablenotemark{5} \\
\enddata

\tablenotetext{}{References - 
(1) \citet{Medhi07};
(2) \citet{Koen10};
(3) \citet{Cutri03};
(4) \citet{vanLeeuwen07};
(5) \citet{Isaacson10};
(6) \citet{Gray03};
(7) \citet{Butler06};
(8) \citet{vanBelle09};
(9) \citet{Noyes97};
(10) \citet{vanBelle09};
(11) \citet{vonBraun14};
(12) \citet{Torres10}
}

\label{tab:stellar_params}
\end{deluxetable*}
\vspace{20pt}

\subsection{HD 164922}
\label{sec:164922_star}
HD 164922, also known as HIP 88348 and Gl 9613, is a bright, chromospherically inactive \citep[$\lrphk=-5.06$,][]{Isaacson10} G9 V dwarf located 22.1 pc away \citep{vanLeeuwen07}. It was previously known to host a single Saturn-mass planet orbiting with a semi-major axis of 2.1 AU \citep{Butler06}. This target was one of several selected for more intensive long-term RV monitoring by Keck/HIRES based on both the stellar properties, and the mass and orbit of the previously detected planet making the system particularly well-suited for detecting additional low-mass planets. It was also on the Eta-Earth target list as part of a deep Doppler survey for low-mass planets \citep{Howard10}.
%We analyzed 5 high SNR Keck spectra using \specmatch to obtain spectroscopic parameters.

\begin{figure}[h]
         \centering
              \includegraphics[width=0.5\textwidth]{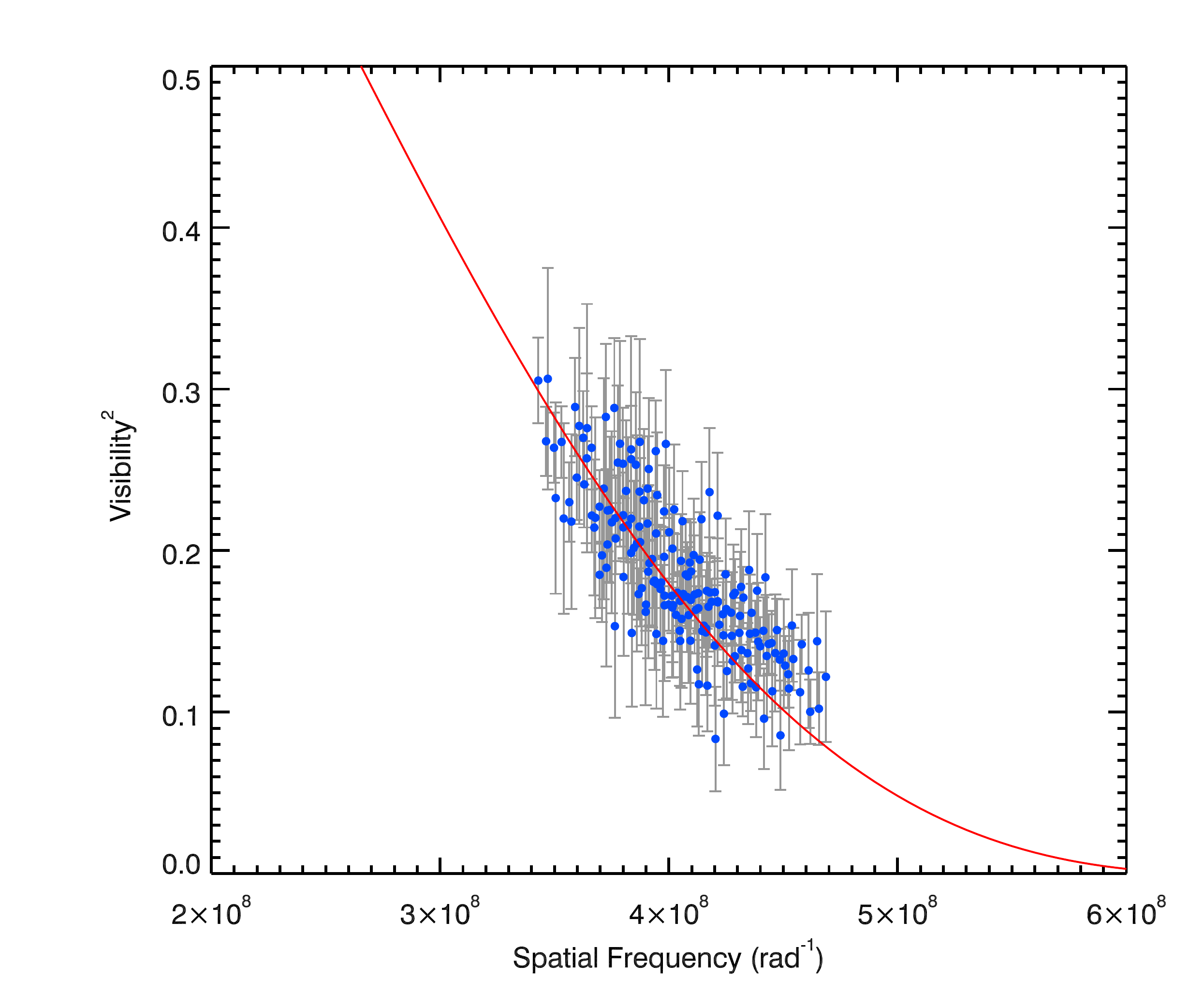}
         \caption{
              Observed squared visibility versus spatial frequency for HD 164922 (blue diamonds). The red line shows the best fit limb-darkened model. See Section \ref{sec:164922_star} for a discussion of the implications of this figure.
         }
         \label{fig:vis}
\end{figure}

We measured the stellar radius for HD 164922 using the CHARA Array. Interferometric observations of HD 164922 were taken on 2012 May 13 and 14 using the Pavo beam combiner \citep{2005ApJ...628..453T, 2008SPIE.7013E..24I}. Observations of the science target were interleaved with the calibrator stars HD\,164900, HD\,161019, and HD\,165373 \citep{2006A&A...456..789B, 2011A&A...535A..53B}.  The data were reduced and calibrated using the standard data reduction pipeline (for details see \citealt{2013MNRAS.433.1262W}).  We use the $R$-band limb-darkening coefficient from \citet{2011A&A...529A..75C}, $\mu_{R} = 0.633$, to determine a limb-darkened angular diameter $\theta_{\rm UD} = $~mas (Figure \ref{fig:vis}).

Combining the angular diameter with the parallax yields a stellar radius of $0.999 \pm 0.017$~\rsun.  We determine a stellar bolometric flux of $F_{\rm Bol} = 4.61 \pm 0.03$~erg/s/cm$^{2}$ by fitting a spectral template from \citet{1998PASP..110..863P} to flux calibrated photometry after applying revised filter profiles from \citet{2015PASP..127..102M}.  This translates to a luminosity $L = 0.703 \pm 0.017$~\lsune.  Lastly, we use our measured angular diameter with the star's bolometric flux to derive an empirical effective temperature $T_{\rm eff} = 5293 \pm 32$~K. %We adopt these empirically derived luminosity and \teff\ values for all future analysis for this star and list them in Table \ref{tab:stellar_params}.

We then use the interferometrically determined parameters to inform a \specmatch analysis of a stack of 5 high SNR APF spectra using an iterative technique. An initial uninformed \specmatch analysis of the APF spectra (all priors uniform) gives \teff$=5318\pm70$ K, \logg$=4.36\pm0.08$ and \feh$=0.17\pm0.05$. We use this spectroscopically measured \feh\ combined with the \teff\ and \rstar\ determined from the CHARA data along with existing $J$,$H$, and $K$ photometry \citep{Cutri03} as Gaussian priors in a fit to Dartmouth isochrones \citep{Dotter08} using the \texttt{isochrones} package \citep{Morton15}\footnote{https://github.com/timothydmorton/isochrones}. This gives the first estimate of the full set of stellar parameters. We then re-run \specmatch with Gaussian priors applied to \teff\ and \logg\ from the \texttt{isochrones} output. The full likelihood with the Gaussian priors is
\begin{scriptsize}
\begin{equation}
\label{eqn:speclike}
\mathcal{L} = \exp{\left[-\frac{1}{2}\left(\chi^2 + \left(\frac{\teff - 5293 \rm{~K}}{32 \rm{~K}}\right)^2 +\left(\frac{\logg - 4.387}{0.014}\right)^2\right)\right]}.
\end{equation}
\end{scriptsize}
In this case the changes to stellar parameters become negligible after two iterations and the process is halted. The resulting stellar parameters are listed in Table \ref{tab:stellar_params}.

Since the star is bright and already a known planet host it has been the subject of many spectroscopic studies. \citet{Santos13} find $\teff=5356\pm45$ K, $\logg=4.34\pm0.08$, and $\feh=+0.14\pm0.03$, all consistent with our analysis to within 1 $\sigma$. \citet{Valenti05} find a significantly higher value for $\logg=4.51\pm0.06$ and $\teff=5385\pm44$ K but a consistent metallicity value of $\feh=+0.17\pm0.03$. \citet{Ghezzi10} measure $\teff=5378\pm50$ K, $\logg=4.30\pm0.22$, and $\feh=+0.21\pm0.03$ also using high-resolution spectroscopy which, except for $\teff$, is also consistent with our analysis to within 1 $\sigma$. As seen by the range of \logg\ values obtained by these various studies, it can be difficult to pin down the stellar gravity from high resolution spectra alone. Since our \logg value is constrained via the direct measurement of the stellar radii from interferometry we adopt the new slightly lower value for \logg. %, and \citet{vanBelle09} interferometrically measure a stellar radius of $\rstar=1.04\pm0.02$ \rsun, consistent with our radius derived using the \citet{Torres10} relation. Although the \citet{Ghezzi10} radius is more precise then ours we choose to adopt the \citet{Torres10} relation value for consistency across all three stars studied in this work. 

 %We calculate the stellar mass for HD 164922 using the relations of \citet{Torres10} and our constraints on \teff, \logg, and \feh.

\subsection{HD 143761}
HD 143761 is the closest and brightest star of the three studied in this work. The star is also known as $\rho$ Corona Borealis, HIP 78459, and Gl 9537. It is a slightly evolved naked eye ($V=5.41$) G0 V star \citep{vanBelle09} located at a distance of 17.236 pc \citep{vanLeeuwen07}. It was previously known to host a warm Jupiter-mass planet with an orbital period of 39 days \citep{Noyes97}. This star was also part of the Eta-Earth survey and was independently selected for intensive long-term RV monitoring based on both the stellar properties, and the mass and orbit of the previously detected planet. Like HD 42618 and HD 164922 this star is chromosherically quiet with $\rphk=-5.05$ \citep{Isaacson10}. As with HD 164922 we performed an iterative interferometric+spectroscopic analysis using the CHARA results from \citet{vonBraun14} and a stack of 5 high SNR APF spectra. The likelihood was essentially the same as Equation \ref{eqn:speclike} but with the \teff\ and \logg\ values for HD 143761 substituted in the last two terms.

%We analyzed 2 high SNR Keck spectra using \specmatch to obtain the mean spectroscopic parameters listed in Table \ref{tab:stellar_params}.

%\citet{vonBraun14} fit a spectral energy distribution (SED) to broadband photometry spanning 0.3 to 2.5 microns for HD 143761 to measure $\teff=5627\pm54$ K, which is slightly more than 1 $\sigma$ discrepant with our spectroscopic measurement.
\citet{Valenti05} measure $\teff=5822\pm44$ K, and \citet{Fuhrmann98} measure $\teff=5821\pm20$ K both using high resolution spectroscopy. Both of these values are significantly hotter then our adopted value of $\teff=5627\pm54$ K from \citet{vonBraun14} . We chose to adopt the value from \citet{vonBraun14} in order to maintain self-consistency with the interferometrically measured stellar radius and luminosity.
%Our analysis gives $\teff=5752\pm60$ K which is consistent within 1 $\sigma$ with the two other spectroscopic \teff\ determinations.
Our metallicity value is also consistent within 1 $\sigma$ to that of \citet{Fuhrmann98} and but is significantly lower than that of \citet{Valenti05}. \citet{Valenti05} again measure a significantly higher value for $\logg=4.36\pm0.06$, but our $\logg=4.121\pm0.018$ value is consistent with $\logg=4.12\pm0.1$ from \citet{Fuhrmann98}.

\section{Keplerian Analysis}
\label{sec:discovery}

\subsection{Discovery}
We discovered each of the three new planets and re-discovered the previously known planets using a technique essentially identical to that of \citet{Fulton15a}. In brief, we calculate a two-dimensional Lomb-Scargle periodogram \citep[2DKLS,][]{Otoole09} to look for significant periodic signals that are well fit by a Keplerian orbital model. Our implementation of the 2DKLS periodogram incorporates arbitrary zero-point offsets between each instrument.
%Since the periodogram represents the improvement in $\chi^2$ relative to a baseline fit, previously known or detected planets are incorporated into the search by calculating the improvement in $\chi^2$ by adding another planet to the fit without subtracting the model for previously detected planets.
The periodogram power ($Z$) represents the improvement to the $\chi^2$ statistic relative to that of a baseline fit. When searching for the first planet in a system the baseline fit is simply a flat line or linear trend. If any significant signals are found after the first iteration the baseline model then becomes the single planet Keplerian model and we calculate the improvement to $\chi^2$ when a second planet is added, without subtracting the first. We repeat this process until no more significant peaks are found in the 2DKLS periodogram. We start the search assuming no known signals in order to ensure that the previously published planets can be automatically detected using our pipeline. An initial jitter term of 2.0 \ms is added in quadrature with the RVs before starting the 2DKLS search in order to ensure fair weighting between the Keck and APF data sets. We expect both data sets to be limited by instrumental/astrophysical systematics rather than photon noise.

The discovery pipeline is completely automated in order to facilitate injection recovery tests that will allow us to characterize the pipeline completeness for future occurrence analysis\citep{Howard16}. We calculate an empirical periodogram false alarm probability (eFAP) by fitting a power law to the distribution of periodogram values between the 50th and 97th quartiles. This fit provides an estimate of the significance of periodogram peaks of a given value. When multiplied by the number of independent test periods, the fit gives the approximate probability that we would find a peak of a given value within any particular periodogram. Any periodogram peak with an eFAP below 0.1\% is automatically considered a viable candidate and the search is continued until no more periodogram peaks fall above the 0.1\% eFAP threshold. Further details of the automated planet detection pipeline can be found in \citet{Howard16}. We note that the eFAP metric is used simply to automatically identify candidates. The significance of the corresponding periodogram peaks are checked using the bootstrapping technique described in Section \ref{sec:fap}. Each of the previously known planets were re-discovered with eFAPs much less than those of the new planets announced in this work.

\begin{figure*}[ht]
         \centering
              \includegraphics[width=0.85\textwidth]{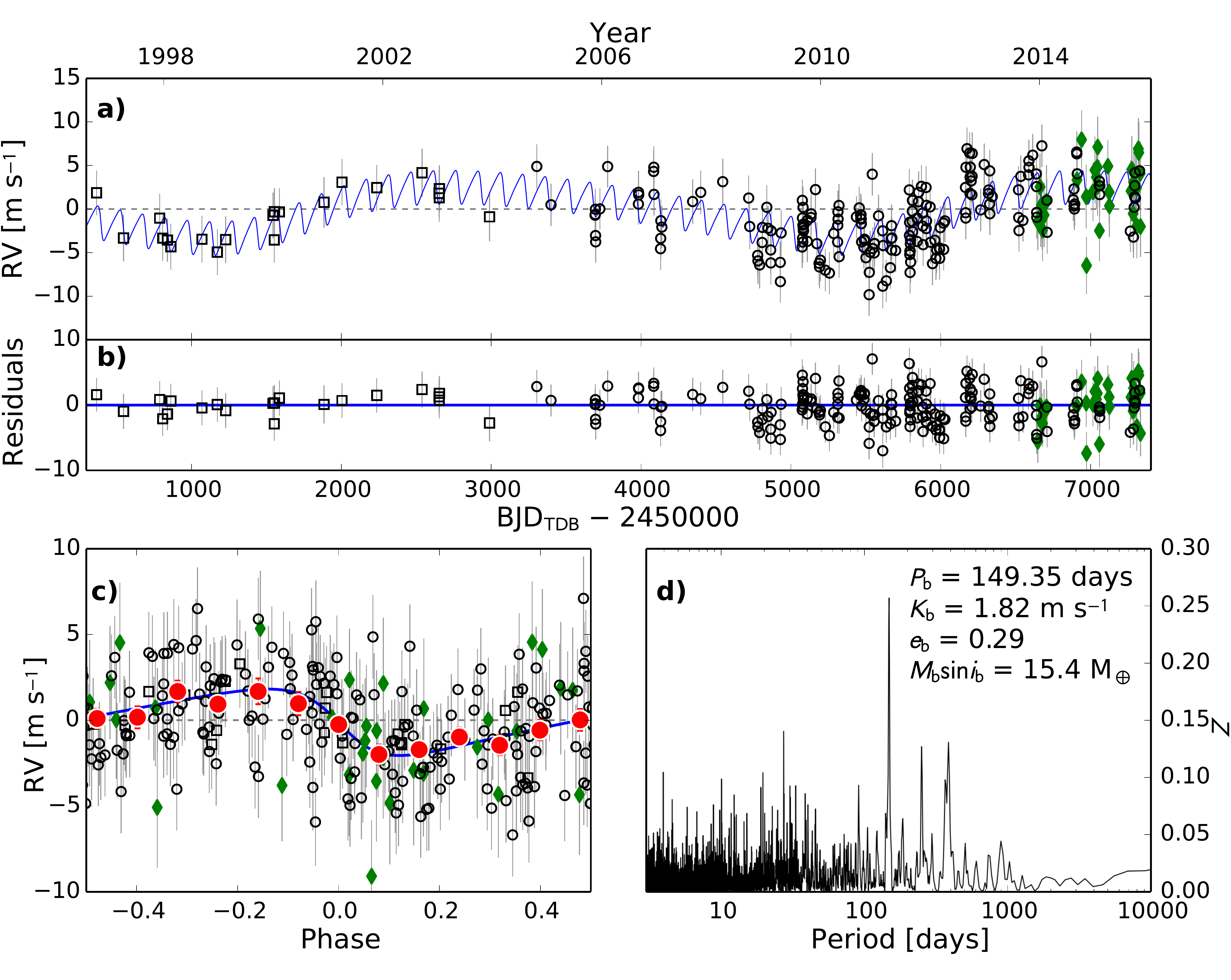}
         \caption{
              One-planet Keplerian orbital model plus one additional long-period Keplerian to model the stellar magnetic activity cycle for HD 42618.
              The most likely model is plotted but the orbital parameters annotated on the figure and listed in Tables \ref{tab:params_42618} and \ref{tab:dparams} are the median values of the posterior distributions. The process used to find the orbital solution is described in \ref{sec:characterization}.
              {\bf a)} Full binned RV time series. Open black squares indicate pre-upgrade Keck/HIRES data (see \S \ref{sec:keckdata}), open black circles are post-upgrade Keck/HIRES data, and filled green diamonds are APF measurements.
              The thin blue line is the most probable 1-planet plus stellar activity model. We add in quadrature the RV jitter term listed in Table \ref{tab:params_42618} with the measurement uncertainties for all RVs.
              {\bf b)} Residuals to the most probable 1-planet plus stellar activity model.
              {\bf c)} Binned RVs phase-folded to the ephemeris of planet b. The long-period stellar activity signal has been subtracted. The small point colors and symbols are the same as in panel
              {\bf a}. For visual clarity, we also bin the velocities in 0.08 units of orbital phase (red circles). The phase-folded model for planet b is shown as the blue line.
              {\bf d)} 2DKLS periodogram showing the improvement to $\chi^2$ for a model including the long period activity signal and a single planet compared to a model that only includes the activity signal.
              %Panels {\bf e)} and {\bf f)}, and panels {\bf g)} and {\bf h)} are the same as panels {\bf c)} and {\bf d)} but for planets HD 7924 c and HD 7924 d respectively.
         }
\label{fig:fitplot_42618}
\end{figure*}

We discover two significant signals in the RV time series of HD 42618. One with a long period of $\sim$4850 days, and a second at a period of 149.6 days. Upon inspection of the \caii\ activity index time series we notice that this index shows a periodicity with a period very similar to that of the long period RV signal. The period of 4850 days is also very similar to the period of the sun's magnetic activity cycle. We conclude that this is likely the signature of the stellar magnetic activity cycle and not the signature of an orbiting planet. We include this long-period signal as an additional eccentric Keplerian in all further modeling. HD 42618 b is easily detected in the Keck data alone and the combined Keck+APF dataset but we don't yet have enough APF measurements to detect it in the APF data alone. Figure \ref{fig:fitplot_42618} shows the most likely model from the posterior of the two-Keplerian model, the 2DKLS periodogram used to discover HD 42618 b, and the RVs phased to the orbital period of planet b.

\begin{figure*}[ht]
	\centering
              \includegraphics[width=0.85\textwidth]{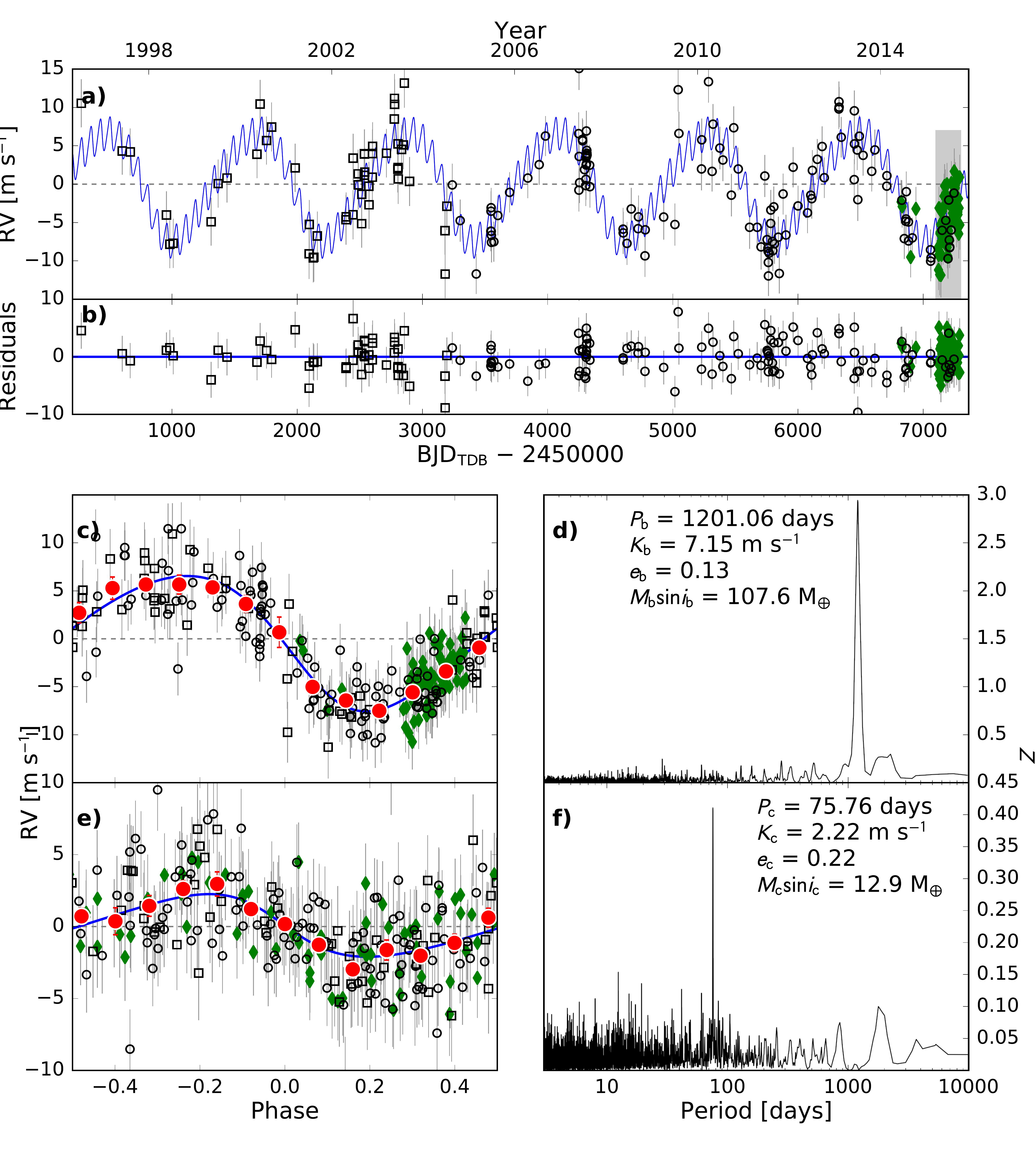}
         \caption{
              Two-planet Keplerian orbital model posterior distributions for HD 164922.
              The most likely model from the posterior distribution is plotted while the orbital parameters annotated on the figure and listed in Tables \ref{tab:params_164922} and \ref{tab:dparams} are the median values of the posterior distributions. The process used to find the orbital solution is described in \ref{sec:characterization}.
              {\bf a)} Full binned RV time series. Open black squares indicate pre-upgrade Keck/HIRES data (see \S \ref{sec:keckdata}), open black circles are post-upgrade Keck/HIRES data, and filled green diamonds are APF measurements.
              The thin blue line is the most probable 2-planet model. We add in quadrature the RV jitter term listed in Table \ref{tab:params_164922} with the measurement uncertainties for all RVs.
              {\bf b)} Residuals to the most probable 2-planet model.
              {\bf c)} Binned RVs phase-folded to the ephemeris of planet b. The Keplerian orbital model for planet c has been subtracted. The small point colors and symbols are the same as in panel {\bf a}. For visual clarity, we also bin the velocities in 0.08 units of orbital phase (red circles). The phase-folded model for planet b is shown as the blue line.
              {\bf d)} 2DKLS periodogram comparing a model including only the long period planet to the two planet model.
              Panels {\bf e)} and {\bf f)} are the same as panels {\bf c)} and {\bf d)} but for planet HD 164922 c.
              The shaded region of panel {\bf a} is re-plotted in Figure \ref{fig:apfzoom}.
              %, and panels {\bf g)} and {\bf h)} 
         }
\label{fig:fitplot_164922}
\end{figure*}

\begin{figure}[h]
         \centering
              \includegraphics[width=0.43\textwidth]{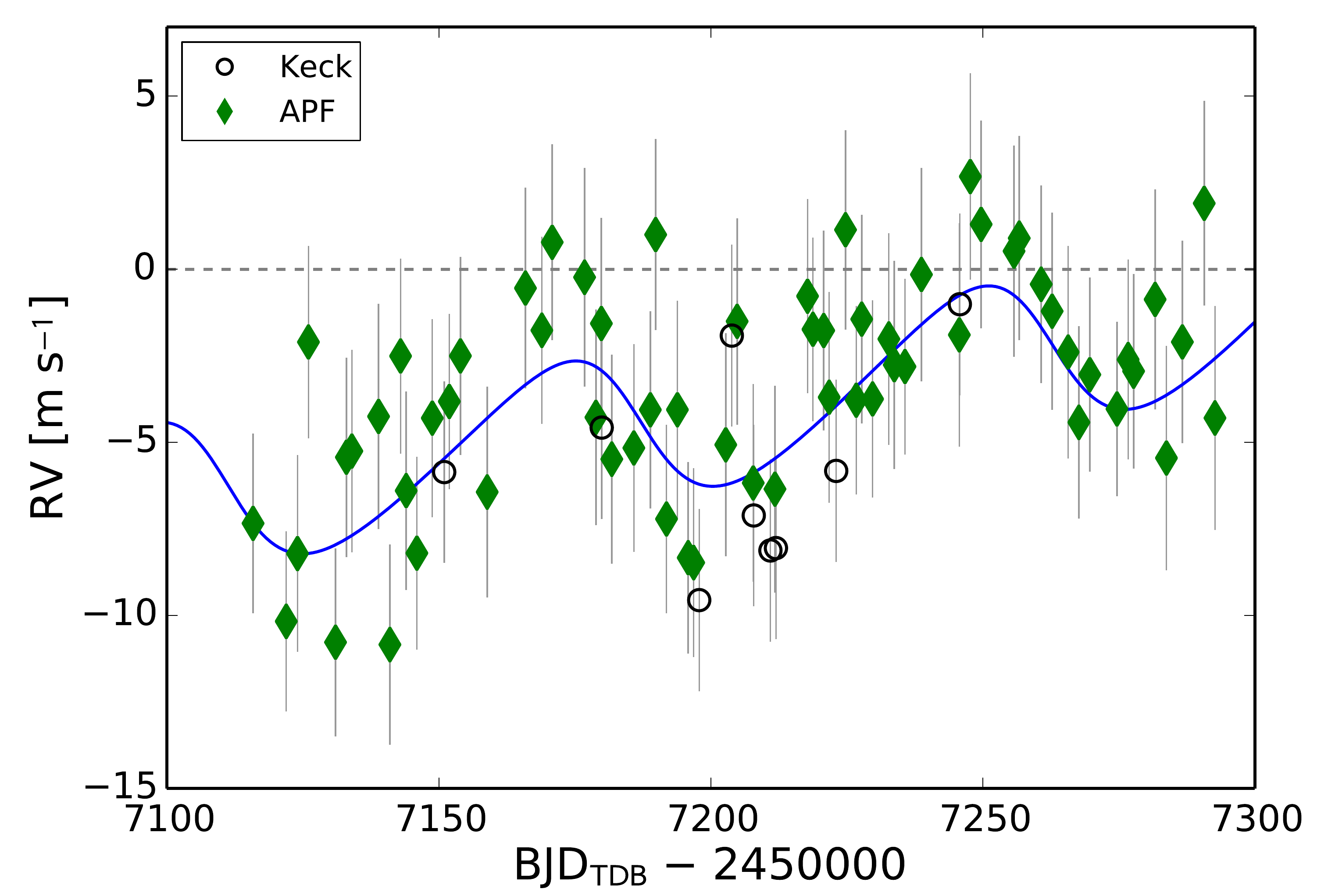}
         \caption{
              Recent RVs for HD 164922 highlighted in the grey box of Figure \ref{fig:fitplot_164922} panel {\bf a)} featuring the high cadence APF observations collected during the most recent observational season.
         }
         \label{fig:apfzoom}
\end{figure}

\citet{Wright07} mentioned a candidate planetary signal with a period of 75.8 d and $K = 3$ \ms orbiting HD 164922 but did not have sufficient data to claim a significant detection. With seven years of additional Keck data, and 2 years of APF data we can firmly establish this signal as being coherent and persistent as expected for the Doppler motion caused by an orbiting planet. The short period planet is easily detected in either the APF or Keck data alone. The long-period planet can only be detected in the long baseline Keck data but we do observe a linear RV trend that emerged during the most recent APF observing season which is a result of the massive outer planet. Figures \ref{fig:fitplot_164922} and \ref{fig:apfzoom} show the most likely model from the posteriors for the two planet Keplerian model and the 2DKLS periodograms used to discover/re-discover each of the two planets.

\begin{figure*}[ht]
         \begin{center}
              \includegraphics[width=0.85\textwidth]{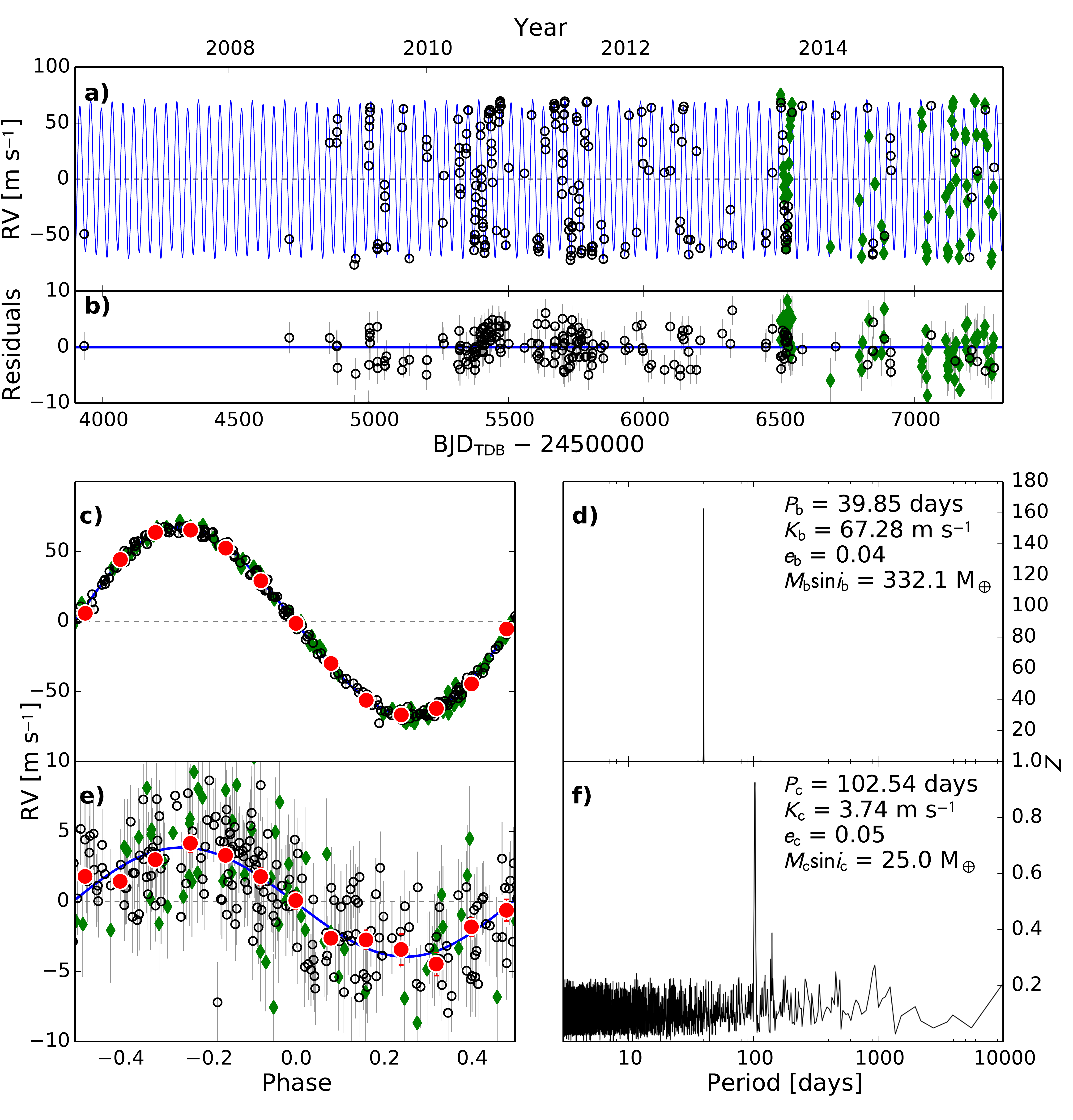}
         \end{center}
         \caption{
              Same as Figure \ref{fig:fitplot_164922} but for planets HD 143761 b and c.
         }
\label{fig:fitplot_143761}
\end{figure*}

We discover a super-Neptune mass planet orbiting HD 143761 exterior to the known Jupiter mass planet that has an orbital period of 39 days. The new planet has an orbital period of 102 days and a semi-amplitude of 3.7 \mse. Each planet is discovered with very high significance in both the Keck and Keck+APF datasets individually. The most likely model from the posteriors for the two planet model and the 2DKLS detection periodograms are shown in Figure \ref{fig:fitplot_143761}.

\subsection{Characterization}
\label{sec:characterization}
We estimated orbital parameters and their associated uncertainties using the \texttt{ExoPy} Differential Evolution Markov Chain Monte Carlo \citep[DE-MCMC,][]{Braak06} modeling code and a technique identical to that of \citet{Fulton13}, \citet{Knutson14}, and \citet{Fulton15a}. 

 Our multi-planet RV model is a sum of Keplerian single-planet models over all planets in the system. For each single-planet Keplerian model (indexed by $i$) we compute posteriors for the orbital period ($P_{i}$), time of inferior conjunction ($T_{\mathrm{conj}, i}$), eccentricity ($e_{i}$), argument of periastron of the star's orbit ($\omega_{i}$), velocity semi-amplitude ($K_{i}$). We also compute posteriors for the offsets between pre-upgrade, post-upgrade, and the APF datasets ($\gamma$), and an RV jitter term with the specific prior described in \citet{Fulton15a} and \citet{Johnson11}.
%In order to speed convergence and minimize biases in parameters that are physically finite we step in modifications to some of the physical parameters (e.g. $\sqrt{e}\cos{\omega}$).
In order to speed convergence, we choose to re-parameterize some of the physical parameters as $\sqrt{e}\cos{\omega}$, $\sqrt{e}\sin{\omega}$, and $\log{K}$. The full likelihood for this Keplerian model is
\begin{scriptsize}
\begin{equation}
\label{eqn:mclike}
\mathcal{L} = \exp{\left(-\frac{1}{2}\sum_n \left[\frac{(\text{v}_n - M_n)^2}{\sigma_{\text{v}, n}^2 + \sigma_{\mathrm{jitt}}^2} + 2 \ln{(\sqrt{2\pi}(\sigma_{\text{v}, n}^2 + \sigma_{\mathrm{jitt}}^2)^{1/4})}\right]\right)}
\end{equation}
\end{scriptsize}
summed over $n$ RV measurements (v$_{n}$) with associated uncertainties $\sigma_{\text{v},n}$. $M_n$ is the Keplerian model for observation $n$. We assign uniform priors to $\log{P}$, $T_{\textrm{conj}}$, $\sqrt{e}\cos{\omega}$,
$\sqrt{e}\sin{\omega}$, $\log{K}$, and $\gamma$ for each instrument. We follow the prescription of \citet{Eastman13} checking for convergence at regular intervals during the
MCMC runs by calculating the Gelmin-Rubin statistic and the number of independent draws \citep[$T_{Z}$,][]{Ford06}.
We consider the chains well mixed and halt the MCMC run when the Gelmin-Rubin statistic is within 1\% of unity and $T_{Z}>1000$ for all free parameters.
All of the adopted median values and 68\% confidence intervals of the posterior distributions are listed in Tables \ref{tab:params_42618}--\ref{tab:params_143761}.

%%%%%%%%%%%%   HD 42618 Orbital parameters %%%%%%%%%%%%%%%%%%%%

\begin{deluxetable}{lrr}
    \tablecaption{Orbital Parameters for HD 42618}
    \tablehead{\colhead{Parameter} & \colhead{Value} & \colhead{Units}}
    \startdata
    \sidehead{\bf{Modified DE-MCMC Step Parameters}\tablenotemark{1}}
log($P_{b}$) & 2.17497 $^{+0.0011}_{-0.00098}$ & log(days)\\
$\sqrt{e_{b}}\cos{\omega_{b}}$ & $-0.03 \pm 0.24$ & \\
$\sqrt{e_{b}}\sin{\omega_{b}}$ & $+0.34 ^{+0.19}_{-0.31}$ & \\
log($K_{b}$) & 0.276 $^{+0.061}_{-0.071}$ & log(\ms)\\
\vspace{5pt}\\

\hline
\sidehead{\bf{Model Parameters}}
$P_{b}$ & 149.61 $^{+0.37}_{-0.34}$ & days\\
$T_{\textrm{conj},b}$ & 2456670.2 $^{+6.1}_{-5.6}$ & \bjdtdb\\
$e_{b}$ & 0.19 $^{+0.15}_{-0.12}$ & \\
$\omega_{b}$ & 101 $^{+69}_{-39}$ & degrees\\
$K_{b}$ & 1.89 $^{+0.29}_{-0.28}$ & \ms\\
\vspace{5pt}\\
$\gamma_{\rm post\mbox{-}upgrade~Keck}$ & 0.61 $\pm 0.21$ & \ms\\
$\gamma_{\rm pre\mbox{-}upgrade~Keck}$ & 0.71 $^{+0.64}_{-0.65}$ & \ms\\
$\dot{\gamma}$ & $\equiv$ 0.0 $\pm 0.0$ & \ms day$^{-1}$\\
$\ddot{\gamma}$ & $\equiv$ 0.0 $\pm 0.0$ & \ms day$^{-2}$\\
$\sigma_{\mathrm{jitt}}$ & 2.34 $^{+0.14}_{-0.12}$ & \ms\\
\vspace{5pt}\\

    \enddata
    \tablenotetext{1}{MCMC jump parameters that were modified from the physical parameters in order to speed convergence and avoid biasing parameters that must physically be finite and positive.}
    \label{tab:params_42618}
    \end{deluxetable}

%%%%%%%%%%%%   HD 164922 Orbital parameters %%%%%%%%%%%%%%%%%%%%
  \begin{deluxetable}{lrr}
    \tablecaption{Orbital Parameters for HD 164922}
    \tablehead{\colhead{Parameter} & \colhead{Value} & \colhead{Units}}
    \startdata
    \sidehead{\bf{Modified DE-MCMC Step Parameters}\tablenotemark{1}}
    log($P_{b}$) & 3.08 $\pm 0.002$ & log(days)\\
$\sqrt{e_{b}}\cos{\omega_{b}}$ & $-0.214 ^{+0.100}_{-0.081}$ & \\
$\sqrt{e_{b}}\sin{\omega_{b}}$ & $+0.264 ^{+0.096}_{-0.130}$ & \\
log($K_{b}$) & 0.854 $\pm 0.019$ & \ms\\

\vspace{5pt}\\
log($P_{c}$) & 1.87947 $^{+0.00033}_{-0.00032}$ & log(days)\\
$\sqrt{e_{c}}\cos{\omega_{c}}$ & $+0.04 ^{+0.29}_{-0.30}$ & \\
$\sqrt{e_{c}}\sin{\omega_{c}}$ & $+0.37 ^{+0.14}_{-0.23}$ & \\
log($K_{c}$) & 0.346 $^{+0.054}_{-0.062}$ & log(\ms)\\
\vspace{5pt}\\

\hline
\sidehead{\bf{Model Parameters}}
$P_{b}$ & 1201.1 $^{+5.6}_{-5.5}$ & days\\
$T_{\textrm{conj},b}$ & 2456778 $^{+18}_{-19}$ & \bjdtdb\\
$e_{b}$ & 0.126 $^{+0.049}_{-0.050}$ & \\
$\omega_{b}$ & 129 $^{+24}_{-20}$ & degrees\\
$K_{b}$ & 7.15 $\pm 0.31$ & \ms\\

\vspace{5pt}\\
$P_{c}$ & 75.765 $^{+0.058}_{-0.056}$ & days\\
$T_{\textrm{conj},c}$ & 2456277.6 $\pm 2.7$ & \bjdtdb\\
$e_{c}$ & 0.22 $\pm 0.13$ & \\
$\omega_{c}$ & 81 $^{+45}_{-49}$ & degrees\\
$K_{c}$ & 2.22 $^{+0.30}_{-0.29}$ & \ms\\

\vspace{5pt}\\
$\gamma_{\rm post\mbox{-}upgrade~Keck}$  & 0.23 $\pm 0.27$ & \ms\\
$\gamma_{\rm pre\mbox{-}upgrade~Keck}$ & 1.02 $\pm 0.54$ & \ms\\
$\gamma_{\rm APF}$ & 0.2 $^{+0.47}_{-0.48}$ & \ms\\
$\dot{\gamma}$ & $\equiv$ 0.0 $\pm 0.0$ & \ms day$^{-1}$\\
$\ddot{\gamma}$ & $\equiv$ 0.0 $\pm 0.0$ & \ms day$^{-2}$\\
$\sigma_{\mathrm{jitt}}$ & 2.63 $^{+0.15}_{-0.14}$ & \ms\\
\vspace{5pt}\\

    \enddata
    \tablenotetext{1}{MCMC jump parameters that were modified from the physical parameters in order to speed convergence and avoid biasing parameters that must physically be finite and positive.}
    \label{tab:params_164922}
    \end{deluxetable}

%\clearpage
%%%%%%%%%%%%   HD 143761 Orbital parameters %%%%%%%%%%%%%%%%%%%%
  \begin{deluxetable}{lrr}
    \tablecaption{Orbital Parameters for HD 143761}
    \tablehead{\colhead{Parameter} & \colhead{Value} & \colhead{Units}}
    \startdata
    \sidehead{\bf{Modified DE-MCMC Step Parameters}\tablenotemark{1}}
log($P_{b}$) & 1.600382 $\pm 1.6e-05$ & log(days)\\
$\sqrt{e_{b}}\cos{\omega_{b}}$ & $+0.002 \pm 0.02$ & \\
$\sqrt{e_{b}}\sin{\omega_{b}}$ & $-0.192 ^{+0.011}_{-0.010}$ & \\
log($K_{b}$) & 1.8279 $\pm 0.0016$ & \ms\\
\vspace{5pt}\\

log($P_{c}$) & 2.01091 $^{+0.00073}_{-0.00070}$ & log(days)\\
$\sqrt{e_{c}}\cos{\omega_{c}}$ & $+0.01 \pm 0.19$ & \\
$\sqrt{e_{c}}\sin{\omega_{c}}$ & $-0.01 \pm 0.19$ & \\
log($K_{c}$) & 0.573 $^{+0.032}_{-0.034}$ & log(\ms)\\
\vspace{5pt}\\

\hline
\sidehead{\bf{Model Parameters}}
$P_{b}$ & 39.8458 $^{+0.0015}_{-0.0014}$ & days\\
$T_{\textrm{conj},b}$ & 2455759.091 $\pm 0.056$ & \bjdtdb\\
$e_{b}$ & 0.0373 $^{+0.0040}_{-0.0039}$ & \\
$\omega_{b}$ & 270.6 $^{+5.9}_{-5.8}$ & degrees\\
$K_{b}$ & 67.28 $\pm 0.25$ & \ms\\
\vspace{5pt}\\

$P_{c}$ & 102.54 $\pm 0.17$ & days\\
$T_{\textrm{conj},c}$ & 2455822 $\pm 2$ & \bjdtdb\\
$e_{c}$ & 0.052 $^{+0.061}_{-0.037}$ & \\
$\omega_{c}$ & 190 $^{+110}_{-140}$ & degrees\\
$K_{c}$ & 3.74 $\pm 0.28$ & \ms\\
\vspace{5pt}\\

$\gamma_{\rm post\mbox{-}upgrade~Keck}$ & -0.6 $\pm 0.2$ & \ms\\
$\gamma_{\rm APF}$ & -0.7 $^{+0.50}_{-0.49}$ & \ms\\
$\dot{\gamma}$ & $\equiv$ 0.0 $\pm 0.0$ & \ms day$^{-1}$\\
$\ddot{\gamma}$ & $\equiv$ 0.0 $\pm 0.0$ & \ms day$^{-2}$\\
$\sigma_{\mathrm{jitt}}$ & 2.57 $^{+0.14}_{-0.13}$ & \ms\\

    \enddata
    \tablenotetext{1}{MCMC jump parameters that were modified from the physical parameters in order to speed convergence and avoid biasing parameters that must physically be finite and positive.}
    \label{tab:params_143761}
    \end{deluxetable}

\begin{deluxetable}{lrr}
\tablewidth{0.35\textwidth}
    \tablecaption{Derived Planet Properties}
    \tablehead{\colhead{Parameter} & \colhead{Value} & \colhead{Units}}
    \startdata
\sidehead{\bf{HD 42618}}
$e_b\cos{\omega_b}$ & -0.009 $^{+0.06}_{-0.076}$ & \\
$e_b\sin{\omega_b}$ & 0.14 $^{+0.13}_{-0.11}$ & \\
$a_{b}$ & 0.554 $\pm 0.011$ & AU\\
$a_{b}/R_{\star}$ & 119.1 $^{+12.0}_{-9.8}$ & \\
$M_{b}\sin{i_b}$ & 14.4 $^{+2.5}_{-2.4}$ & $M_{\oplus}$\\
$S_{b}$\tablenotemark{*} & 3.16 $^{+0.61}_{-0.55}$ & $S_{\oplus}$\\
$T_{eq, b}$\tablenotemark{**} & 337 $^{+15}_{-16}$ & K\\

\hline
\vspace{3pt}\\
\sidehead{\bf{HD 164922}}
$e_b\cos{\omega_b}$ & -0.025 $^{+0.016}_{-0.019}$ & \\
$e_b\sin{\omega_b}$ & 0.032 $^{+0.028}_{-0.022}$ & \\
$a_{b}$ & 2.115 $\pm 0.012$ & AU\\
$a_{b}/R_{\star}$ & 454.9 $^{+8.3}_{-7.9}$ & \\
$M_{b}\sin{i_b}$ & 107.6 $^{+4.9}_{-4.8}$ & $M_{\oplus}$ \\
$S_{b}$\tablenotemark{*} & 0.1578 $^{+0.0069}_{-0.0067}$ & $S_{\oplus}$\\
$T_{eq, b}$\tablenotemark{**} & 159.4 $\pm 1.7$ & K \\
\vspace{3pt}\\
$e_c\cos{\omega_c}$ & 0.003 $^{+0.073}_{-0.063}$ & \\
$e_c\sin{\omega_c}$ & 0.079 $^{+0.089}_{-0.066}$ & \\
$a_{c}$ & 0.3351 $\pm 0.0015$ & AU \\
$a_{c}/R_{\star}$ & 72.1$^{+1.3}_{-1.2}$ & \\
$M_{c}\sin{i_c}$ & 12.9 $\pm 1.6$ & $M_{\oplus}$\\
$S_{c}$\tablenotemark{*} & 6.29 $^{+0.27}_{-0.26}$ & $S_{\oplus}$\\
$T_{eq, c}$\tablenotemark{**} & 400.5 $\pm 4.3$ & K\\

\hline
\vspace{3pt}\\
\sidehead{\bf{HD 143761}}
$e_b\cos{\omega_b}$ & 7e-05 $^{+0.00072}_{-0.00073}$ & \\
$e_b\sin{\omega_b}$ & -0.0072 $^{+0.0011}_{-0.0012}$ & \\
$a_{b}$ & 0.2196 $^{+0.0024}_{-0.0025}$ & AU\\
$a_{b}/R_{\star}$ & 34.66 $^{+0.78}_{-0.76}$ & \\
$M_{b}\sin{i_b}$ & 332.1 $^{+7.5}_{-7.6}$ & $M_{\oplus}$\\
$S_{b}$\tablenotemark{*} &  34.7 $^{+2.1}_{-2.0}$  & $S_{\oplus}$\\
$T_{eq, b}$\tablenotemark{**} & 614.0 $^{+9.1}_{-9.0}$ & K\\
\vspace{3pt}\\
$e_c\cos{\omega_c}$ & 0.0001 $^{+0.013}_{-0.011}$ & \\
$e_c\sin{\omega_c}$ & -0.0001 $^{+0.011}_{-0.015}$ & \\
$a_{c}$ & 0.4123 $^{+0.0046}_{-0.0047}$ & AU\\
$a_{c}/R_{\star}$ & 65.1 $^{+1.5}_{-1.4}$ & \\
$M_{c}\sin{i_c}$ & 25 $\pm 2$ & $M_{\oplus}$\\
$S_{c}$\tablenotemark{*} & 9.85 $^{+0.6}_{-0.56}$ & $S_{\oplus}$\\
$T_{eq, c}$\tablenotemark{**} & 448.1 $\pm 6.6$  & K\\

    \enddata
    \tablenotetext{*}{Stellar irradiance received at the planet relative to the Earth.}
    \tablenotetext{**}{Assuming a bond albedo of 0.32 \citep{Demory14}.}
    \label{tab:dparams}
    \end{deluxetable}

\subsection{Bootstrap False Alarm Assessment}
\label{sec:fap}
\begin{figure}
\centering
\includegraphics[width=0.45\textwidth]{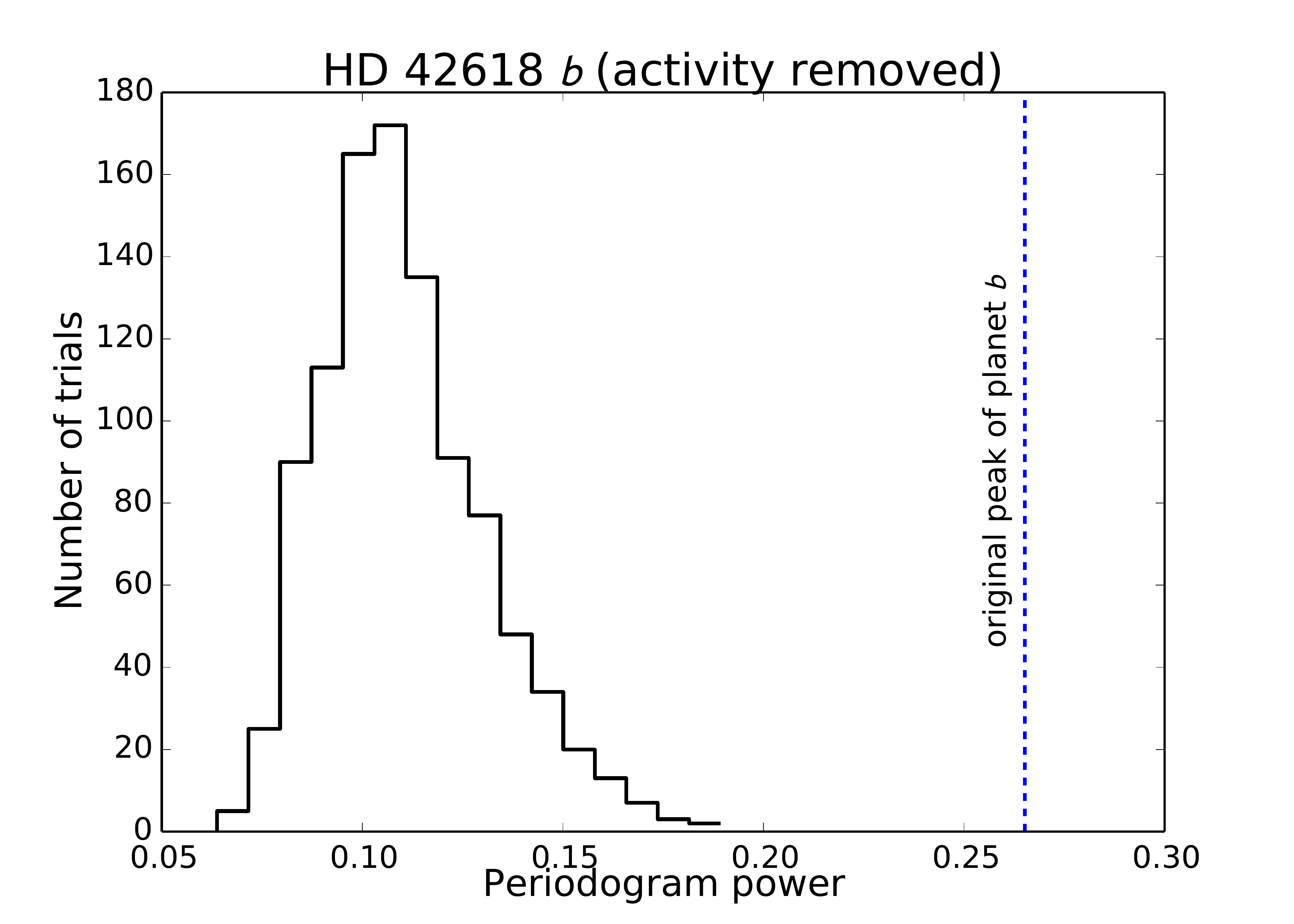}
\includegraphics[width=0.45\textwidth]{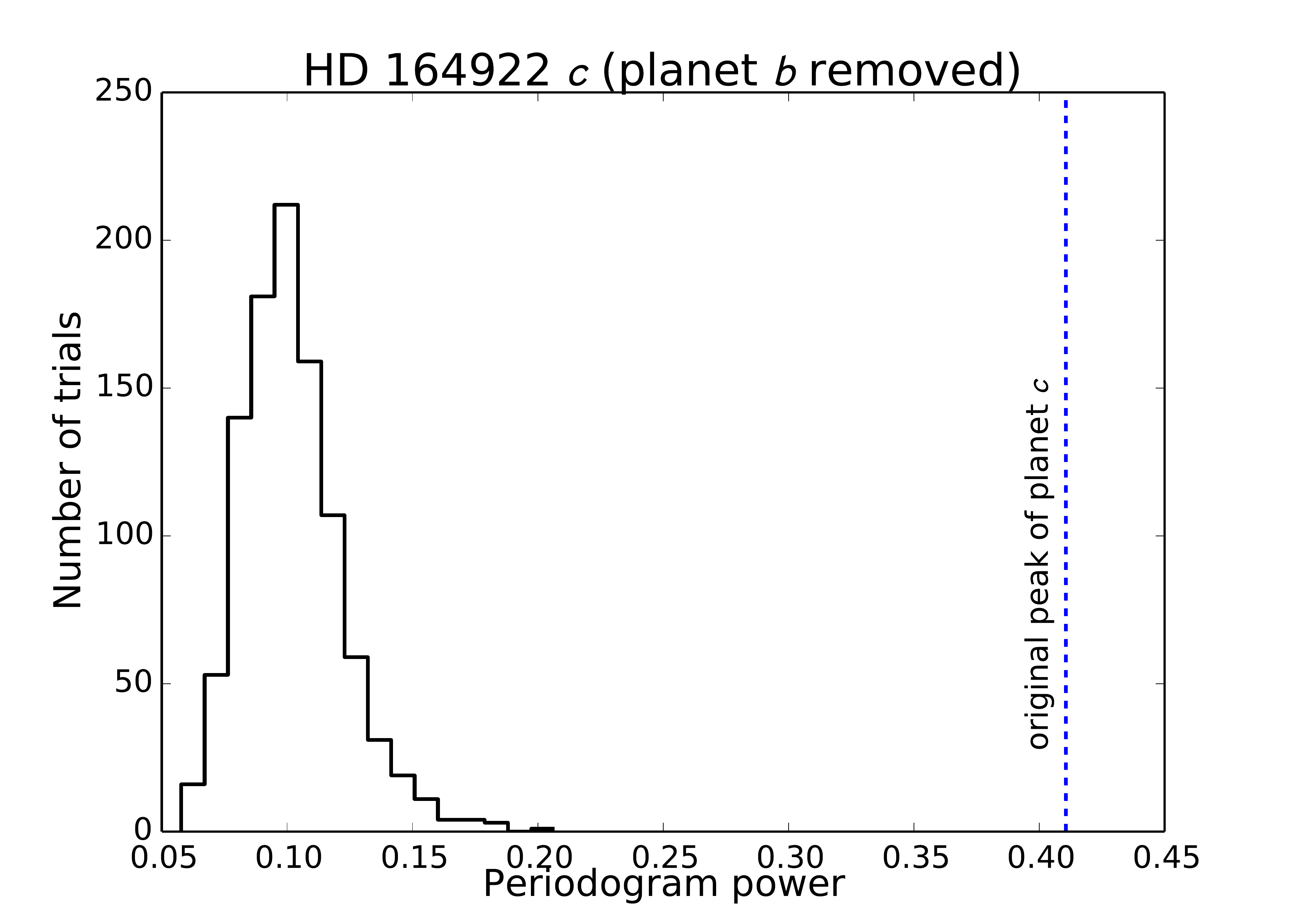}
\includegraphics[width=0.45\textwidth]{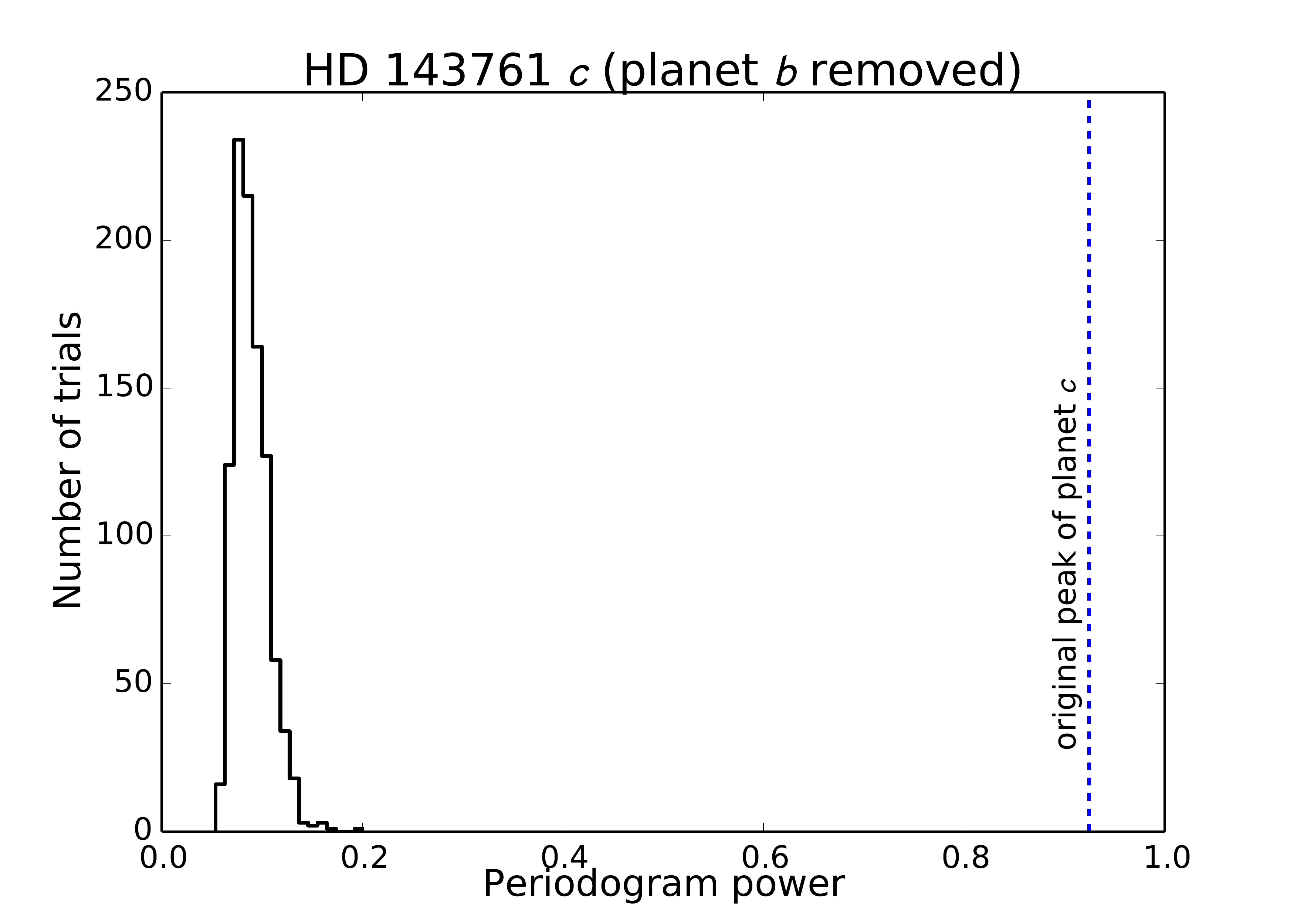}
\caption{
	     Graphical representation of the bootstrap false alarm tests described in Section \ref{sec:fap}.
              \emph{Top:} Distribution of maximum periodogram peak heights for 1000 2DKLS periodograms of scrambled RV time series for HD 42618. The long period activity signal was subtracted before scrambling the data set. The vertical dashed blue line marks the height of the original periodogram peak for planet b which is clearly separated from the distribution of peaks caused by random fluctuations.
              \emph{Middle:} Same as the top panel for planet HD 164922 c.
              \emph{Bottom:} Same as the top panel for planet HD 143761 c. In each case, the scrambled RVs generate peaks with significantly lower power than the power observed from the new planetary signals.
}
\label{fig:boot}
\end{figure}

%We estimate the probability that the 2DKLS periodogram peak heights corresponding to each of the planets are the result of random noise instead of a periodic Keplerian-like signal by performing a bootstrap false alarm assessment.
We conduct a bootstrap false alarm assessment to verify and double check that the periodogram peaks with low eFAPs are indeed statistically significant periodic signals and not caused by random fluctuations of noise.
For all three stars we scramble the RV time series 1000 times and recalculate the 2DKLS periodogram searching for N+1 planets where N is the number of previously published planets in the system. We record the highest periodogram value from each trial and plot the distribution of periodogram peak heights relative to the periodogram peak values corresponding to the newly discovered planets. These distributions are plotted in Figure \ref{fig:boot}. The periodogram peak heights corresponding to each of the new planets are well separated from the distribution of peaks in the scrambled RV trials. This indicates that the probability that random noise could conspire to create the periodogram peaks used to detect the planets are  $<0.1\%$. However, a visual inspection of the distribution of periodogram peak heights in Figure \ref{fig:boot} suggests that the FAPs are likely much lower.

\subsection{ Brown Dwarf Companion to HD 143761?}
HD 143761 b was one of the first exoplanets discovered \citep{Noyes97}. \citet{Gatewood01} later reported that they detected the signature of HD 143761 b in astrometric data from the Multichannel Astrometric Photometer and Hipparcos. \citet{Zucker01} were quick to point out that the statistical significance of this astrometric detection is only 2 $\sigma$. Over a decade after the discovery of HD 143761 b, \citet{Reffert11} claimed to detect the same astrometric signal of the warm Jupiter in a re-reduction of Hipparcos data. The astrometric orbit suggests that the system is nearly face on with an inclination between 0.4 and 0.7 degrees. This would imply that, after correcting for the viewing angle, HD 143761 b is not a planet but instead a low mass M star with $100<M_b<200$ M$_J$.

\newpage
\subsubsection{Interferometry of HD 143761}
Long-baseline interferometry is sensitive to some stellar binaries. HD 143761 was observed interferometrically with the CHARA Array \citep{vonBraun14}. If we assume a face-on orbit of HD 143761 b at a distance of 17.2 pc, then the angular distance between it and the principal component is 14 milliarcseconds independent of phase angle. This is detectable as a separated fringe packet at $H$-band with CHARA \citep{Farrington10}, provided the brightness contrast is not larger than $\Delta H\sim 2$ \citep{Farrington10,Raghavan12}. 

Using the mass constraints from the astrometric orbit, a comparison with the Dartmouth isochrones for metallicities spanning the 1 $\sigma$ \feh\ uncertainty of HD 143761 from Table \ref{tab:stellar_params} corresponds to an apparent $H$ magnitude at a distance of 17.2 pc of around 10. This implies a $\Delta H \simeq 6$, which is significantly below what could be detected as a separate component in CHARA data \citep[cf. equation A5 in][]{Boyajian08}. Since component c's apparent $H$ magnitude is much fainter than component b's, its detection is impossible in the CHARA data. As expected, an inspection of the CHARA data used in \citet{vonBraun14} did not yield any indication of additional fringe packets.

\begin{figure}[h]
         \begin{center}
              \includegraphics[width=0.45\textwidth]{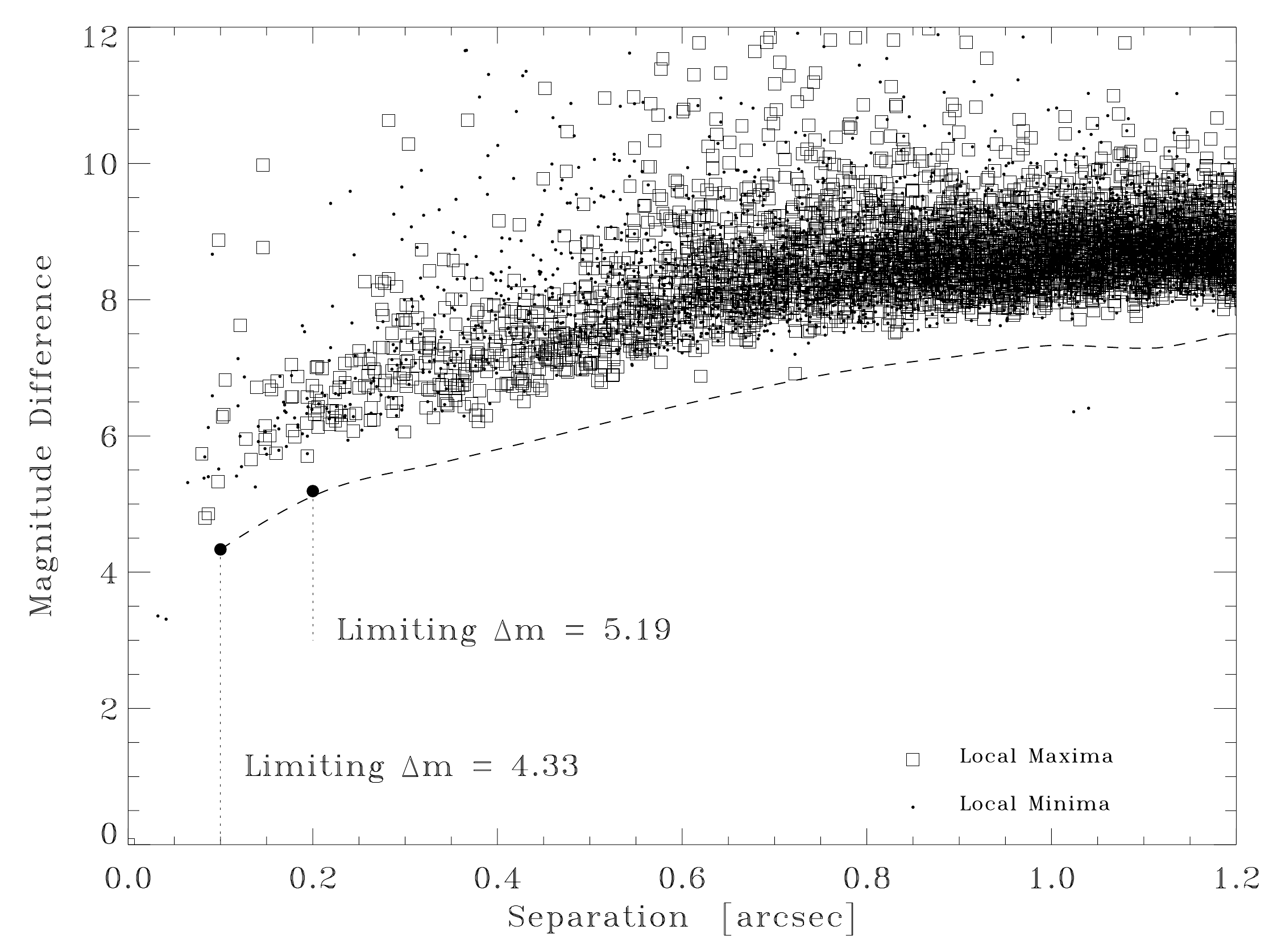}
         \end{center}
         \caption{
              Limiting magnitude as a function of separation from the $\rho$ CrB. Also shown are a cubic spline interpolation of the 5$\sigma$ detection limit (dashed line) and limiting magnitudes for 0\farcs1 and 0\farcs2.
         }
         \label{fig:contrast}
\end{figure}

We also obtained imaging observations of $\rho$ CrB using the Differential Speckle Survey Instrument (DSSI) on Gemini-North during the nights of July 19, 24, and 25 in 2014. The DSSI camera is a dual-channel speckle imaging system, expounded upon in more detail by \citet{Horch09,Horch11}. Observations were acquired using red and near-infrared filters centered on 692 nm and 880 nm respectively. Our instrument setup is the same as that described in \citet{Horch12} and our analysis methodology is outlined by \citet{Kane14}. Briefly, we estimate the limiting magnitude $\Delta m$ (difference between local image maxima and minima) as a function of target separation resulting in a 5$\sigma$ detection curve. More details on the derivation of the DSSI detection limits can be found in \citet{Howell11}. All of our DSSI $\rho$ CrB observations show no evidence of a stellar companion to the host star. Figure \ref{fig:contrast} shows the detection curve from the 880 nm image acquired for $\rho$ CrB on the night of July 25 2014. The dashed curve is the cubic spline interpolation of the 5$\sigma$ detection limit from 0\farcs1 to 1\farcs{2}. The results exclude companions with $\Delta m \sim 5.2$ and $\Delta m \sim 7.5$ at separations of 0\farcs1 and 1\farcm4 respectively. Given the distance of $\rho$ CrB of 17.236 pc, these angular separations correspond to a physical exclusion range of 1.7--24.1 AU. We can thus rule out stellar companions in close proximity to the host star, supporting the evidence that the system is not a face-on triple star system, but instead a multi-planet system viewed at moderate to high inclination.

\subsubsection{Stability of the HD 143761 System}
%We also assessed the potential large mass of HD 143761 b using a stability analysis.
Two planet systems on circular orbits are likely to be unstable if $\Delta<2\sqrt{3}$ \citet{Gladman93} for all mutual inclinations, where
%\begin{equation}
$\Delta = \frac{a_{in} - a_{out}}{R_H}$
%\end{equation}
and R$_{H}$ is the Hill radius,
\begin{equation}
R_H = \left(\frac{M_{in} + M_{out}}{3M_\star}\right)^{1/3}\frac{a_{in}+a_{out}}{2}.
\end{equation}
Sky-projected inclinations smaller than 4 degrees combined with the \msini\ constraints listed in Table \ref{tab:dparams} imply large companion masses and push $\Delta$ below the $2\sqrt{3}$ stability threshold. If we assume that the system is dynamically stable then the presence of HD 143761 c rules out the low inclination orbit found by \citet{Gatewood01}, and \citet{Reffert11}.

HD 164922 b and c are widely separated and intuitively we would expect them to be in a stable configuration. For completeness, we calculate $\Delta=26.3$ for the minimum masses which suggests that this system configuration is likely to be dynamically stable for a long time.

\subsection{Additional Planet Candidates}
There is an additional significant periodic signal in the RV data for HD 42618 at a period of 388 days and a velocity semi-amplitude of 2 m/s. This would be a \msini = 22\mearth planet orbiting just outside 1 AU. The periodogram peak for this candidate falls above the eFAP threshold automatically calculated by our discovery pipeline. However, we do not believe that we have enough evidence to claim a concrete detection of a bona fide planet due to the proximity of this period to 1 year and incomplete phase coverage of the orbit. Telluric contamination of the template or problems with the barycentric correction could inject a false signal with a period near 1 year \citep[][S. Wang personal communication]{Wright14, Fischer16}. The amplitude and periodicity of this signal depends on the outcomes of modeling HD 42618 b, and MCMC runs for models including this candidate fail to converge. We will continue to monitor this star intensively with both Keck and the APF to confirm or refute this planet candidate in the upcoming years.

We find a candidate periodicity in the HD 164922 system with a period of 41.7 d and an amplitude of 1.9 \mse. The eFAP of this 2DKLS periodogram peak is 0.00098 and falls just above our 0.1\% eFAP threshold. However, we do not consider this to be a viable planet candidate due to its marginal detection and proximity to the expected rotation period for this star \citep[44 days,][]{Isaacson10}. Further monitoring and a detailed analysis that includes the effects of rotational modulation of starspots is needed to determine the nature of this signal. 

There is no evidence for significant periodic signals from other candidates in the periodograms for HD 143761. However, visually there appears to be some long-period structure in the residuals to our most probable model (see Figure \ref{fig:fitplot_143761}). This marginal variability, if real, likely has a period of $\ge10$ years and an amplitude of only a few \ms and it appears to be at a shorter period than the stellar magnetic activity cycle as seen in the \shk\ values for this star. Long-term monitoring of this target is required to determine if this signal is real and the signature of a planetary companion.

\subsection{Chromospheric Activity}
\label{sec:activity}
\begin{figure}[h]
         \begin{center}
              \includegraphics[width=0.45\textwidth]{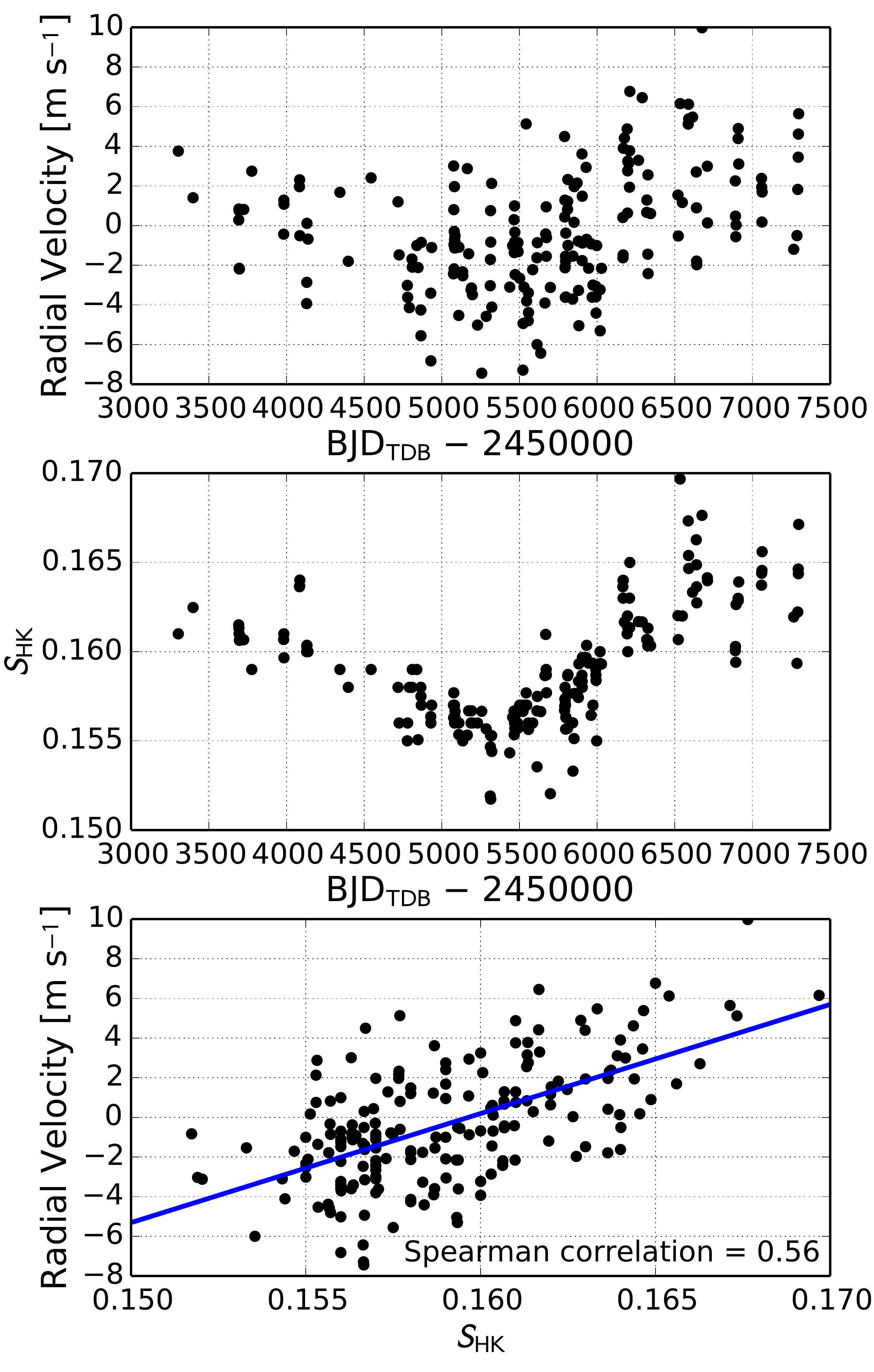}
         \end{center}
         \caption{
              Velocity-activity correlation for HD 42618. A discussion of the chromospheric activity of each of the three stars can be found in Section \ref{sec:activity}.
              \emph{Top:} Binned RV time series of the post-upgrade Keck data with planet b.
              \emph{Middle:} Binned \shk\ time series of the post-upgrade Keck data only. Note the similarities between the variability in the top and middle panels.
              \emph{Bottom:} Spearman rank correlation test of the velocities with \shk\ values \citep{Spearman1904}. We do not subtract this correlation from the RVs of HD 42618 but instead model the magnetic activity cycle as an additional long-period Keplerian (see Section \ref{sec:characterization}).
         }
         \label{fig:svals}
\end{figure}

These stars were all selected to be part of the APF-50 survey of nearby stars due, in part, to their extremely low mean chromospheric activity of $\rphk \le -4.95$. However, in the case of HD 42618 we do detect significant long-period variability in the \shk\ values that is strongly correlated with the RVs (see Figure \ref{fig:svals}) that is likely the signature of the stellar magnetic activity cycle. We do not find any significant periodic signals in the \shk\ values after removal of this long-period trend that might be the signature of rotation. However, we clearly identify the rotation period of the star to be 16.9 days in CoRoT photometry (see Section \ref{sec:corot_phot}). We account for the activity cycle in the RV data of HD 42618 by including an additional long-period Keplerian signal in the model.

\begin{figure}[h]
         \begin{center}
              \includegraphics[width=0.45\textwidth]{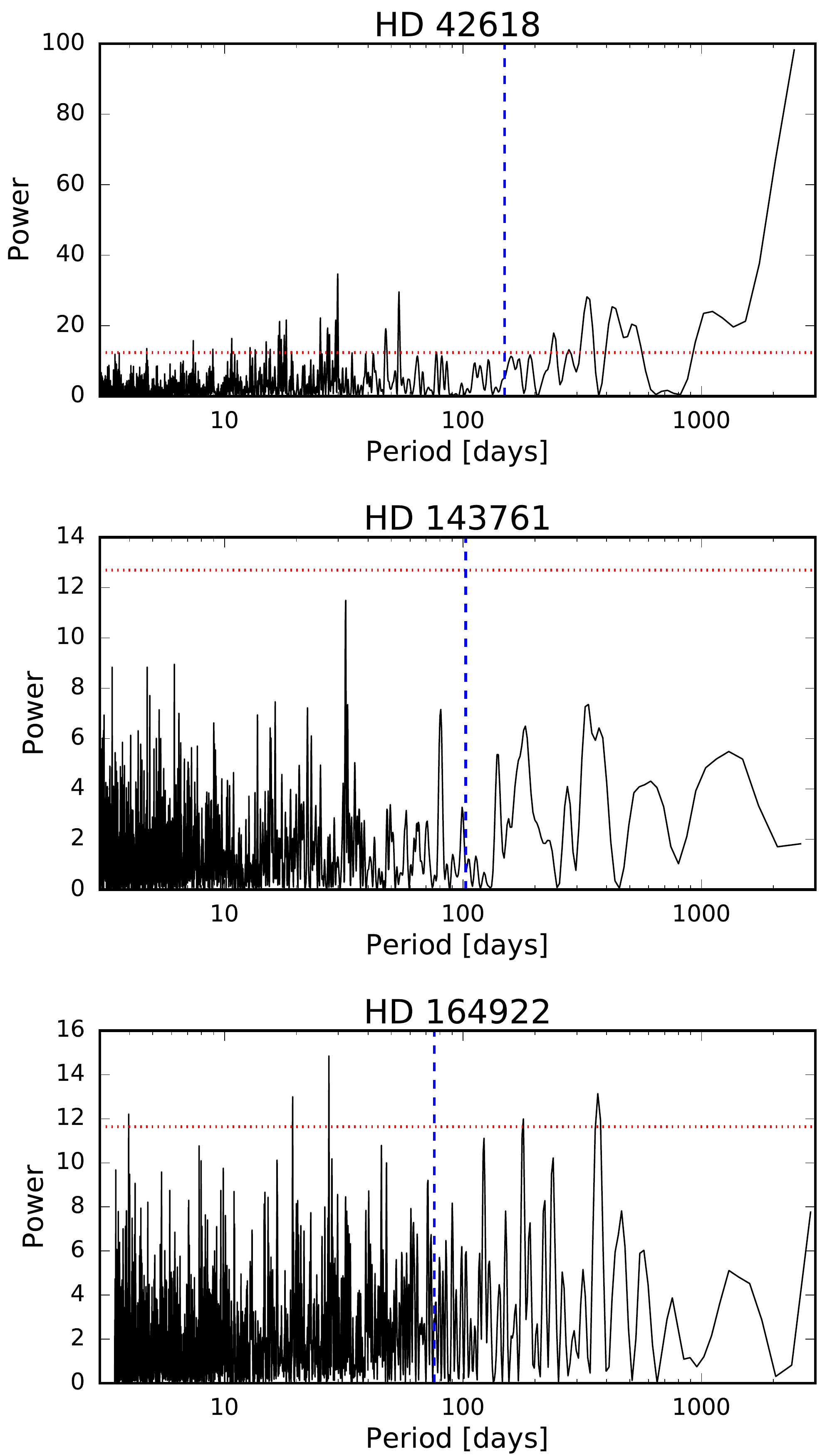}
         \end{center}
         \caption{ 
              Lomb-Scargle periodograms of \shk\ chromospheric activity. In each panel the period of the planet announced in this work is marked by the blue dashed line
              and the power corresponding to an analytical false alarm probability of 1\% is marked by the red dotted line \citep{Schwarzenberg-Czerny98}. \shk\ values measured from spectra with SNR$<$40 per pixel or exposure times $>$25\% longer then the median exposure time (due to clouds and/or seeing) can be badly contaminated by the solar spectrum and cause our \shk\ extraction pipeline to produce large outliers. These measurements were excluded before calculating the periodograms. No significant periodicity is detected in any of the stars at the orbital periods of the new planets. 
              \emph{Top:} Periodogram of \shk\ values for HD 42618.
              \emph{Middle:} Periodogram of \shk\ values for HD 143761.
              \emph{Bottom:} Periodogram of \shk\ values for HD 164922.
         }
         \label{fig:sval_multi}
\end{figure}

HD 164922 shows only a linear trend in the \shk\ values but we do not detect the effect of this change in chromospheric activity in the residuals to the two planet fit. There is also a very weak peak in the Lomb-Scargle periodogram \citep[L-S,][]{Lomb76, Scargle82} of the \shk\ values at 37.8 days (see Figure \ref{fig:sval_multi}). This may be the signature of stellar rotation since this is near the expected period for a star of this type and age \citep{Isaacson10}. However, this rotation period is well separated from the orbital periods of the two planets and does not influence our two planet fits.

We do not detect any long term variability in the \shk\ values of HD 143761 but we see a clear peak in a periodogram of the \shk\ values at 18.5 days that is likely caused by the rotational modulation of star spots (see Figure \ref{fig:sval_multi}). Since the rotation period is well separated from the periods of either of the planets orbiting HD 143761 this does not affect our Keplerian modeling and is likely absorbed into the stellar jitter term. 

\section{Photometry}
\label{sec:photometry}

\subsection{CoRoT Photometry of HD 42618}
\label{sec:corot_phot}
HD 42618 was the target of high cadence, high precision, continuous photometric monitoring for $\approx$0.5 years with the purpose of detecting solar like oscillations \citep{Baglin12}. We perform a simple polynomial detrending of the space-based photometry. After removing large ramp-shaped features at the start of two long observing campaigns we then fit an 8th order polynomial to all continuous segments of the data. These segments are 2-20 days in length.

\begin{figure}
         \begin{center}
              \includegraphics[width=0.45\textwidth]{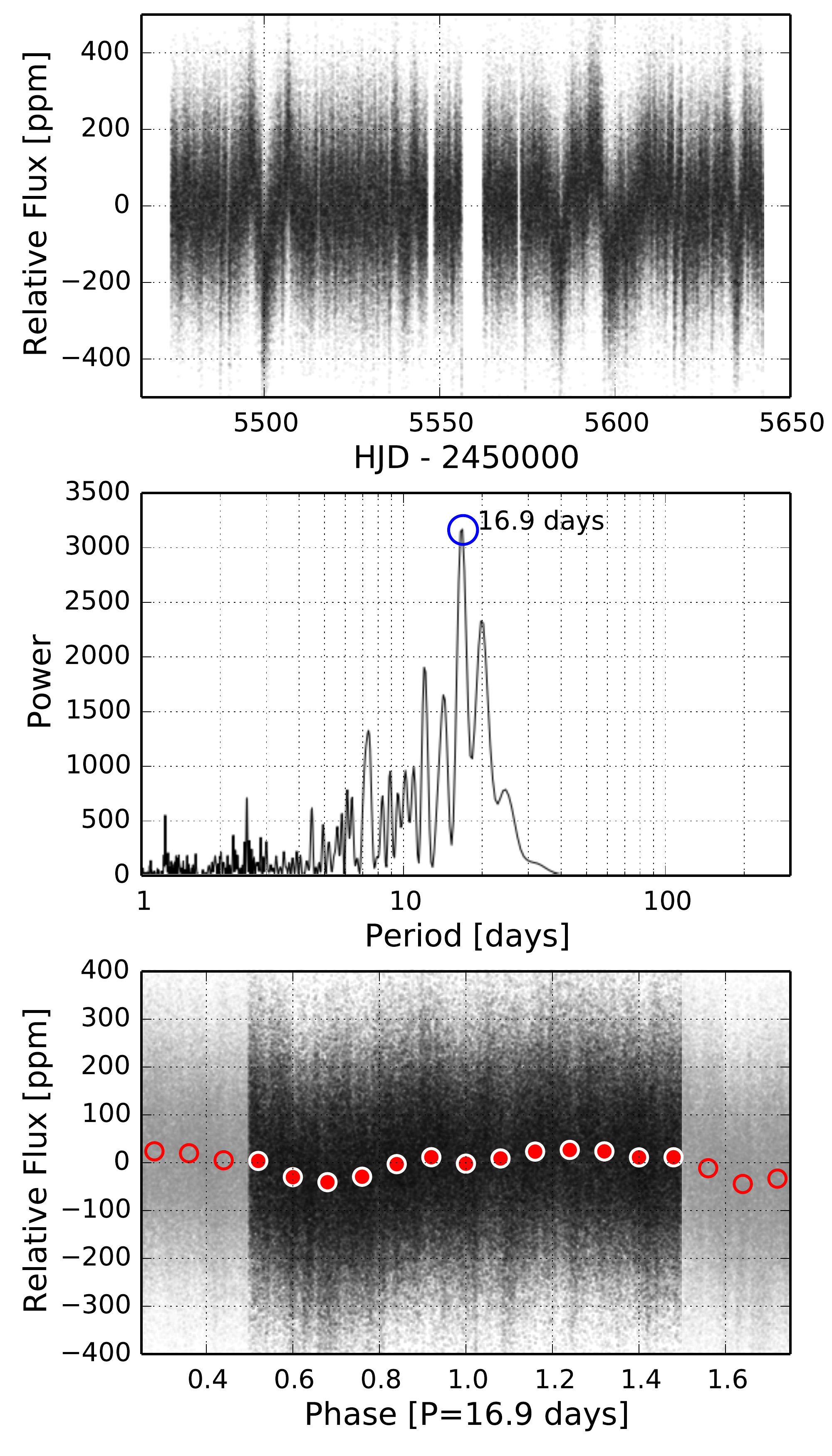}
         \end{center}
         \caption{
              CoRoT photometry of HD 42618 discussed in Section \ref{sec:corot_phot}.
              \emph{Top:} Detrended light curve.
              \emph{Middle:} Lomb-scargle periodogram of the light curve.
              \emph{Bottom:} Photometry phase-folded to the period corresponding to the highest peak in the Lomb-Scargle periodogram (16.9 days). We also bin the photometry with bin widths of 0.05 units of phase (red circles).
         }
         \label{fig:corot_peri}
\end{figure}

We use the detrended photometry to look for periodic photometric variability that might be caused by rotationally modulated star spots. We detect significant variability with a period of 16.9 days but with a broad distribution of periodogram power around the highest peak (see Figure \ref{fig:corot_peri}). This is a clear signature of stellar rotation with slightly changing phase and/or differential rotation which creates a broad distribution of increased power in Fourier space near the true rotation period. This star is very similar to the Sun in mass, age, and chemical abundance so the fact that the rotation period is also similar to that of the Sun (26 days) is not surprising. However, we note that the precise location of the highest peak in the Lomb-Scargle periodogram of HD 42618 depends on the polynomial order used to detrend the CoRoT photometry. We also tried high-pass filtering the CoRoT photometry using running median filters with window widths of 20-50 days and only analyzing continuous segments of data longer than 20 days. We found that the period of highest power is somewhat variable but always falls between 12-18 days. Since the periodogram period is dependent on the detrending algorithm we can't determine the rotation period of HD 42618 precisely, but we estimate that it falls within the range of 12-18 days. We do not detect any significant periodic signal in the RV data near the photometric period.

\begin{figure}
         \begin{center}
              \includegraphics[width=0.45\textwidth]{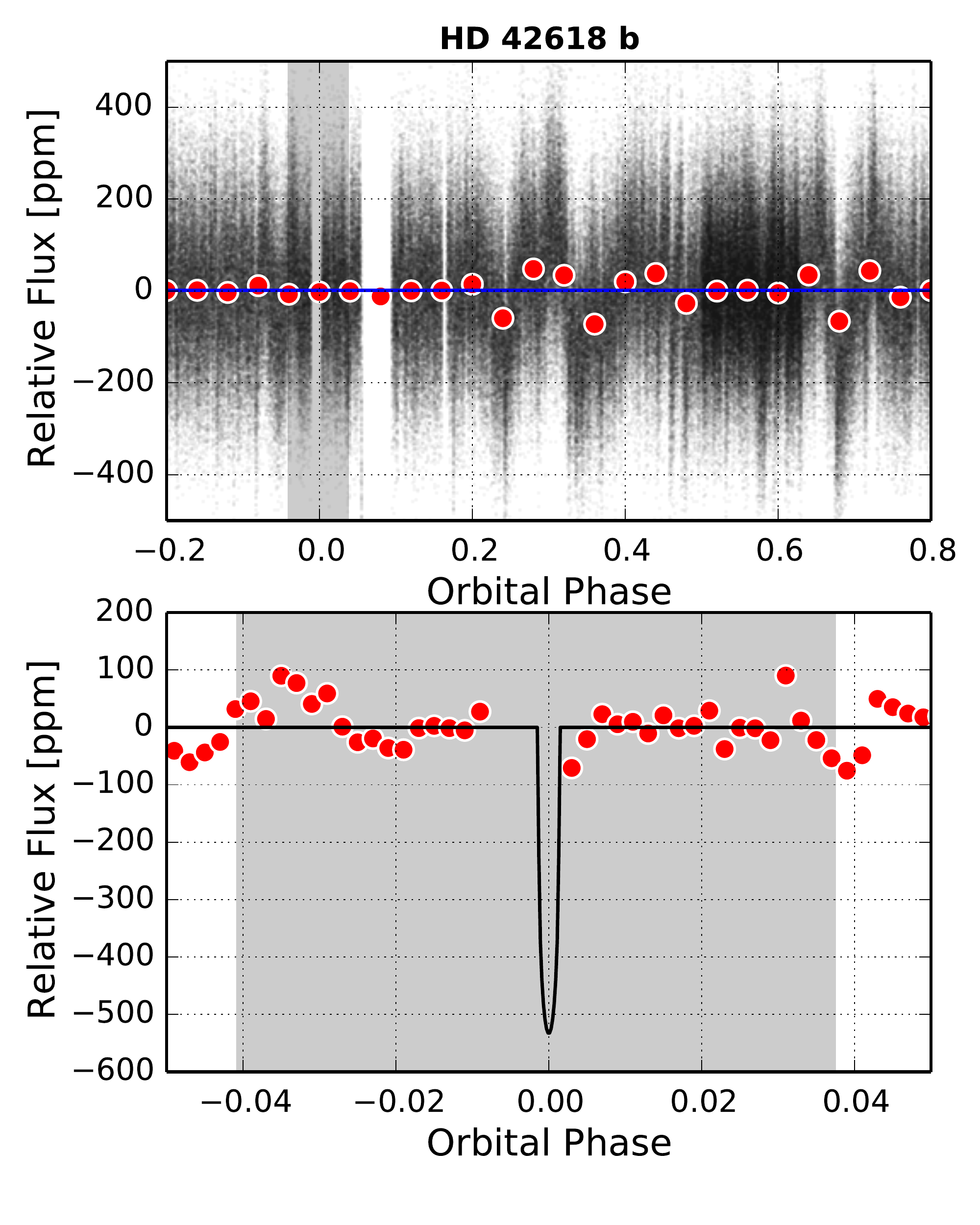}
         \end{center}
         \caption{
              CoRoT photometry of HD 42618 phase-folded to the orbital period of planet b. The transit search for HD 42618 b is discussed in Section \ref{sec:corot_phot}.
              \emph{Top:} Photometry over the full orbital phase of planet b. The red circles are binned photometric measurements with bin widths of 0.04 units of orbital phase. The grey shaded region shows the 1 $\sigma$ uncertainty on the time of inferior conjunction derived from the RV modeling.
              \emph{Bottom:} Same as top panel with the x-axis zoomed-in near the time of inferior conjunction. In this panel we only plot the measurements binned with bin widths of 0.002 units of orbital phase. Again, the shaded region represents the 1 $\sigma$ uncertainty on the time of inferior conjunction. The black transit model shows the predicted transit depth for a solid iron planet using the mass-radius relation of \citet{Weiss14}.
         }
         \label{fig:corot_phase}
\end{figure}

We searched through the detrended CoRoT light curve using the \texttt{TERRA} planet detection algorithm \citep{Petigura13b,Petigura13a}. We did not find any periodic box-shaped dimmings with signal-to-noise ratios (SNR) greater than 7. Searches of Kepler photometry commonly require SNR$~>7$ \citep{Jenkins13} or SNR$~>12$ \citep{Petigura13a}, though the SNR threshold depends on the noise structure of the photometry. We conclude that there are no transiting planets having periods between 0.5 and 60 days with transits that are detectable above Poisson, stellar, and instrumental noise. Given the photometric noise properties of HD 42618, we can rule out planets with transits deeper than $\approx$150 ppm ($\approx$1.3 \rearthe) at $\approx$5 day orbital periods and transits deeper than $\approx$300 ppm ($\approx$1.9 \rearthe) for $\approx$50 day orbital periods. The \emph{a priori} transit probability for HD 42618 b is only 0.8\% so it is not surprising that we do not detect transits.

\subsubsection{Asteroseismic Mass Determination}
\label{sec:asteroseismology}
Convection in the outer layers of a star excites stochastic oscillations, which can be observed on the stellar surface. In the case of main sequence stars, these oscillations manifest themselves as periodic variations on the order of cm/s in radial velocity data or $\sim$ppm in photometric data. Photometric space telescopes such as CoRoT proved to be quite effective for measuring and characterizing these oscillations, which can be used to derive global stellar properties (such as radius, mass and age), as well as to constrain the stellar interior \citep[e.g.][]{Michel08,Chaplin13}.

\begin{figure}
         \begin{center}
              \includegraphics[width=0.5\textwidth]{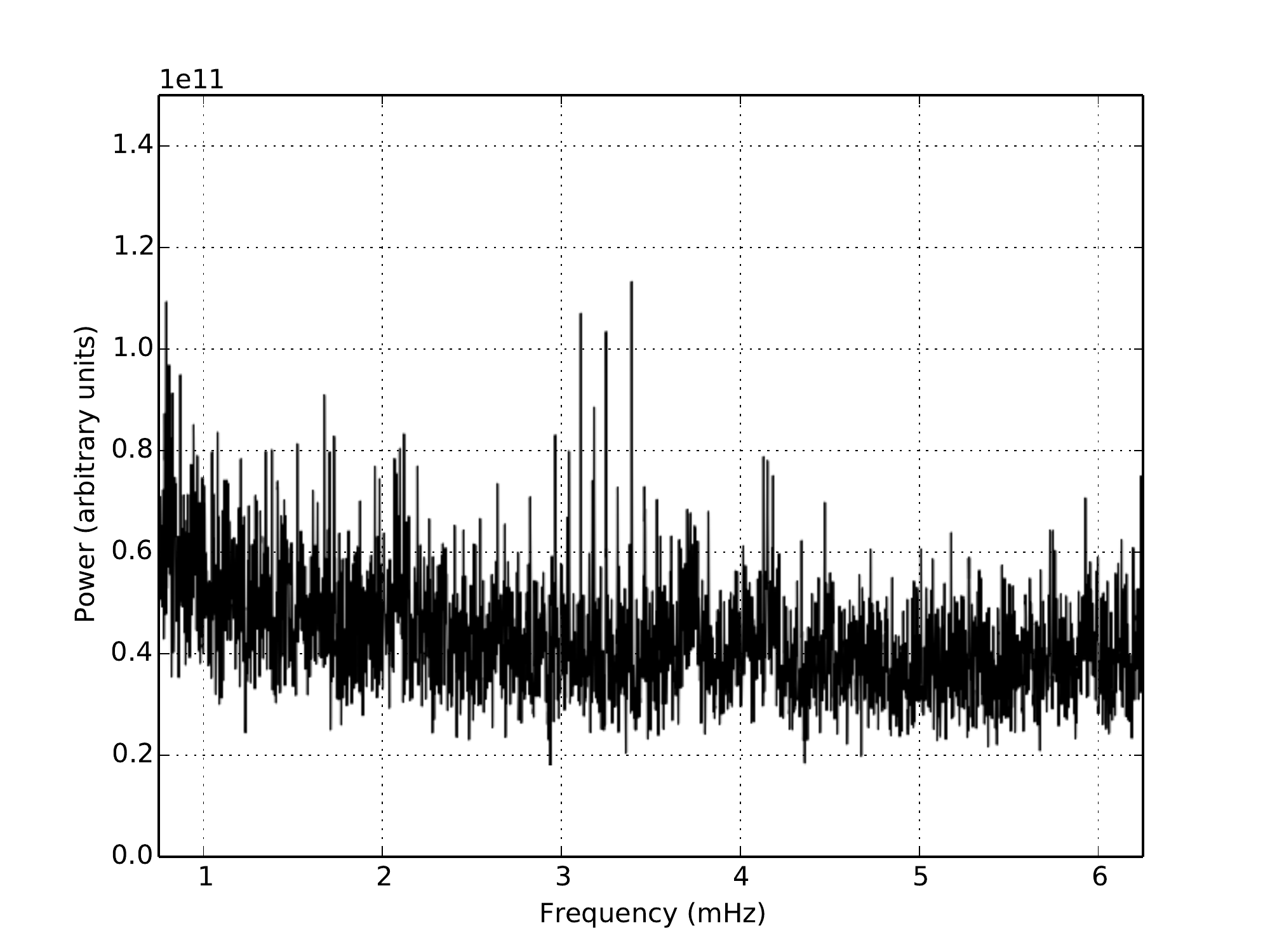}
         \end{center}
         \caption{
              Smoothed one-dimensional power spectrum of HD 42618 from the CoRoT data. The comb of peaks in the power spectrum near a frequency of 3 mHz is the signature of solar-like asteroseismic oscillations. Our asteroseismic analysis of HD 42618 is described in Section \ref{sec:asteroseismology}.
         }
         \label{fig:powspec}
\end{figure}

We measured the mass of HD42618 from CoRoT photometry obtained during two long observing runs spanning 79 and 94 days, respectively.  Through a Fourier analysis of the CoRoT lightcurve, we produced the power spectral density function shown in Figure \ref{fig:powspec}. We then stacked the power spectrum in equally sized pieces to create an echelle diagram, revealing the distinct $l$=0, 1, and 2 latitudinal modes of oscillation. We then collapsed this echelle diagram, effectively creating a binned power spectrum, and fit the power excess with a Gaussian to measure a maximum oscillation power frequency $\nu_{\mathrm{max}}$ of 3.16 $\pm$ 0.10 mHz. We then collapsed the echelle diagram along the perpendicular axis to preserve the frequency spacing, computed its autocorrelation, and fit the autocorrelation with a Gaussian to measure a large oscillation frequency spacing $\Delta\nu$ of 141.6 $\pm$ 0.8 $\mu$Hz. Using scaling relations \citep{christensen1983, kjeldsen1995, kallinger2010, huber2011}, solar parameters taken from \citet{huber2011}, and an effective temperature equivalent to the Sun's within errors \citep{morel2013} we measure an asteroseismic radius of 0.95 $\pm$ 0.05 R$_{\odot}$, and an asteroseismic mass of 0.93 $\pm$ 0.13 M$_{\odot}$. This is in agreement with a previous CoRoT asteroseismic analysis \citep{Barban13} and our estimate of the stellar mass and radius of HD42618 using our spectroscopic constraints and the \citet{Torres10} relations. The precision on the asteroseismic mass is lower compared to our spectroscopic+isochrone mass but it is much less model-dependent. If we were to fit the spectroscopic parameters to isochrones derived using different input physics we may find that the error on the spectroscopic mass is much larger. We adopt the higher precision, spectroscopic mass for all calculations of planet minimum masses and orbital separations.

\subsection{APT Photometry}
\label{sec:APT}
%We obtained long-term photometric monitoring data for each of the three stars from the 0.8 m automated photometric telescope (APT) at Fairborn observatory to look for photometric signatures of stellar activity and search for transits of each planet. In each case we measured the flux using photomultiplier tubes in Str\"{o}mgren $b$ and $y$ filters simultaneously and later combine the measurements into a single ($b+y$)/2 passband to improve the SNR. The automated telescope observes the target star and several comparison stars in short succession to measure the flux of the target relative to the companion stars to correct for atmospheric transparency fluctuations. A detailed description of the observing procedure and analysis techniques can be found in \citet{Henry13}.
Long-term photometric observations of HD~42618, HD~143761, and HD~164922 were collected with Tennessee State University's T11 0.80~m, T4 0.75~m, and T12 0.80~m APTs at Fairborn Observatory.  These three stars are among a collection of more than 300 being observed by the APTs to study magnetic cycles in solar-type stars (e.g., \citet{Lockwood13} and references therein) and have APT observational histories between 15 and 23 years.  At the beginning of the APF survey, the vast majority of the target stars were already being observed by the APTs.  The remaining few have been added so that all 51 stars in the APF survey are also being observed nightly by the APTs.  

The APTs are equipped with two-channel precision photometers that use a dichroic filter and two EMI 9124QB bi-alkali photomultiplier tubes to measure the Str\"omgren $b$ and $y$ pass bands simultaneously.  The APTs are programmed to make differential brightness measurements of a program star with respect to three comparison stars.  For the APF project, we use the two best comparison stars (C1 and C2) and compute the differential magnitudes  $P-C1$, $P-C2$, and $C2-C1$, correct them for atmospheric extinction, and transform them to the Str\"omgren system.  To maximize the precision of the nightly observations, we combine the differential $b$ and $y$ observations into a single $(b+y)/2$ ``passband" and also compute the differential magnitudes of the program star against the mean brightness of the two comparison stars.  The resulting precision of the individual $P-(C1+C2)/2_{by}$ differential magnitudes ranges between $\sim0.0010$ mag and $\sim0.0015$ mag on good nights.  Further details of our automatic telescopes, precision photometers, and observing and data reduction procedures can be found in \citet{Henry99}, \citet{Eaton03}, and \citet{Henry13}.

\begin{figure}
         \begin{center}
              \includegraphics[width=0.45\textwidth]{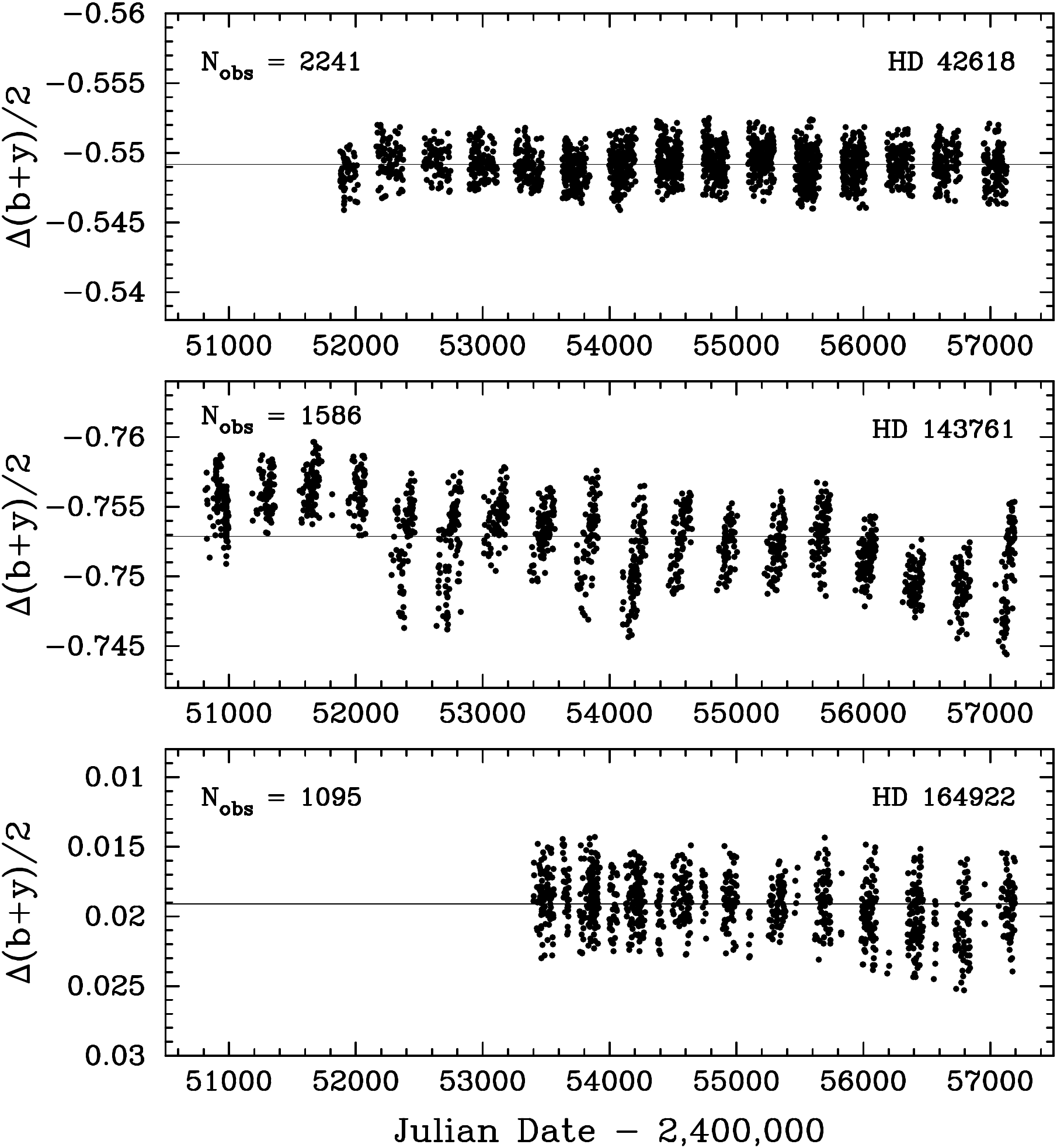}
         \end{center}
         \caption{
             Long-term photometric observations of the planetary candidate host 
stars HD 42618 (top), HD 143761 (middle), and HD 164922 (bottom) acquired
with TSU's T11 0.80~m, T4 0.75~m, and T12 0.80~m APTs at Fairborn Observatory
in southern Arizona.  All three stars are plotted with identical x and y 
scales.  The horizonal line in each panel marks the mean of each data set. The APT photometry and analysis are described in Section \ref{sec:APT}.
         }
         \label{fig:apt_phot_all}
\end{figure}

\subsubsection{APT Photometry of HD 42618}
We collected 2241 relative flux measurements of HD 42618 over the past 15 years. We search for photometric variability on short timescales by first subtracting the mean magnitude from each observing season to remove seasonal offsets. This removes all astrophysical and systematic instrumental variability on timescales longer then one year. %The RMS of the 15 year dataset is 0.125 mmag after removing the seasonal offsets.
A L-S period search returns a very weak periodicity with a period of 16.5 days and an amplitude of 0.3 mmag. This may be the same signature of stellar rotation as detected in the CoRoT data but it is too close to the precision limit of the ground-based dataset to be certain.

\begin{figure}
         \begin{center}
              \includegraphics[width=0.45\textwidth]{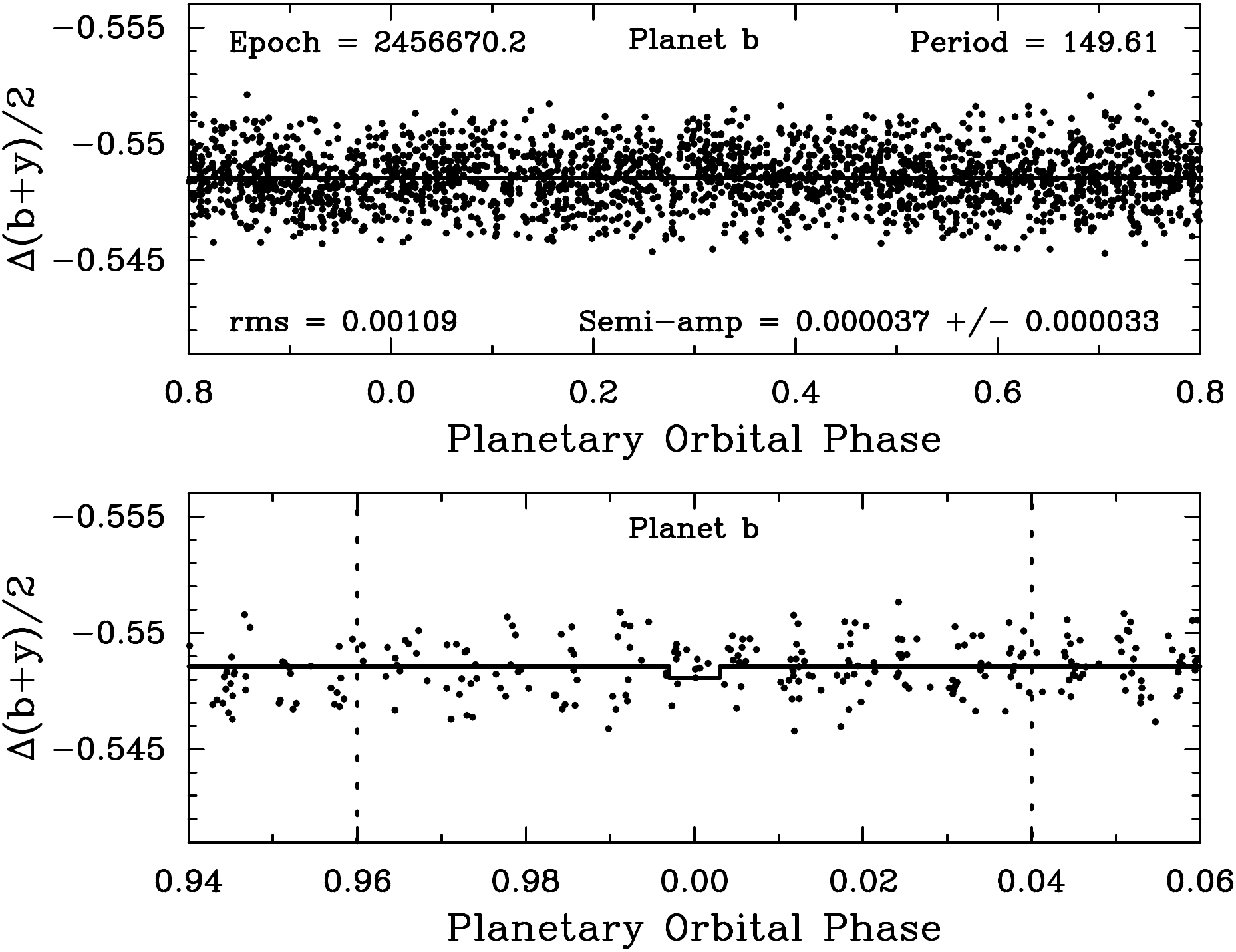}
         \end{center}
         \caption{
             \emph{Top:} Fifteen years of photometric observations of HD~42618 from the
top panel of Fig. \ref{fig:apt_phot_all} plotted against the 149.6-day planetary orbital period 
and time of conjunction derived from the radial velocity observations. A 
least-squares sine fit on the radial velocity period gives a semi-amplitude 
of just $0.000037~\pm~0.000033$ mag, firmly establishing the lack of stellar 
activity on the radial velocity period and thus confirming the presence of 
stellar reflex motion caused by an orbiting planet.
	\emph{Bottom:} Closeup of the
observations near the time of planetary conjunction at phase 0.0.  The solid
line shows a toy model transit of a sphere of constant 1.0 g cm$^{3}$ density and radius determined by the relation of \citet{Weiss14}. The
vertical lines mark the uncertainty in the predicted transit times.  Our
current photometric observations provide no evidence for transits.
         }
         \label{fig:42618_transit}
\end{figure}

Photometric variability on the timescale of the orbital period may indicate that the RV fluctuations are the result of rotational modulation of star spots \citep{Queloz01}. We find no evidence of photometric variability at the orbital period of the planet to the limit of our photometric precision. A least-squares sine fit on the orbital period of HD 42618 gives a semi-amplitude of just 0.000037 mag, showing the complete absence of any surface activity that could affect the radial velocities. Figure \ref{fig:apt_phot_all} shows the full photometric dataset and Figure \ref{fig:42618_transit} shows the photometry phase folded to the orbital period of planet b. The lack of variability at the orbital period is consistent with the results of the CoRoT analysis and strengthens our claim that the RV fluctuations are caused by a Neptune-mass planet orbiting HD 42618. We also find no evidence of the transit of HD 42618 b in the APT data.

\begin{figure}
         \begin{center}
              \includegraphics[width=0.45\textwidth]{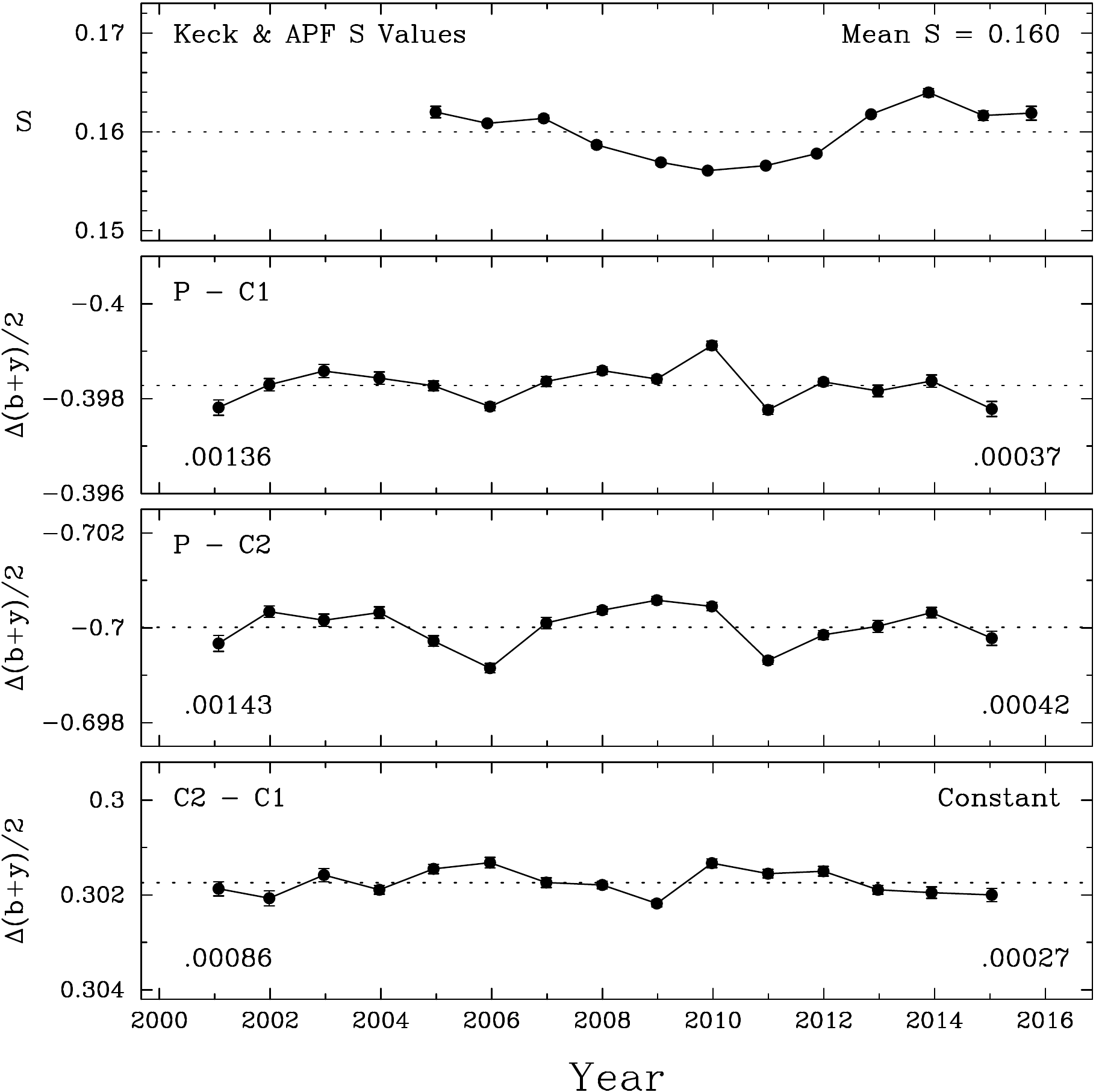}
         \end{center}
         \caption{
             \emph{Top:} Yearly means of the Mt. Wilson-calibrated 
S values acquired along with the radial velocity measurements. 
             \emph{Bottom three panels:} Yearly means of HD~42618's $P-C1$, $P-C2$, and 
$C2-C1$ differential magnitudes.  The horizontal dotted lines designate the 
grand means of the observations while the numbers in the lower-left and 
lower-right give the total range and standard deviation of each data set,
respectively. It is evident that we have resolved low-level brightness 
variability in HD~42618 compared to the two comparison stars, $C1$ and $C2$. Low-amplitude
cycles of roughly 0.001~mag over 5 years are seen in both the $P-C1$ and 
$P-C2$ light curves. There appears to be little or no correlation of S values with photometric brightness.
%The S values 
%are anti-correlated with brightness in the sense that the star gets fainter
%as it becomes more active.  This is typical of stars much younger than 
%HD~42618 whose brightness variability is dominated by cool starspots.
         }
         \label{fig:42618_means}
\end{figure}

Figure \ref{fig:42618_means} shows the mean \shk\ values from Keck and APF and the long-term photometric variability of HD 42618 by plotting the seasonal means of both the S values and APT photometry. For old solar type stars we expect a positive correlation of chromospheric activity as measured by the \shk\ values with the mean brightness since the number of bright faculae regions on the star increases during more active periods. However, in this case we see no correlation of mean brightness with \shk. Young stars typically show a negative correlation of brightness with \shk\ because their photometric variations are spot-dominated instead of faculae dominated. While somewhat unusual, this behavior is not unprecedented among similar stars \citep{Hall09}.
%HD 42618 may be in a transition period between its spot-dominated youth and faculae-dominated old age.

\subsubsection{APT Photometry of HD 164922}
\begin{figure}
         \begin{center}
              \includegraphics[width=0.45\textwidth]{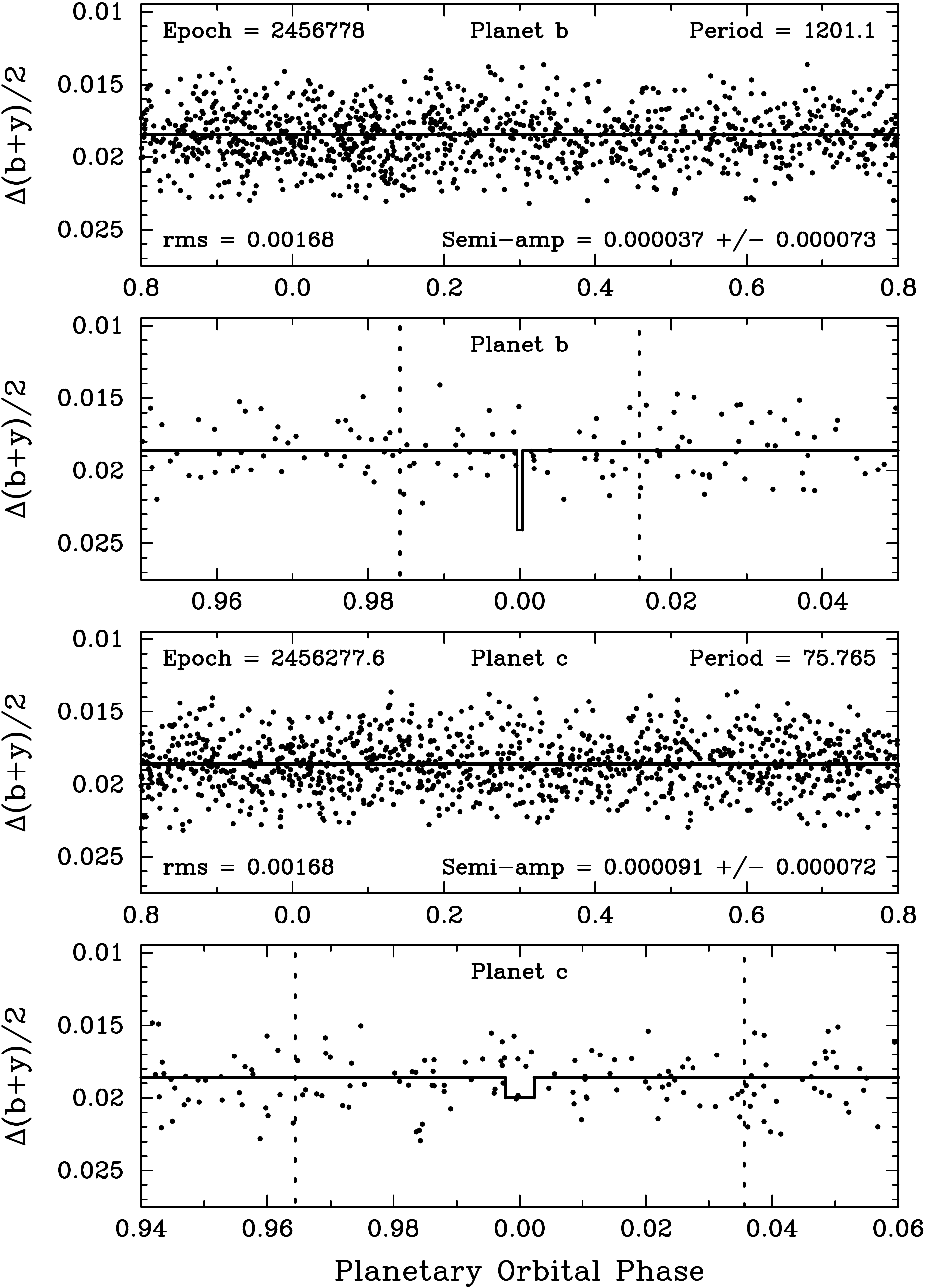}
         \end{center}
         \caption{
             Same as Figure \ref{fig:42618_transit} but for HD 164922 b and c.
             }
         \label{fig:164922_transit}
\end{figure}

We collected a total of 1095 photometric measurements for HD 164922 over the past 11 observing seasons from 2005 to 2015. As with HD 42618 we remove seasonal offsets from the photometry to search for short period variability and search for transits of HD164922 b and c.
%The RMS of the offset-corrected dataset is 1.7 mmag.
We find no significant periodic variability with a period between 1 and 100 days and do not detect the rotation period of the star. We also find no evidence of transits for either planet b or c or periodic photometric variability at the orbital period of either planet (Figure \ref{fig:164922_transit}). 

\begin{figure}
         \begin{center}
              \includegraphics[width=0.45\textwidth]{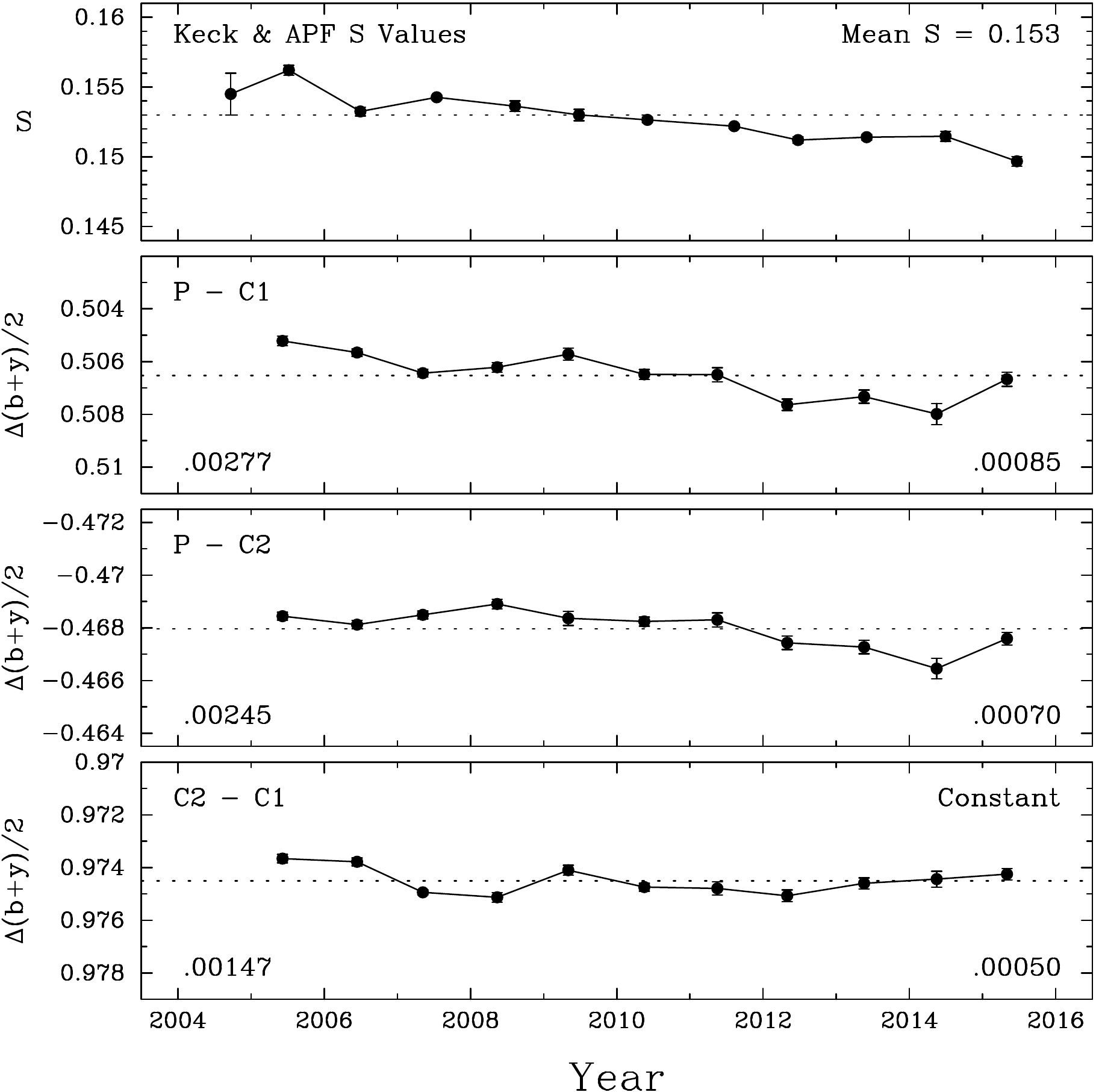}
         \end{center}
         \caption{
             Same as Figure \ref{fig:42618_means} but for HD 164922. In this case we see a positive correlation of the brightness of HD 164922 with the \shk\ index.
         }
         \label{fig:164922_means}
\end{figure}

We study the long-term photometric variability of HD 164922 by comparing the mean brightness of the star to the \shk\ activity index (Figure \ref{fig:164922_means}). In contrast to the results for HD 42618, in this case we see a clear positive correlation of the brightness of HD 164922 with the \shk\ index. It is interesting that we do not see a RV vs. \shk\ correlation for HD 164922, but we do find that the RV is strongly correlated with \shk\ for HD 42618 where the photometry is not. In other cases we have seen a correlation in both the photometry and RV data \citep[e.g.][]{Fulton15a}.
% It is possible that the long-period radial velocity signal in the HD 42618 system is due to a planet, and is correlated with \shh\ only by chance. Monitoring the star through the completion of the solar cycle should help to break the degeneracy between these two scenarios.

\subsubsection{APT Photometry of HD 143761}
\begin{figure}
         \begin{center}
              \includegraphics[width=0.45\textwidth]{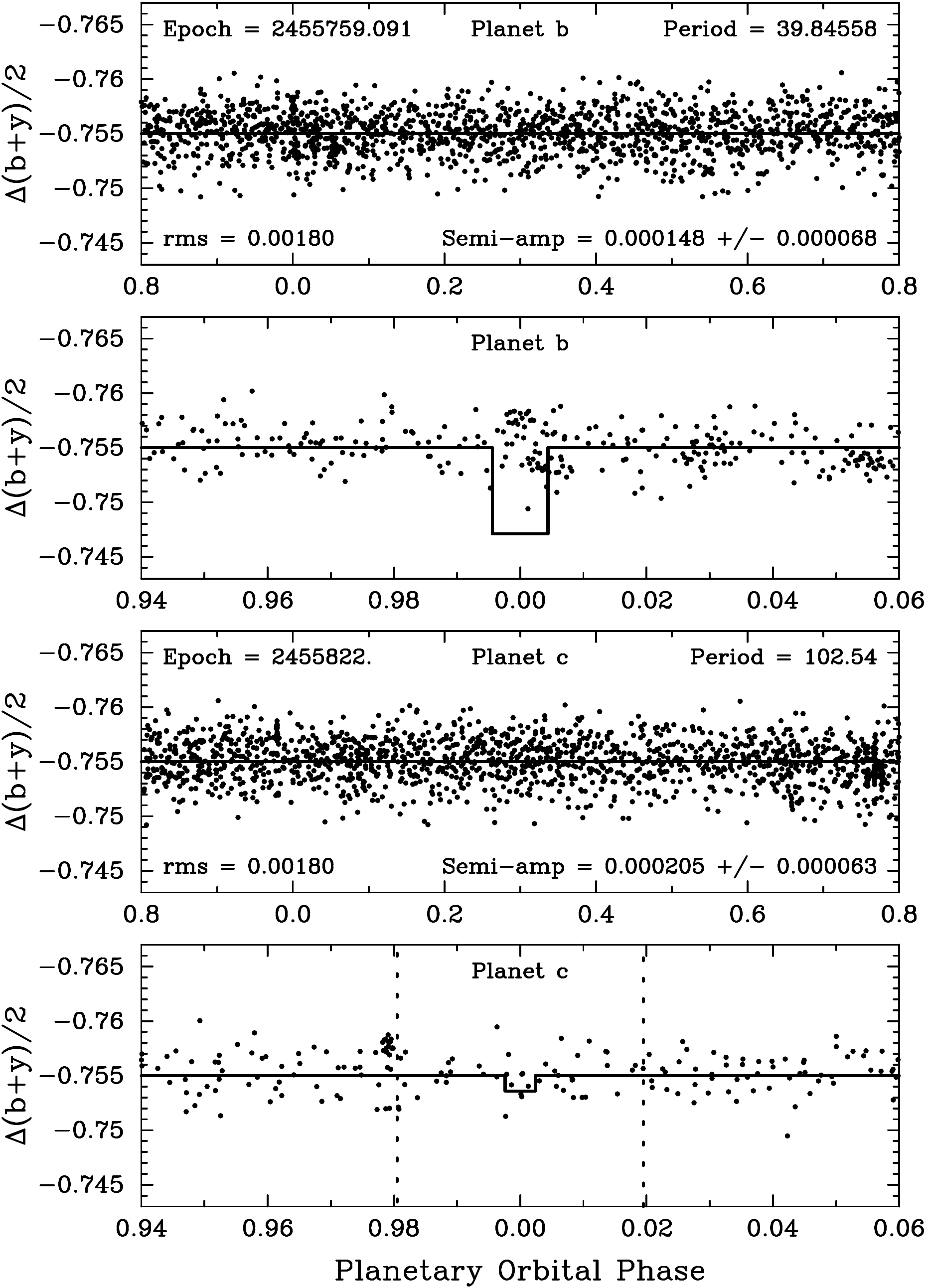}
         \end{center}
         \caption{
             Same as Figure \ref{fig:42618_transit} but for HD 143761 b and c.
             }
         \label{fig:143761_transit}
\end{figure}

We collected 1586 photometric measurements of HD 143761 over the past 18 observing seasons from 1997 to 2015 (Figure \ref{fig:apt_phot_all}). Our reduction and analysis techniques are the same as for HD 42618 and HD 164922 discussed in the previous two sections. We find no evidence of the photometric signature of rotationally modulated star spots or photometric variability at the orbital periods of HD 143761 b or c. There is no evidence of transits of either planet b or c (Figure \ref{fig:143761_transit}), however shallow transits of a rocky planet c can not be ruled out by this dataset.

\begin{figure}
         \begin{center}
              \includegraphics[width=0.45\textwidth]{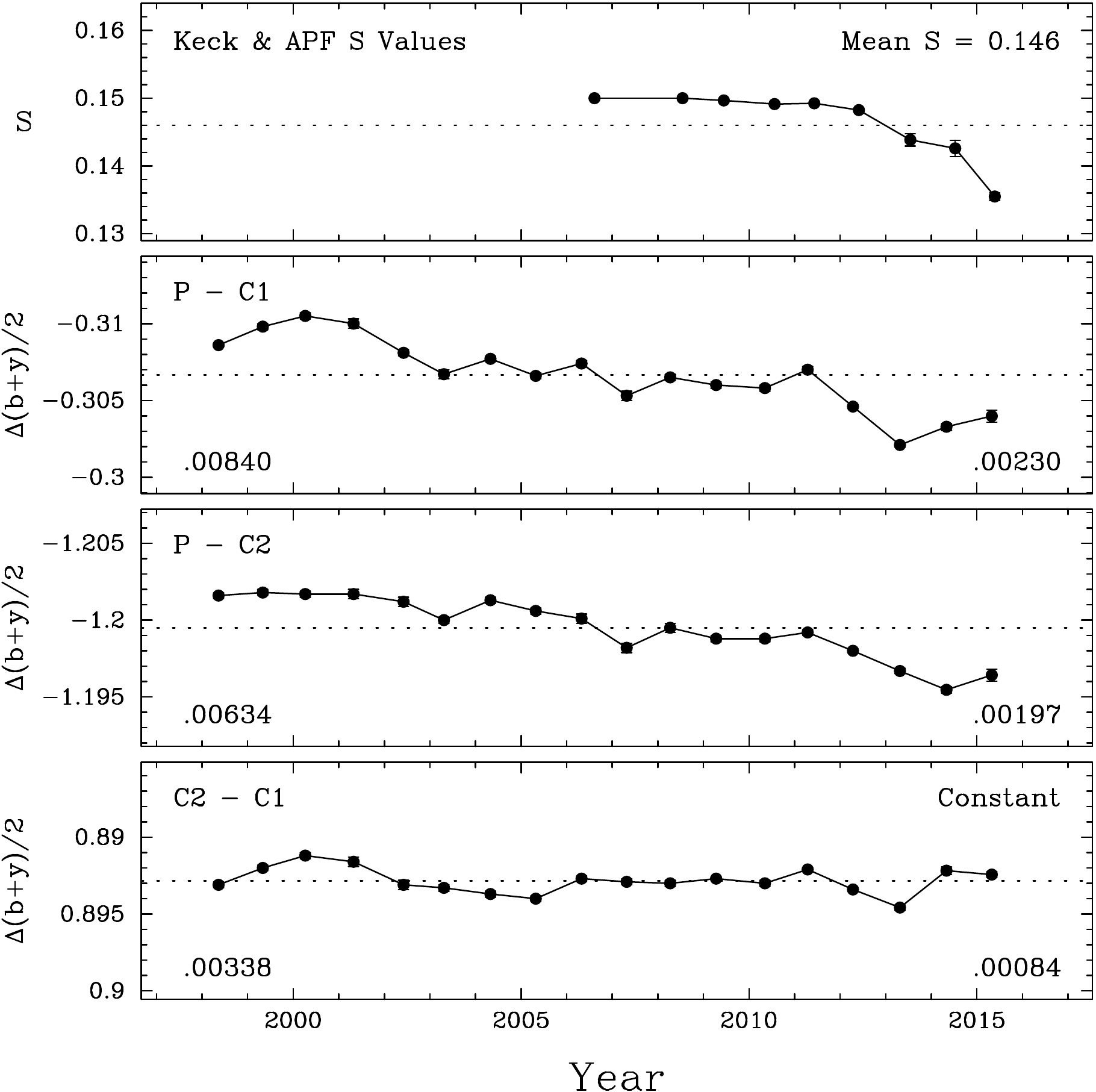}
         \end{center}
         \caption{
             Same as Figure \ref{fig:42618_means} but for HD 143761. In this case we again see a positive correlation of the brightness of HD 143761 with the \shk\ index.
         }
         \label{fig:143761_means}
\end{figure}

The mean photometric brightness binned by observing season is well correlated with the \shk\ values measured using Keck and APF as expected for an old solar type star.
%We see a slow decline of \shk\ values and a small decrease in overall brightness of the star.
As with HD 164922, we do not see a correlation of \shk\ with RV but there is a positive correlation of \shk\ with mean brightness (Figure \ref{fig:143761_means}). 

\newpage
\section{Discussion \& Summary}
\label{sec:discussion}
We present the discovery of three approximately Neptune mass planets orbiting three bright, nearby stars. The planet orbiting HD 42618 has a minimum mass of \msini\ = \mfourtwo\ and is the first discovered to orbit this star. There has been some discussion in the literature that stellar abundance patterns similar to the Sun might be evidence of the formation of terrestrial planets similar to those that exist in our solar system \citep[e.g.][]{Melendez09,Gonzalez10}. While we can not rule out the existence of terrestrial planets in the HD 42618 system, the presence of a temperate Neptune mass planet orbiting at 0.554 AU with an orbital period of 149 days shows that, at the present time, this system is not a close analogue to our own solar system. We cannot determine the initial planetary architecture of this system because migration may have played an important role to sculpt the current configuration.

We also detect the signature of the stellar magnetic activity cycle with a period of $\sim$12 years. This activity cycle manifests as a 3.1 \ms amplitude signal in the RV time series. We identify the rotation period of the star to be $\approx$17 days using public data from the CoRoT space telescope. Transits of HD 42618 b are expected to be extremely unlikely and we do not find any evidence for transits of this planet in the CoRoT data. This is a temperate planet receiving only 3.1 times the radiation that the Earth receives from the Sun. The planet's equilibrium temperature, assuming a bond albedo of 0.32 \citep{Demory14}, is 337 K. We perform an asteroseismic study of HD 42618 to detect solar like oscillations and measure a precise stellar radius and mass.

HD 164922 c is the second planet in a system previously known to host one Jupiter mass planet orbiting at 2.1 AU. The new planet announced in this work is a sub-Neptune mass planet with \msini\ =\ \monesix\ orbiting at a distance of $a=0.34$ AU and an orbital period of 75 days. This planet is also temperate with an equilibrium temperature of 401 K and receiving 6.3 times the flux received by the Earth from the Sun.

HD 143761 c is the second planet in a system previously known to host a warm Jupiter mass planet orbiting with a period of 39 days. The new planet is a super-Neptune with \msini\ =\ \monefour\ orbiting with a period of 102 days. This planet is the warmest of the three with a stellar irradiance 9.6 times that of the Earth-Sun system and an equilibrium temperature of 445 K. We find that the previous low inclination orbit for HD 143761 detected in Hipparcos astrometry can not be stable with the presence of HD 143761 c. 

These three planets are some of the nearest long period Neptune mass planets yet discovered. They demonstrate the capabilities of the combined Keck$+$APF-50 survey and are the beginning of a complete census of small planets in the local neighborhood.

{\it Facilities:} \facility{Automated Planet Finder (Levy)}, \facility{Keck:I (HIRES)}, \facility{CoRoT}

\acknowledgments{
We thank the many observers who contributed to the measurements reported here.
We thank Kyle Lanclos, Matt Radovan, Will Deich and the rest of the UCO Lick staff for their invaluable help shepherding, planning,
and executing observations, in addition to writing the low-level software that made the automated APF observations possible.
We thank Gail Schaefer for her help with the calculations related to the CHARA interferometric observations.
We thank Debra Fischer, Jason Wright, and John Johnson for their many nights of observing that contributed to the Keck data presented in this work.
We gratefully acknowledge the efforts and dedication of the Keck Observatory staff, 
especially Scott Dahm, Greg Doppman, Hien Tran, and Grant Hill for support of HIRES 
and Greg Wirth for support of remote observing.  
We are grateful to the time assignment committees of the University of Hawai`i, the University of California, and NASA 
for their generous allocations of observing time.  
Without their long-term commitment to RV monitoring, these planets would likely remain unknown.  
We acknowledge R.\ Paul Butler and S.\,S.\ Vogt for many years
of contributing to the data presented here.
A.\,W.\,H.\ acknowledges NSF grant AST-1517655 and NASA grant NNX12AJ23G.
L.\,M.\,W.\ gratefully acknowledges support from Ken and Gloria Levy.
D.H. acknowledges support by the Australian Research Council's Discovery Projects funding scheme
(project number DE140101364) and support by the National Aeronautics and Space Administration under
Grant NNX14AB92G issued through the Kepler Participating Scientist Program.
This material is based upon work supported by the National Science Foundation Graduate Research
Fellowship under Grant No. 2014184874. Any opinion, findings, and conclusions or recommendations
expressed in this material are those of the authors and do not necessarily reflect the views of the
National Science Foundation.
The Center for Exoplanets and Habitable Worlds is supported by the Pennsylvania State University,
the Eberly College of Science, and the Pennsylvania Space Grant Consortium.
J.T.W.\ and E.B.F.\ acknowledge support from multiple NASA Keck PI Data Awards, administered by the
NASA Exoplanet Science Institute, to follow multiple exoplanets systems including HD 164922 and
HD 143761 from semester 2010A to through 2012B (semester 2010B excluded).
J.T.W.\ acknowledges support from NSF grant AST-1211441
G.\,W.\,H.\ acknowledges support from NASA, NSF, Tennessee State University, and
the State of Tennessee through its Centers of Excellence program.
This research made use of the Exoplanet Orbit Database
and the Exoplanet Data Explorer at exoplanets.org.
Based on observations obtained at the Gemini Observatory, which is operated by the Association of Universities for
Research in Astronomy, Inc., under a cooperative agreement with the NSF on behalf of the Gemini partnership: the 
National Science Foundation (United States), the National Research Council (Canada), CONICYT (Chile), Ministerio de 
Ciencia, Tecnolog�a e Innovaci�n Productiva (Argentina), and Minist�rio da Ci�ncia, Tecnologia e Inova��o (Brazil).
This work made use of the SIMBAD database (operated at CDS, Strasbourg, France), 
and NASA's Astrophysics Data System Bibliographic Services.
Finally, the authors wish to extend special thanks to those of Hawai`ian ancestry 
on whose sacred mountain of Maunakea we are privileged to be guests.  
Without their generous hospitality, the Keck observations presented herein
would not have been possible.
}

%\newpage
\bibliographystyle{apj}
\bibliography{references}

\begin{thebibliography}{112}
\expandafter\ifx\csname natexlab\endcsname\relax\def\natexlab#1{#1}\fi

\bibitem[{{Asplund} {et~al.}(2009){Asplund}, {Grevesse}, {Sauval}, \&
  {Scott}}]{Asplund09}
{Asplund}, M., {Grevesse}, N., {Sauval}, A.~J., \& {Scott}, P. 2009, \araa, 47,
  481

\bibitem[{{Baglin} {et~al.}(2009){Baglin}, {Auvergne}, {Barge}, {Deleuil},
  {Michel}, \& {CoRoT Exoplanet Science Team}}]{Baglin09}
{Baglin}, A., {Auvergne}, M., {Barge}, P., {Deleuil}, M., {Michel}, E., \&
  {CoRoT Exoplanet Science Team}. 2009, in IAU Symposium, Vol. 253, IAU
  Symposium, ed. F.~{Pont}, D.~{Sasselov}, \& M.~J. {Holman}, 71--81

\bibitem[{{Baglin} {et~al.}(2012){Baglin}, {Michel}, \& {CoRoT
  Team}}]{Baglin12}
{Baglin}, A., {Michel}, E., \& {CoRoT Team}. 2012, in Astronomical Society of
  the Pacific Conference Series, Vol. 462, Progress in Solar/Stellar Physics
  with Helio- and Asteroseismology, ed. H.~{Shibahashi}, M.~{Takata}, \& A.~E.
  {Lynas-Gray}, 492

\bibitem[{{Barban} {et~al.}(2013){Barban}, {Deheuvels}, {Goupil}, {Lebreton},
  {Mathur}, {Michel}, {Morel}, {Ballot}, {Baudin}, {Belkacem}, {Benomar},
  {Boumier}, {Davies}, {Garc{\'{\i}}a}, {Hall}, {Mosser}, {Poretti},
  {R{\'e}gulo}, {Roxburgh}, {Samadi}, {Verner}, \& {the CoRoT Team}}]{Barban13}
{Barban}, C., {et~al.} 2013, Journal of Physics Conference Series, 440, 012030

\bibitem[{{Bonneau} {et~al.}(2011){Bonneau}, {Delfosse}, {Mourard}, {Lafrasse},
  {Mella}, {Cetre}, {Clausse}, \& {Zins}}]{2011A&A...535A..53B}
{Bonneau}, D., {Delfosse}, X., {Mourard}, D., {Lafrasse}, S., {Mella}, G.,
  {Cetre}, S., {Clausse}, J.-M., \& {Zins}, G. 2011, \aap, 535, A53

\bibitem[{{Bonneau} {et~al.}(2006){Bonneau}, {Clausse}, {Delfosse}, {Mourard},
  {Cetre}, {Chelli}, {Cruzal{\`e}bes}, {Duvert}, \&
  {Zins}}]{2006A&A...456..789B}
{Bonneau}, D., {et~al.} 2006, \aap, 456, 789

\bibitem[{{Boyajian} {et~al.}(2008){Boyajian}, {McAlister}, {Baines}, {Gies},
  {Henry}, {Jao}, {O'Brien}, {Raghavan}, {Touhami}, {ten Brummelaar},
  {Farrington}, {Goldfinger}, {Sturmann}, {Sturmann}, {Turner}, \&
  {Ridgway}}]{Boyajian08}
{Boyajian}, T.~S., {et~al.} 2008, \apj, 683, 424

\bibitem[{{Butler} {et~al.}(1996){Butler}, {Marcy}, {Williams}, {McCarthy},
  {Dosanjh}, \& {Vogt}}]{Butler96b}
{Butler}, R.~P., {Marcy}, G.~W., {Williams}, E., {McCarthy}, C., {Dosanjh}, P.,
  \& {Vogt}, S.~S. 1996, \pasp, 108, 500

\bibitem[{{Butler} {et~al.}(2006){Butler}, {Wright}, {Marcy}, {Fischer},
  {Vogt}, {Tinney}, {Jones}, {Carter}, {Johnson}, {McCarthy}, \&
  {Penny}}]{Butler06}
{Butler}, R.~P., {et~al.} 2006, \apj, 646, 505

\bibitem[{{Chaplin} \& {Miglio}(2013)}]{Chaplin13}
{Chaplin}, W.~J., \& {Miglio}, A. 2013, \araa, 51, 353

\bibitem[{{Chatterjee} \& {Tan}(2014)}]{Chatterjee14}
{Chatterjee}, S., \& {Tan}, J.~C. 2014, \apj, 780, 53

\bibitem[{{Chiang} \& {Laughlin}(2013)}]{Chiang13}
{Chiang}, E., \& {Laughlin}, G. 2013, \mnras, 431, 3444

\bibitem[{{Christensen-Dalsgaard} \& {Frandsen}(1983)}]{christensen1983}
{Christensen-Dalsgaard}, J., \& {Frandsen}, S. 1983, \solphys, 82, 469

\bibitem[{{Claret} \& {Bloemen}(2011)}]{2011A&A...529A..75C}
{Claret}, A., \& {Bloemen}, S. 2011, \aap, 529, A75

\bibitem[{{Coelho}(2014)}]{Coelho14}
{Coelho}, P.~R.~T. 2014, \mnras, 440, 1027

\bibitem[{{Cumming} {et~al.}(2008){Cumming}, {Butler}, {Marcy}, {Vogt},
  {Wright}, \& {Fischer}}]{Cumming08}
{Cumming}, A., {Butler}, R.~P., {Marcy}, G.~W., {Vogt}, S.~S., {Wright}, J.~T.,
  \& {Fischer}, D.~A. 2008, \pasp, 120, 531

\bibitem[{{Cutri} {et~al.}(2003){Cutri}, {Skrutskie}, {van Dyk}, {Beichman},
  {Carpenter}, {Chester}, {Cambresy}, {Evans}, {Fowler}, {Gizis}, {Howard},
  {Huchra}, {Jarrett}, {Kopan}, {Kirkpatrick}, {Light}, {Marsh}, {McCallon},
  {Schneider}, {Stiening}, {Sykes}, {Weinberg}, {Wheaton}, {Wheelock}, \&
  {Zacarias}}]{Cutri03}
{Cutri}, R.~M., {et~al.} 2003, VizieR Online Data Catalog, 2246, 0

\bibitem[{Demory(2014)}]{Demory14}
Demory, B.-O. 2014, The Astrophysical Journal, 789, L20

\bibitem[{{Dotter} {et~al.}(2008){Dotter}, {Chaboyer}, {Jevremovi{\'c}},
  {Kostov}, {Baron}, \& {Ferguson}}]{Dotter08}
{Dotter}, A., {Chaboyer}, B., {Jevremovi{\'c}}, D., {Kostov}, V., {Baron}, E.,
  \& {Ferguson}, J.~W. 2008, \apjs, 178, 89

\bibitem[{{Dumusque} {et~al.}(2011){Dumusque}, {Udry}, {Lovis}, {Santos}, \&
  {Monteiro}}]{Dumusque11}
{Dumusque}, X., {Udry}, S., {Lovis}, C., {Santos}, N.~C., \& {Monteiro},
  M.~J.~P.~F.~G. 2011, \aap, 525, A140

\bibitem[{{Eastman} {et~al.}(2013){Eastman}, {Gaudi}, \& {Agol}}]{Eastman13}
{Eastman}, J., {Gaudi}, B.~S., \& {Agol}, E. 2013, \pasp, 125, 83

\bibitem[{{Eaton} {et~al.}(2003){Eaton}, {Henry}, \& {Fekel}}]{Eaton03}
{Eaton}, J.~A., {Henry}, G.~W., \& {Fekel}, F.~C. 2003, in Astrophysics and
  Space Science Library, Vol. 288, Astrophysics and Space Science Library, ed.
  T.~D. {Oswalt}, 189

\bibitem[{{Fabrycky} {et~al.}(2014){Fabrycky}, {Lissauer}, {Ragozzine}, {Rowe},
  {Steffen}, {Agol}, {Barclay}, {Batalha}, {Borucki}, {Ciardi}, {Ford},
  {Gautier}, {Geary}, {Holman}, {Jenkins}, {Li}, {Morehead}, {Morris},
  {Shporer}, {Smith}, {Still}, \& {Van Cleve}}]{Fabrycky14}
{Fabrycky}, D.~C., {et~al.} 2014, \apj, 790, 146

\bibitem[{{Farrington} {et~al.}(2010){Farrington}, {ten Brummelaar}, {Mason},
  {Hartkopf}, {McAlister}, {Raghavan}, {Turner}, {Sturmann}, {Sturmann}, \&
  {Ridgway}}]{Farrington10}
{Farrington}, C.~D., {et~al.} 2010, \aj, 139, 2308

\bibitem[{{Fischer} {et~al.}(2016){Fischer}, {Anglada-Escude}, {Arriagada},
  {Baluev}, {Bean}, {Bouchy}, {Buchhave}, {Carroll}, {Chakraborty}, {Crepp},
  {Dawson}, {Diddams}, {Dumusque}, {Eastman}, {Endl}, {Figueira}, {Ford},
  {Foreman-Mackey}, {Fournier}, {F{\H u}r{\'e}sz}, {Gaudi}, {Gregory},
  {Grundahl}, {Hatzes}, {H{\'e}brard}, {Herrero}, {Hogg}, {Howard}, {Johnson},
  {Jorden}, {Jurgenson}, {Latham}, {Laughlin}, {Loredo}, {Lovis}, {Mahadevan},
  {McCracken}, {Pepe}, {Perez}, {Phillips}, {Plavchan}, {Prato}, {Quirrenbach},
  {Reiners}, {Robertson}, {Santos}, {Sawyer}, {Segransan}, {Sozzetti},
  {Steinmetz}, {Szentgyorgyi}, {Udry}, {Valenti}, {Wang}, {Wittenmyer}, \&
  {Wright}}]{Fischer16}
{Fischer}, D.~A., {et~al.} 2016, \pasp, 128, 066001

\bibitem[{{Ford}(2006)}]{Ford06}
{Ford}, E.~B. 2006, \apj, 642, 505

\bibitem[{{Fressin} {et~al.}(2013){Fressin}, {Torres}, {Charbonneau}, {Bryson},
  {Christiansen}, {Dressing}, {Jenkins}, {Walkowicz}, \& {Batalha}}]{Fressin13}
{Fressin}, F., {et~al.} 2013, \apj, 766, 81

\bibitem[{{Fuhrmann} {et~al.}(1998){Fuhrmann}, {Pfeiffer}, \&
  {Bernkopf}}]{Fuhrmann98}
{Fuhrmann}, K., {Pfeiffer}, M.~J., \& {Bernkopf}, J. 1998, \aap, 336, 942

\bibitem[{{Fulton} {et~al.}(2013){Fulton}, {Howard}, {Winn}, {Albrecht},
  {Marcy}, {Crepp}, {Bakos}, {Johnson}, {Hartman}, {Isaacson}, {Knutson}, \&
  {Zhao}}]{Fulton13}
{Fulton}, B.~J., {et~al.} 2013, \apj, 772, 80

\bibitem[{{Fulton} {et~al.}(2015){Fulton}, {Weiss}, {Sinukoff}, {Isaacson},
  {Howard}, {Marcy}, {Henry}, {Holden}, \& {Kibrick}}]{Fulton15a}
---. 2015, \apj, 805, 175

\bibitem[{{Gatewood} {et~al.}(2001){Gatewood}, {Han}, \& {Black}}]{Gatewood01}
{Gatewood}, G., {Han}, I., \& {Black}, D.~C. 2001, \apjl, 548, L61

\bibitem[{{Ghezzi} {et~al.}(2010){Ghezzi}, {Cunha}, {Smith}, {de Ara{\'u}jo},
  {Schuler}, \& {de la Reza}}]{Ghezzi10}
{Ghezzi}, L., {Cunha}, K., {Smith}, V.~V., {de Ara{\'u}jo}, F.~X., {Schuler},
  S.~C., \& {de la Reza}, R. 2010, \apj, 720, 1290

\bibitem[{{Gladman}(1993)}]{Gladman93}
{Gladman}, B. 1993, Icarus, 106, 247

\bibitem[{{Gonz{\'a}lez Hern{\'a}ndez} {et~al.}(2010){Gonz{\'a}lez
  Hern{\'a}ndez}, {Israelian}, {Santos}, {Sousa}, {Delgado-Mena}, {Neves}, \&
  {Udry}}]{Gonzalez10}
{Gonz{\'a}lez Hern{\'a}ndez}, J.~I., {Israelian}, G., {Santos}, N.~C., {Sousa},
  S., {Delgado-Mena}, E., {Neves}, V., \& {Udry}, S. 2010, \apj, 720, 1592

\bibitem[{{Gray} {et~al.}(2003){Gray}, {Corbally}, {Garrison}, {McFadden}, \&
  {Robinson}}]{Gray03}
{Gray}, R.~O., {Corbally}, C.~J., {Garrison}, R.~F., {McFadden}, M.~T., \&
  {Robinson}, P.~E. 2003, \aj, 126, 2048

\bibitem[{{Hall} {et~al.}(2009){Hall}, {Henry}, {Lockwood}, {Skiff}, \&
  {Saar}}]{Hall09}
{Hall}, J.~C., {Henry}, G.~W., {Lockwood}, G.~W., {Skiff}, B.~A., \& {Saar},
  S.~H. 2009, \aj, 138, 312

\bibitem[{{Han} {et~al.}(2014){Han}, {Wang}, {Wright}, {Feng}, {Zhao},
  {Fakhouri}, {Brown}, \& {Hancock}}]{Han14}
{Han}, E., {Wang}, S.~X., {Wright}, J.~T., {Feng}, Y.~K., {Zhao}, M.,
  {Fakhouri}, O., {Brown}, J.~I., \& {Hancock}, C. 2014, \pasp, 126, 827

\bibitem[{{Hansen} \& {Murray}(2012)}]{Hansen12}
{Hansen}, B.~M.~S., \& {Murray}, N. 2012, \apj, 751, 158

\bibitem[{{Henry}(1999)}]{Henry99}
{Henry}, G.~W. 1999, \pasp, 111, 845

\bibitem[{{Henry} {et~al.}(2013){Henry}, {Kane}, {Wang}, {Wright}, {Boyajian},
  {von Braun}, {Ciardi}, {Dragomir}, {Farrington}, {Fischer}, {Hinkel},
  {Howard}, {Jensen}, {Laughlin}, {Mahadevan}, \& {Pilyavsky}}]{Henry13}
{Henry}, G.~W., {et~al.} 2013, \apj, 768, 155

\bibitem[{{Horch} {et~al.}(2011){Horch}, {Gomez}, {Sherry}, {Howell}, {Ciardi},
  {Anderson}, \& {van Altena}}]{Horch11}
{Horch}, E.~P., {Gomez}, S.~C., {Sherry}, W.~H., {Howell}, S.~B., {Ciardi},
  D.~R., {Anderson}, L.~M., \& {van Altena}, W.~F. 2011, \aj, 141, 45

\bibitem[{{Horch} {et~al.}(2012){Horch}, {Howell}, {Everett}, \&
  {Ciardi}}]{Horch12}
{Horch}, E.~P., {Howell}, S.~B., {Everett}, M.~E., \& {Ciardi}, D.~R. 2012,
  \aj, 144, 165

\bibitem[{{Horch} {et~al.}(2009){Horch}, {Veillette}, {Baena Gall{\'e}},
  {Shah}, {O'Rielly}, \& {van Altena}}]{Horch09}
{Horch}, E.~P., {Veillette}, D.~R., {Baena Gall{\'e}}, R., {Shah}, S.~C.,
  {O'Rielly}, G.~V., \& {van Altena}, W.~F. 2009, \aj, 137, 5057

\bibitem[{{Howard} \& {Fulton}(2016)}]{Howard16}
{Howard}, A.~W., \& {Fulton}, B.~J. 2016, ArXiv e-prints

\bibitem[{{Howard} {et~al.}(2009){Howard}, {Johnson}, {Marcy}, {Fischer},
  {Wright}, {Henry}, {Giguere}, {Isaacson}, {Valenti}, {Anderson}, \&
  {Piskunov}}]{Howard09}
{Howard}, A.~W., {et~al.} 2009, \apj, 696, 75

\bibitem[{Howard {et~al.}(2010)Howard, Marcy, Johnson, Fischer, Wright,
  Isaacson, Valenti, Anderson, Lin, \& Ida}]{Howard10}
Howard, A.~W., {et~al.} 2010, Science, 330, 653

\bibitem[{{Howard} {et~al.}(2011{\natexlab{a}}){Howard}, {Johnson}, {Marcy},
  {Fischer}, {Wright}, {Henry}, {Isaacson}, {Valenti}, {Anderson}, \&
  {Piskunov}}]{Howard11a}
{Howard}, A.~W., {et~al.} 2011{\natexlab{a}}, \apj, 726, 73

\bibitem[{{Howard} {et~al.}(2011{\natexlab{b}}){Howard}, {Johnson}, {Marcy},
  {Fischer}, {Wright}, {Henry}, {Isaacson}, {Valenti}, {Anderson}, \&
  {Piskunov}}]{Howard11b}
---. 2011{\natexlab{b}}, \apj, 730, 10

\bibitem[{{Howard} {et~al.}(2012){Howard}, {Marcy}, {Bryson}, {Jenkins},
  {Rowe}, {Batalha}, {Borucki}, {Koch}, {Dunham}, {Gautier}, {Van Cleve},
  {Cochran}, {Latham}, {Lissauer}, {Torres}, {Brown}, {Gilliland}, {Buchhave},
  {Caldwell}, {Christensen-Dalsgaard}, {Ciardi}, {Fressin}, {Haas}, {Howell},
  {Kjeldsen}, {Seager}, {Rogers}, {Sasselov}, {Steffen}, {Basri},
  {Charbonneau}, {Christiansen}, {Clarke}, {Dupree}, {Fabrycky}, {Fischer},
  {Ford}, {Fortney}, {Tarter}, {Girouard}, {Holman}, {Johnson}, {Klaus},
  {Machalek}, {Moorhead}, {Morehead}, {Ragozzine}, {Tenenbaum}, {Twicken},
  {Quinn}, {Isaacson}, {Shporer}, {Lucas}, {Walkowicz}, {Welsh}, {Boss},
  {Devore}, {Gould}, {Smith}, {Morris}, {Prsa}, {Morton}, {Still}, {Thompson},
  {Mullally}, {Endl}, \& {MacQueen}}]{Howard12}
---. 2012, \apjs, 201, 15

\bibitem[{{Howard} {et~al.}(2014){Howard}, {Marcy}, {Fischer}, {Isaacson},
  {Muirhead}, {Henry}, {Boyajian}, {von Braun}, {Becker}, {Wright}, \&
  {Johnson}}]{Howard14}
---. 2014, \apj, 794, 51

\bibitem[{{Howell} {et~al.}(2011){Howell}, {Everett}, {Sherry}, {Horch}, \&
  {Ciardi}}]{Howell11}
{Howell}, S.~B., {Everett}, M.~E., {Sherry}, W., {Horch}, E., \& {Ciardi},
  D.~R. 2011, \aj, 142, 19

\bibitem[{{Huber} {et~al.}(2011){Huber}, {Bedding}, {Stello}, {Hekker},
  {Mathur}, {Mosser}, {Verner}, {Bonanno}, {Buzasi}, {Campante}, {Elsworth},
  {Hale}, {Kallinger}, {Silva Aguirre}, {Chaplin}, {De Ridder},
  {Garc{\'{\i}}a}, {Appourchaux}, {Frandsen}, {Houdek}, {Molenda-{\.Z}akowicz},
  {Monteiro}, {Christensen-Dalsgaard}, {Gilliland}, {Kawaler}, {Kjeldsen},
  {Broomhall}, {Corsaro}, {Salabert}, {Sanderfer}, {Seader}, \&
  {Smith}}]{huber2011}
{Huber}, D., {et~al.} 2011, \apj, 743, 143

\bibitem[{{Ida} \& {Lin}(2004)}]{Ida04}
{Ida}, S., \& {Lin}, D.~N.~C. 2004, \apj, 604, 388

\bibitem[{{Ida} \& {Lin}(2008)}]{Ida08}
---. 2008, \apj, 673, 487

\bibitem[{{Ireland} {et~al.}(2008){Ireland}, {M{\'e}rand}, {ten Brummelaar},
  {Tuthill}, {Schaefer}, {Turner}, {Sturmann}, {Sturmann}, \&
  {McAlister}}]{2008SPIE.7013E..24I}
{Ireland}, M.~J., {et~al.} 2008, in Society of Photo-Optical Instrumentation
  Engineers (SPIE) Conference Series, Vol. 7013, Society of Photo-Optical
  Instrumentation Engineers (SPIE) Conference Series, 24

\bibitem[{Isaacson \& Fischer(2010)}]{Isaacson10}
Isaacson, H., \& Fischer, D. 2010, The Astrophysical Journal, 725, 875

\bibitem[{{Jenkins} {et~al.}(2010){Jenkins}, {Chandrasekaran}, {McCauliff},
  {Caldwell}, {Tenenbaum}, {Li}, {Klaus}, {Cote}, \& {Middour}}]{Jenkins13}
{Jenkins}, J.~M., {et~al.} 2010, in \procspie, Vol. 7740, Software and
  Cyberinfrastructure for Astronomy, 77400D

\bibitem[{{Johnson} {et~al.}(2011){Johnson}, {Clanton}, {Howard}, {Bowler},
  {Henry}, {Marcy}, {Crepp}, {Endl}, {Cochran}, {MacQueen}, {Wright}, \&
  {Isaacson}}]{Johnson11}
{Johnson}, J.~A., {et~al.} 2011, \apjs, 197, 26

\bibitem[{{Kallinger} {et~al.}(2010){Kallinger}, {Weiss}, {Barban}, {Baudin},
  {Cameron}, {Carrier}, {De Ridder}, {Goupil}, {Gruberbauer}, {Hatzes},
  {Hekker}, {Samadi}, \& {Deleuil}}]{kallinger2010}
{Kallinger}, T., {et~al.} 2010, \aap, 509, A77

\bibitem[{{Kane} {et~al.}(2014){Kane}, {Howell}, {Horch}, {Feng}, {Hinkel},
  {Ciardi}, {Everett}, {Howard}, \& {Wright}}]{Kane14}
{Kane}, S.~R., {et~al.} 2014, \apj, 785, 93

\bibitem[{{Kjeldsen} \& {Bedding}(1995)}]{kjeldsen1995}
{Kjeldsen}, H., \& {Bedding}, T.~R. 1995, \aap, 293, 87

\bibitem[{{Knutson} {et~al.}(2014){Knutson}, {Fulton}, {Montet}, {Kao}, {Ngo},
  {Howard}, {Crepp}, {Hinkley}, {Bakos}, {Batygin}, {Johnson}, {Morton}, \&
  {Muirhead}}]{Knutson14}
{Knutson}, H.~A., {et~al.} 2014, \apj, 785, 126

\bibitem[{{Koen} {et~al.}(2010){Koen}, {Kilkenny}, {van Wyk}, \&
  {Marang}}]{Koen10}
{Koen}, C., {Kilkenny}, D., {van Wyk}, F., \& {Marang}, F. 2010, \mnras, 403,
  1949

\bibitem[{{Lee} \& {Chiang}(2015)}]{Lee15}
{Lee}, E.~J., \& {Chiang}, E. 2015, ArXiv e-prints

\bibitem[{{Lee} {et~al.}(2014){Lee}, {Chiang}, \& {Ormel}}]{Lee14}
{Lee}, E.~J., {Chiang}, E., \& {Ormel}, C.~W. 2014, \apj, 797, 95

\bibitem[{{Lockwood} {et~al.}(2013){Lockwood}, {Henry}, {Hall}, \&
  {Radick}}]{Lockwood13}
{Lockwood}, G.~W., {Henry}, G.~W., {Hall}, J.~C., \& {Radick}, R.~R. 2013, in
  Astronomical Society of the Pacific Conference Series, Vol. 472, New Quests
  in Stellar Astrophysics III: A Panchromatic View of Solar-Like Stars, With
  and Without Planets, ed. M.~{Chavez}, E.~{Bertone}, O.~{Vega}, \& V.~{De la
  Luz}, 203

\bibitem[{{Lomb}(1976)}]{Lomb76}
{Lomb}, N.~R. 1976, \apss, 39, 447

\bibitem[{{Lopez}(2014)}]{Lopez14}
{Lopez}, E.~D. 2014, PhD thesis, University of California, Santa Cruz

\bibitem[{{Lopez} \& {Fortney}(2014)}]{Lopez13}
{Lopez}, E.~D., \& {Fortney}, J.~J. 2014, \apj, 792, 1

\bibitem[{{Mann} \& {von Braun}(2015)}]{2015PASP..127..102M}
{Mann}, A.~W., \& {von Braun}, K. 2015, \pasp, 127, 102

\bibitem[{{Mayor} {et~al.}(2011){Mayor}, {Marmier}, {Lovis}, {Udry},
  {S{\'e}gransan}, {Pepe}, {Benz}, {Bertaux}, {Bouchy}, {Dumusque}, {Lo Curto},
  {Mordasini}, {Queloz}, \& {Santos}}]{Mayor11}
{Mayor}, M., {et~al.} 2011, arXiv:1109.2497

\bibitem[{{Medhi} {et~al.}(2007){Medhi}, {Messina}, {Parihar}, {Pagano},
  {Muneer}, \& {Duorah}}]{Medhi07}
{Medhi}, B.~J., {Messina}, S., {Parihar}, P.~S., {Pagano}, I., {Muneer}, S., \&
  {Duorah}, K. 2007, \aap, 469, 713

\bibitem[{{Mel{\'e}ndez} {et~al.}(2009){Mel{\'e}ndez}, {Asplund}, {Gustafsson},
  \& {Yong}}]{Melendez09}
{Mel{\'e}ndez}, J., {Asplund}, M., {Gustafsson}, B., \& {Yong}, D. 2009, \apjl,
  704, L66

\bibitem[{{Michel} {et~al.}(2008){Michel}, {Baglin}, {Auvergne}, {Catala},
  {Samadi}, {Baudin}, {Appourchaux}, {Barban}, {Weiss}, {Berthomieu},
  {Boumier}, {Dupret}, {Garcia}, {Fridlund}, {Garrido}, {Goupil}, {Kjeldsen},
  {Lebreton}, {Mosser}, {Grotsch-Noels}, {Janot-Pacheco}, {Provost},
  {Roxburgh}, {Thoul}, {Toutain}, {Tiph{\`e}ne}, {Turck-Chieze}, {Vauclair},
  {Vauclair}, {Aerts}, {Alecian}, {Ballot}, {Charpinet}, {Hubert},
  {Ligni{\`e}res}, {Mathias}, {Monteiro}, {Neiner}, {Poretti}, {Renan de
  Medeiros}, {Ribas}, {Rieutord}, {Cort{\'e}s}, \& {Zwintz}}]{Michel08}
{Michel}, E., {et~al.} 2008, Science, 322, 558

\bibitem[{{Mordasini} {et~al.}(2009){Mordasini}, {Alibert}, \&
  {Benz}}]{Mordasini09}
{Mordasini}, C., {Alibert}, Y., \& {Benz}, W. 2009, \aap, 501, 1139

\bibitem[{{Morel} {et~al.}(2013{\natexlab{a}}){Morel}, {Rainer}, {Poretti},
  {Barban}, \& {Boumier}}]{Morel13}
{Morel}, T., {Rainer}, M., {Poretti}, E., {Barban}, C., \& {Boumier}, P.
  2013{\natexlab{a}}, \aap, 552, A42

\bibitem[{{Morel} {et~al.}(2013{\natexlab{b}}){Morel}, {Rainer}, {Poretti},
  {Barban}, \& {Boumier}}]{morel2013}
---. 2013{\natexlab{b}}, \aap, 552, A42

\bibitem[{{Morton}(2015)}]{Morton15}
{Morton}, T.~D. 2015, {isochrones: Stellar model grid package}, Astrophysics
  Source Code Library

\bibitem[{{Motalebi} {et~al.}(2015){Motalebi}, {Udry}, {Gillon}, {Lovis},
  {S{\'e}gransan}, {Buchhave}, {Demory}, {Malavolta}, {Dressing}, {Sasselov},
  {Rice}, {Charbonneau}, {Collier Cameron}, {Latham}, {Molinari}, {Pepe},
  {Affer}, {Bonomo}, {Cosentino}, {Dumusque}, {Figueira}, {Fiorenzano},
  {Gettel}, {Harutyunyan}, {Haywood}, {Johnson}, {Lopez}, {Lopez-Morales},
  {Mayor}, {Micela}, {Mortier}, {Nascimbeni}, {Philips}, {Piotto}, {Pollacco},
  {Queloz}, {Sozzetti}, {Vanderburg}, \& {Watson}}]{Motalebi15}
{Motalebi}, F., {et~al.} 2015, \aap, 584, A72

\bibitem[{{Noyes} {et~al.}(1997){Noyes}, {Jha}, {Korzennik}, {Krockenberger},
  {Nisenson}, {Brown}, {Kennelly}, \& {Horner}}]{Noyes97}
{Noyes}, R.~W., {Jha}, S., {Korzennik}, S.~G., {Krockenberger}, M., {Nisenson},
  P., {Brown}, T.~M., {Kennelly}, E.~J., \& {Horner}, S.~D. 1997, \apjl, 483,
  L111

\bibitem[{{O'Toole} {et~al.}(2009){O'Toole}, {Jones}, {Tinney}, {Butler},
  {Marcy}, {Carter}, {Bailey}, \& {Wittenmyer}}]{Otoole09}
{O'Toole}, S.~J., {Jones}, H.~R.~A., {Tinney}, C.~G., {Butler}, R.~P., {Marcy},
  G.~W., {Carter}, B., {Bailey}, J., \& {Wittenmyer}, R.~A. 2009, \apj, 701,
  1732

\bibitem[{{Pepe} {et~al.}(2004){Pepe}, {Mayor}, {Queloz}, {Benz}, {Bonfils},
  {Bouchy}, {Lo Curto}, {Lovis}, {M{\'e}gevand}, {Moutou}, {Naef}, {Rupprecht},
  {Santos}, {Sivan}, {Sosnowska}, \& {Udry}}]{Pepe04}
{Pepe}, F., {et~al.} 2004, \aap, 423, 385

\bibitem[{{Petigura}(2015)}]{Petigura15}
{Petigura}, E.~A. 2015, arXiv:1510.03902

\bibitem[{{Petigura} {et~al.}(2013{\natexlab{a}}){Petigura}, {Howard}, \&
  {Marcy}}]{Petigura13b}
{Petigura}, E.~A., {Howard}, A.~W., \& {Marcy}, G.~W. 2013{\natexlab{a}},
  Proceedings of the National Academy of Science, 110, 19273

\bibitem[{{Petigura} {et~al.}(2013{\natexlab{b}}){Petigura}, {Marcy}, \&
  {Howard}}]{Petigura13a}
{Petigura}, E.~A., {Marcy}, G.~W., \& {Howard}, A.~W. 2013{\natexlab{b}}, \apj,
  770, 69

\bibitem[{{Pickles}(1998)}]{1998PASP..110..863P}
{Pickles}, A.~J. 1998, \pasp, 110, 863

\bibitem[{{Queloz} {et~al.}(2001){Queloz}, {Henry}, {Sivan}, {Baliunas},
  {Beuzit}, {Donahue}, {Mayor}, {Naef}, {Perrier}, \& {Udry}}]{Queloz01}
{Queloz}, D., {et~al.} 2001, \aap, 379, 279

\bibitem[{{Radovan} {et~al.}(2014){Radovan}, {Lanclos}, {Holden}, {Kibrick},
  {Allen}, {Deich}, {Rivera}, {Burt}, {Fulton}, {Butler}, \&
  {Vogt}}]{Radovan14}
{Radovan}, M.~V., {et~al.} 2014, in Society of Photo-Optical Instrumentation
  Engineers (SPIE) Conference Series, Vol. 9145, Society of Photo-Optical
  Instrumentation Engineers (SPIE) Conference Series, 2

\bibitem[{{Raghavan} {et~al.}(2012){Raghavan}, {Farrington}, {ten Brummelaar},
  {McAlister}, {Ridgway}, {Sturmann}, {Sturmann}, \& {Turner}}]{Raghavan12}
{Raghavan}, D., {Farrington}, C.~D., {ten Brummelaar}, T.~A., {McAlister},
  H.~A., {Ridgway}, S.~T., {Sturmann}, L., {Sturmann}, J., \& {Turner}, N.~H.
  2012, \apj, 745, 24

\bibitem[{{Ram{\'{\i}}rez} {et~al.}(2012){Ram{\'{\i}}rez}, {Fish}, {Lambert},
  \& {Allende Prieto}}]{Ramirez12}
{Ram{\'{\i}}rez}, I., {Fish}, J.~R., {Lambert}, D.~L., \& {Allende Prieto}, C.
  2012, \apj, 756, 46

\bibitem[{{Ram{\'{\i}}rez} {et~al.}(2014){Ram{\'{\i}}rez}, {Mel{\'e}ndez},
  {Bean}, {Asplund}, {Bedell}, {Monroe}, {Casagrande}, {Schirbel}, {Dreizler},
  {Teske}, {Tucci Maia}, {Alves-Brito}, \& {Baumann}}]{Ramirez14}
{Ram{\'{\i}}rez}, I., {et~al.} 2014, \aap, 572, A48

\bibitem[{{Reffert} \& {Quirrenbach}(2011)}]{Reffert11}
{Reffert}, S., \& {Quirrenbach}, A. 2011, \aap, 527, A140

\bibitem[{{Santos} {et~al.}(2013){Santos}, {Sousa}, {Mortier}, {Neves},
  {Adibekyan}, {Tsantaki}, {Delgado Mena}, {Bonfils}, {Israelian}, {Mayor}, \&
  {Udry}}]{Santos13}
{Santos}, N.~C., {et~al.} 2013, \aap, 556, A150

\bibitem[{{Scargle}(1982)}]{Scargle82}
{Scargle}, J.~D. 1982, \apj, 263, 835

\bibitem[{{Schwarzenberg-Czerny}(1998)}]{Schwarzenberg-Czerny98}
{Schwarzenberg-Czerny}, A. 1998, \mnras, 301, 831

\bibitem[{Spearman(1904)}]{Spearman1904}
Spearman, C. 1904, The American Journal of Psychology, 15, pp. 72

\bibitem[{{Swift} {et~al.}(2015){Swift}, {Bottom}, {Johnson}, {Wright},
  {McCrady}, {Wittenmyer}, {Plavchan}, {Riddle}, {Muirhead}, {Herzig}, {Myles},
  {Blake}, {Eastman}, {Beatty}, {Barnes}, {Gibson}, {Lin}, {Zhao}, {Gardner},
  {Falco}, {Criswell}, {Nava}, {Robinson}, {Sliski}, {Hedrick}, {Ivarsen},
  {Hjelstrom}, {de Vera}, \& {Szentgyorgyi}}]{Swift15}
{Swift}, J.~J., {et~al.} 2015, Journal of Astronomical Telescopes, Instruments,
  and Systems, 1, 027002

\bibitem[{{ten Brummelaar} {et~al.}(2005){ten Brummelaar}, {McAlister},
  {Ridgway}, {Bagnuolo}, {Turner}, {Sturmann}, {Sturmann}, {Berger}, {Ogden},
  {Cadman}, {Hartkopf}, {Hopper}, \& {Shure}}]{2005ApJ...628..453T}
{ten Brummelaar}, T.~A., {et~al.} 2005, \apj, 628, 453

\bibitem[{Ter~Braak(2006)}]{Braak06}
Ter~Braak, C. 2006, Statistics and Computing

\bibitem[{{Torres} {et~al.}(2010){Torres}, {Andersen}, \&
  {Gim{\'e}nez}}]{Torres10}
{Torres}, G., {Andersen}, J., \& {Gim{\'e}nez}, A. 2010, \aapr, 18, 67

\bibitem[{{Valenti} \& {Fischer}(2005)}]{Valenti05}
{Valenti}, J.~A., \& {Fischer}, D.~A. 2005, \apjs, 159, 141

\bibitem[{{van Belle} \& {von Braun}(2009)}]{vanBelle09}
{van Belle}, G.~T., \& {von Braun}, K. 2009, \apj, 694, 1085

\bibitem[{{van Leeuwen}(2007)}]{vanLeeuwen07}
{van Leeuwen}, F. 2007, \aap, 474, 653

\bibitem[{{Veras} \& {Ford}(2012)}]{Veras12}
{Veras}, D., \& {Ford}, E.~B. 2012, \mnras, 420, L23

\bibitem[{{Vogt} {et~al.}(1994){Vogt}, {Allen}, {Bigelow}, {Bresee}, {Brown},
  {Cantrall}, {Conrad}, {Couture}, {Delaney}, {Epps}, {Hilyard}, {Hilyard},
  {Horn}, {Jern}, {Kanto}, {Keane}, {Kibrick}, {Lewis}, {Osborne},
  {Pardeilhan}, {Pfister}, {Ricketts}, {Robinson}, {Stover}, {Tucker}, {Ward},
  \& {Wei}}]{Vogt94}
{Vogt}, S.~S., {et~al.} 1994, in Proc. SPIE Instrumentation in Astronomy VIII,
  David L. Crawford; Eric R. Craine; Eds., 2198, 362

\bibitem[{{Vogt} {et~al.}(2014){Vogt}, {Radovan}, {Kibrick}, {Butler},
  {Alcott}, {Allen}, {Arriagada}, {Bolte}, {Burt}, {Cabak}, {Chloros},
  {Cowley}, {Deich}, {Dupraw}, {Earthman}, {Epps}, {Faber}, {Fischer}, {Gates},
  {Hilyard}, {Holden}, {Johnston}, {Keiser}, {Kanto}, {Katsuki}, {Laiterman},
  {Lanclos}, {Laughlin}, {Lewis}, {Lockwood}, {Lynam}, {Marcy}, {McLean},
  {Miller}, {Misch}, {Peck}, {Pfister}, {Phillips}, {Rivera}, {Sandford},
  {Saylor}, {Stover}, {Thompson}, {Walp}, {Ward}, {Wareham}, {Wei}, \&
  {Wright}}]{Vogt14}
{Vogt}, S.~S., {et~al.} 2014, \pasp, 126, 359

\bibitem[{{von Braun} {et~al.}(2014){von Braun}, {Boyajian}, {van Belle},
  {Kane}, {Jones}, {Farrington}, {Schaefer}, {Vargas}, {Scott}, {ten
  Brummelaar}, {Kephart}, {Gies}, {Ciardi}, {L{\'o}pez-Morales}, {Mazingue},
  {McAlister}, {Ridgway}, {Goldfinger}, {Turner}, \& {Sturmann}}]{vonBraun14}
{von Braun}, K., {et~al.} 2014, \mnras, 438, 2413

\bibitem[{{Weiss} \& {Marcy}(2014)}]{Weiss14}
{Weiss}, L.~M., \& {Marcy}, G.~W. 2014, \apjl, 783, L6

\bibitem[{{White} {et~al.}(2013){White}, {Huber}, {Maestro}, {Bedding},
  {Ireland}, {Baron}, {Boyajian}, {Che}, {Monnier}, {Pope}, {Roettenbacher},
  {Stello}, {Tuthill}, {Farrington}, {Goldfinger}, {McAlister}, {Schaefer},
  {Sturmann}, {Sturmann}, {ten Brummelaar}, \& {Turner}}]{2013MNRAS.433.1262W}
{White}, T.~R., {et~al.} 2013, \mnras, 433, 1262

\bibitem[{{Wright} \& {Eastman}(2014)}]{Wright14}
{Wright}, J.~T., \& {Eastman}, J.~D. 2014, \pasp, 126, 838

\bibitem[{{Wright} {et~al.}(2007){Wright}, {Marcy}, {Fischer}, {Butler},
  {Vogt}, {Tinney}, {Jones}, {Carter}, {Johnson}, {McCarthy}, \&
  {Apps}}]{Wright07}
{Wright}, J.~T., {et~al.} 2007, \apj, 657, 533

\bibitem[{{Zucker} \& {Mazeh}(2001)}]{Zucker01}
{Zucker}, S., \& {Mazeh}, T. 2001, arXiv, astro-ph/0104098

\end{thebibliography}

\enddocument